\documentclass[aps,rmp,groupedaddress,showpacs,showkeys]{revtex4}

\usepackage{amsmath,amssymb,amsbsy}
\usepackage{graphicx}
\usepackage{color}

\newcommand{\cor}[1]{#1}
\newcommand{\del}[1]{}
\newcommand{\red}[1]{#1}

\newcommand{\fig}[1]{fig.~#1}

\newcommand{\smeson}{$\sigma$-meson }
\newcommand{\pc}{$\pi^{\pm}\pi^0$ }
\newcommand{\pn}{$\pi^0\pi^0$ }
\newcommand{\BG}{\text{\begin{tiny}BG\end{tiny}}}
\newcommand{\M}{\mathcal{M}}

\newcommand{\MeV}{\; \mathrm{MeV} }
\newcommand{\GeV}{\; \mathrm{GeV} }

\newcommand{\BUU}{GiBUU }

\newcommand{\fullBUU}{Giessen BUU (\BUU\hspace{-4pt}) }

\begin{document}


\title{Low-energy pions in nuclear matter and \boldmath{$\pi \pi$} photoproduction within a BUU transport model
\footnote[6]{Work supported by DFG}
} 


\author{Oliver Buss}
\email{oliver.buss@theo.physik.uni-giessen.de}
\homepage{http://theorie.physik.uni-giessen.de/~oliver}
\thanks{Work supported by DFG.}
\author{Luis Alvarez-Ruso}
\author{Pascal M\"uhlich}
\author{Ulrich Mosel}
\affiliation{Institut f\"ur Theoretische Physik, Universit\"at Giessen, Germany}

\date{\today}
\begin{abstract}
A description of low-energy scattering of pions and nuclei within a BUU transport model is presented. 
Implementing different scenarios of medium modifications, the mean free path of pions in nuclear matter at low momenta and pion absorption reactions on nuclei have been studied and compared to data and to results obtained via quantum mechanical scattering theory. We show that even in a regime of a long pionic wave length the semi-classical \cor{transport model is still a reliable framework} for pion kinetic energies greater than $\approx20-30 \MeV$. Results are presented on $\pi$-absorption cross sections in the regime of $10 \MeV\leq T^{\pi}_{kin}\leq 130 \MeV$ and on photon-induced $\pi \pi$ production at incident beam energies of $400-500 \MeV$.
\end{abstract}
\maketitle

\section{Introduction and Motivation}\label{intro}
According to our present knowledge, the properties of hadrons and their interactions can be described in terms of Quantum Chromo Dynamics (QCD). In the limit of vanishing quark masses QCD incorporates chiral symmetry, which is spontaneously broken in vacuum. The order parameter $\langle \bar{q}q\rangle$ of this symmetry breaking is expected to decrease by about 30\% already at normal nuclear matter density~\cite{Drukarev:1988kd,Cohen:1991nk,Brockmann:1996iv}. Therefore, signals for partial chiral symmetry restoration should be observable in nuclear reaction experiments and, in particular, photon induced processes are highly suited due to two key \cor{reasons}. The reaction leaves the nucleus close to its ground state, so the reaction takes place under well defined conditions. And as a second point, the photon penetrates deeply into the nucleus, giving rise to a high effective density.

The modification of the so-called $\sigma$ or $f_0(600)$ meson inside the nuclear medium was proposed \cor{as a} signal for such a partial symmetry restoration. Theoretical models predict a shift of its spectral strength to lower masses and a more narrow width due to the onset of the restoration~\cite{Bernard:1987im,Hatsuda:1999kd}. The \smeson is a very short-lived state with a width of roughly $600-1000 \MeV$~\cite{PDGdata}, decaying predominantly into a $\pi \pi$ final state in $S$-wave. Therefore, the experimental aim has been to find \cor{modifications of this state} in $\pi\pi$ production reactions in finite nuclear systems close to threshold. 

Such experiments have been performed with incident pions by the CHAOS collaboration~\cite{Bonutti1,Bonutti2} and with photons by the TAPS collaboration~\cite{Messch,SchadmandHabil,Schadmand:2005ji,Schadmand:2005xy}. Both experiments have shown an accumulation of strength near the $\pi \pi$ threshold in the decay channel of the $\sigma$ in large nuclei. A possible interpretation of this effect is the in-medium modification of this resonance due to partial symmetry restoration. Already in an earlier work\,\cite{Muhlich:2004zj}, \cor{we have, however,} pointed out the importance of conventional final-state effects in the analysis \red{of} the experiment performed by the TAPS collaboration. 

To analyse the experiment performed by the TAPS collaboration\,\cite{Messch,SchadmandHabil,Schadmand:2005ji,Schadmand:2005xy}, we need to consider pions with very low energy in nuclear matter. Aiming to employ a BUU transport model, we first need to establish that it is meaningful to discuss pions with a long wave length in such a semi-classical picture. Actually, this question can only be answered by comparison to quantum calculations and experimental data. Therefore, in this paper we investigate the validity of our model for the treatment of the $\pi \pi$ final state with the description of pion-induced scattering and absorption experiments within the BUU-framework. 

Already in earlier works of Salcedo et al.~\cite{osetSimulation}, with a simulation of pion propagation in nuclear matter, and of Engel et al.~\cite{Engel:1993jh}, with a precursor of our present simulation,  pions with kinetic energies in the regime of $85 - 300 \MeV$ have been investigated in transport models. As motivated above, we now investigate even less energetic pions. \cor{Therefore, we have} carefully accounted for Coulomb corrections in the initial channel of $\pi$-induced processes and improved \cor{the description} of the threshold behavior of the cross sections in the model.

This article is structured in the following way. First we introduce our \fullBUU transport model for the treatment of photon- and pion-induced nuclear reactions emphasizing the most relevant issues. Next, we present results on the pion mean-free-path in nuclear matter and discuss the consequences of medium modifications of the pion. Hereafter we compare our simulation to experimental results on pion scattering off complex nuclei to validate our model. Finally, we conclude with results concerning the $\pi\pi$ photoproduction at incident beam energies of $400-500 \MeV$.

\section{The \BUU transport model}\label{buu}
Boltzmann-Uehling-Uhlenbeck (BUU) transport models are based on the Boltzmann equation, which was modified in the 1930's  by Nordheim, Uehling and Uhlenbeck to incorporate quantum statistics. It is of semi-classical nature, but can be derived as a gradient-expansion of the fully quantum Kadanoff-Baym-equations\,\cite{kadanoffBaym}. Nowadays this model is widely used in heavy-ion collisions, electro- and pion-induced reactions to describe these processes in a coupled-channel approach. 

The Giessen BUU-model has been utilized in various applications including heavy-ion-collisions, photon-, electron-, pion- and neutrino-induced processes. In contrast to earlier works\,\cite{Effenberger:1996rc,Teis:1996kx,Muhlich:2004zj} we are now using a new implementation, named GiBUU~\cite{GiBUUWebpage}. It is in its algorithms based upon the old implementation, therefore also reproduces the former results. For a detailed discussion concerning the physical input and the algorithms we refer the reader to\,\cite{diplom,Effenberger:1996rc,Teis:1996kx} and the references therein.
For the reader interested in technical details, we note that the new code is now written in modular manner in modern Fortran\,\cite{F2003} utilizing Subversion\,\cite{subversion} version control in a multi-user environment. It represents a major leap forward to a simplified handling of the transport code and a more transparent implementation of future projects.

\subsection{The BUU equation}\label{buuEQ}
The BUU equation actually consists of a series of coupled differential equations, which describe the time evolution of the one-particle phase-space-densities $f^a_1(\vec{p},\vec{r},t)$ in the limit of low particle densities. The index $a=\pi,\omega,N,\Delta,\ldots$ denotes the different particle species in our model. 

For a particle of species X with negligible width, its time evolution is given by
\begin{eqnarray}
\frac{d f^X_{1}(\vec{r},\vec{p},t)}{d t}&=&\frac{\partial f^X_{1}(\vec{r},\vec{p},t)}{\partial t} +\frac{\partial H_X}{\partial \vec{p}}\frac{\partial f^X_{1}(\vec{r},\vec{p},t)}{\partial \vec{r}} 
-\frac{\partial H_X}{\partial \vec{r}}\frac{\partial f^{X}_1(\vec{r},\vec{p},t)}{\partial \vec{p}}  =I_{coll}\left(f^X_{1},f^a_{1},f^b_{1},\ldots\right)
\label{BUUEquation}
\end{eqnarray}
with the one-body Hamilton function 
\begin{eqnarray}
H_X(\vec{r},\vec{p})&=&\sqrt{\left(\vec{p}+\vec{A}_X(\vec{r},\vec{p})\right)^{2}+m_X^{2}+U_X(\vec{r},\vec{p})}\nonumber +A_X^{0}(\vec{r},\vec{p}) \; .
\label{HamiltonFunc}
\end{eqnarray}
The scalar potential $U_X$ and the vector potential $A^\mu_X$ of species X may in principle depend upon the phase space densities of all other species. Hence, the differential equations are already coupled through the mean fields. In the limit of $I_{coll}=0$, equation (\ref{BUUEquation}) becomes the well-known Vlasov equation. The collision term $I_{coll}\left(\left\lbrace f^a_{1}, a=\pi,\omega,N,\Delta,\ldots  \right\rbrace \right)$ on the right-hand side of equation (\ref{BUUEquation}) incorporates explicitly all scattering processes among the particles. The  reaction probabilities used in this collision term are chosen to match the elementary collisions among the particles in vacuum. 
Within the BUU framework the total reaction cross sections are given by an incoherent sum over all partial cross sections. Interference effects are therefore neglected.

\subsection{Elementary processes}
The elementary processes of interest in the present study are $\pi N \rightarrow \pi N$ and $\pi N N \leftrightarrow N N$ reactions. Both are modeled within a resonance picture. Therefore, we express the relevant cross sections as an incoherent sum of resonance and background contributions. 
\begin{eqnarray}
\sigma_{\pi N \rightarrow \pi N}&=&\sigma_{\pi N \rightarrow R \rightarrow \pi N}+\sigma^{\BG}_{\pi N \rightarrow \pi N} \nonumber \\
\sigma_{N N \rightarrow N N \pi}&=&\sigma_{N N \rightarrow R N \rightarrow N N \pi}+\sigma^{\BG}_{N N\rightarrow N N \pi}
\label{backDef}
\end{eqnarray}
The resonance cross sections are obtained from the partial wave analysis of ref.~\cite{ManleySaleski}. The background cross sections denoted by $\sigma^{\BG}$ are chosen in such a manner, that the elementary cross section data in the vacuum are reproduced~\cite{diplom}. Within our model, background contributions are instantaneous in space-time, whereas reactions through resonance channels are delayed by the lifetime of the resonance. 

\begin{figure}[ht!]
\centering
\includegraphics[angle=0,width=8cm]{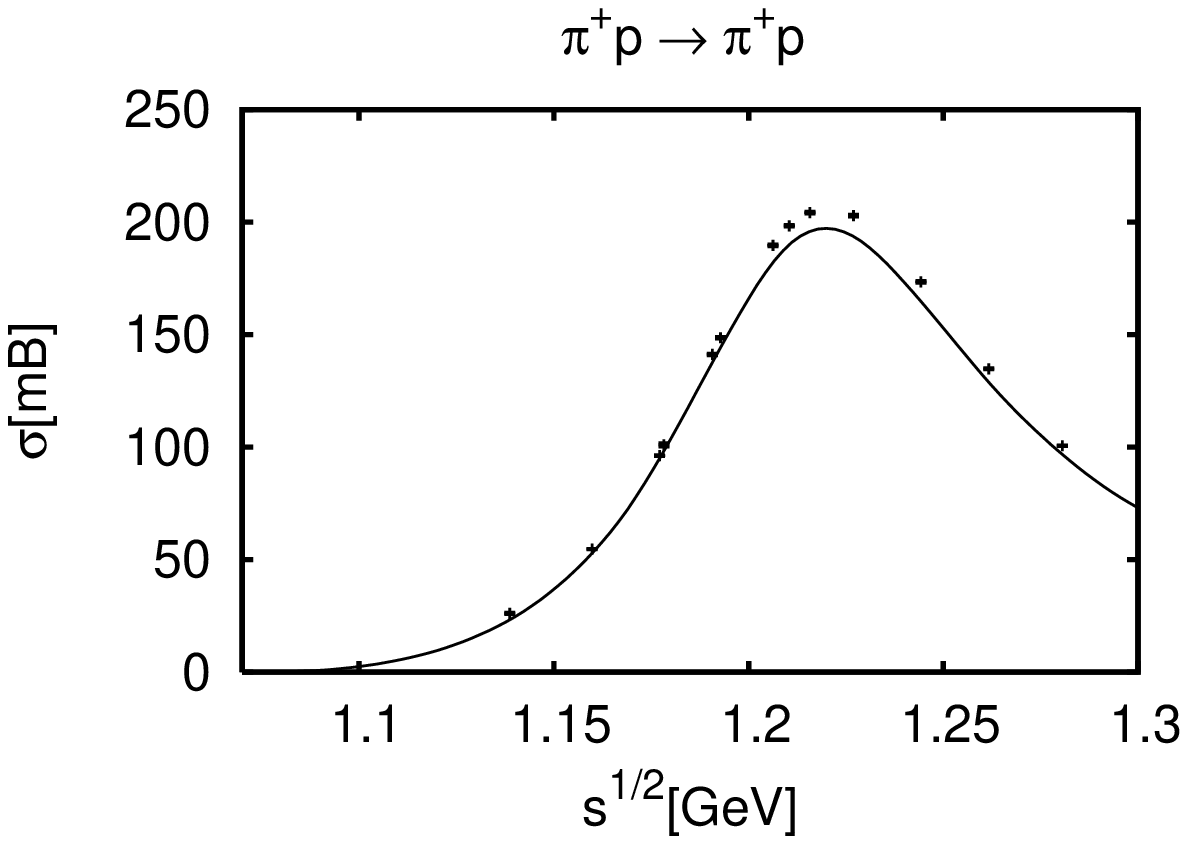}
\includegraphics[angle=0,width=8cm]{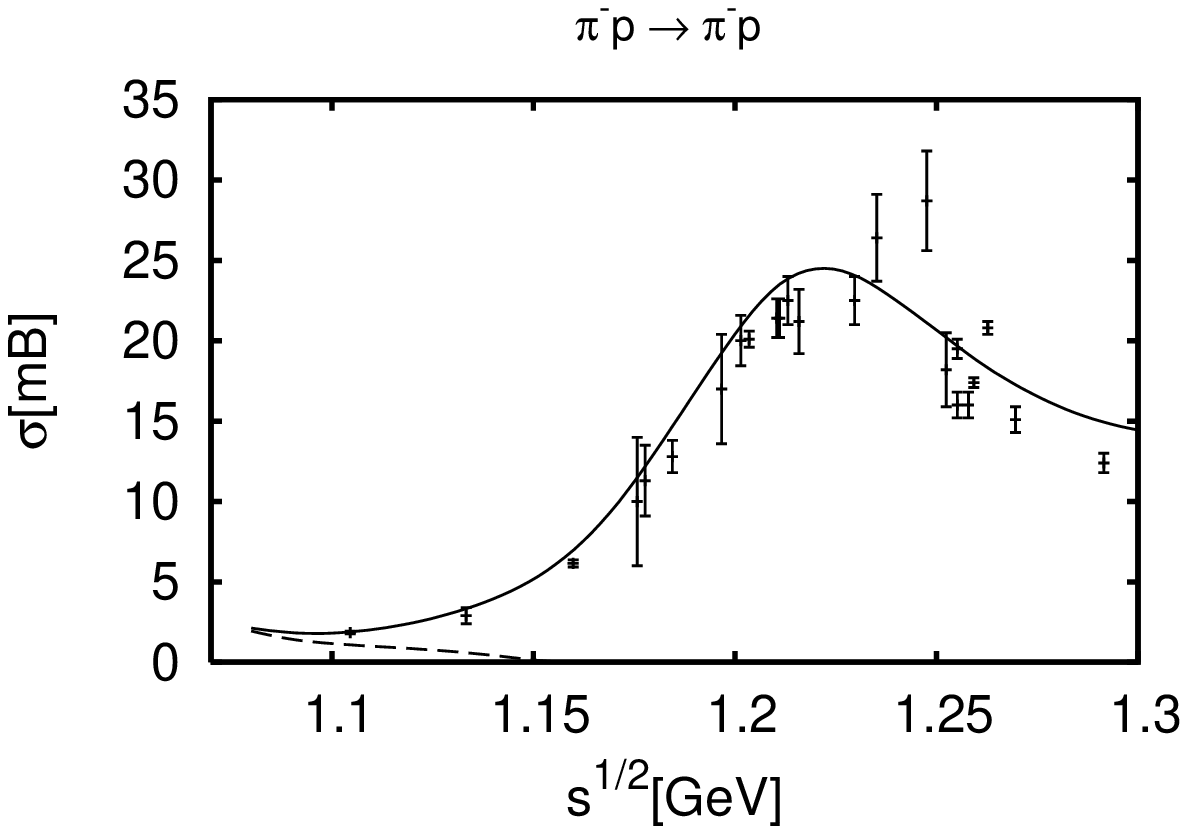}
\includegraphics[angle=0,width=8cm]{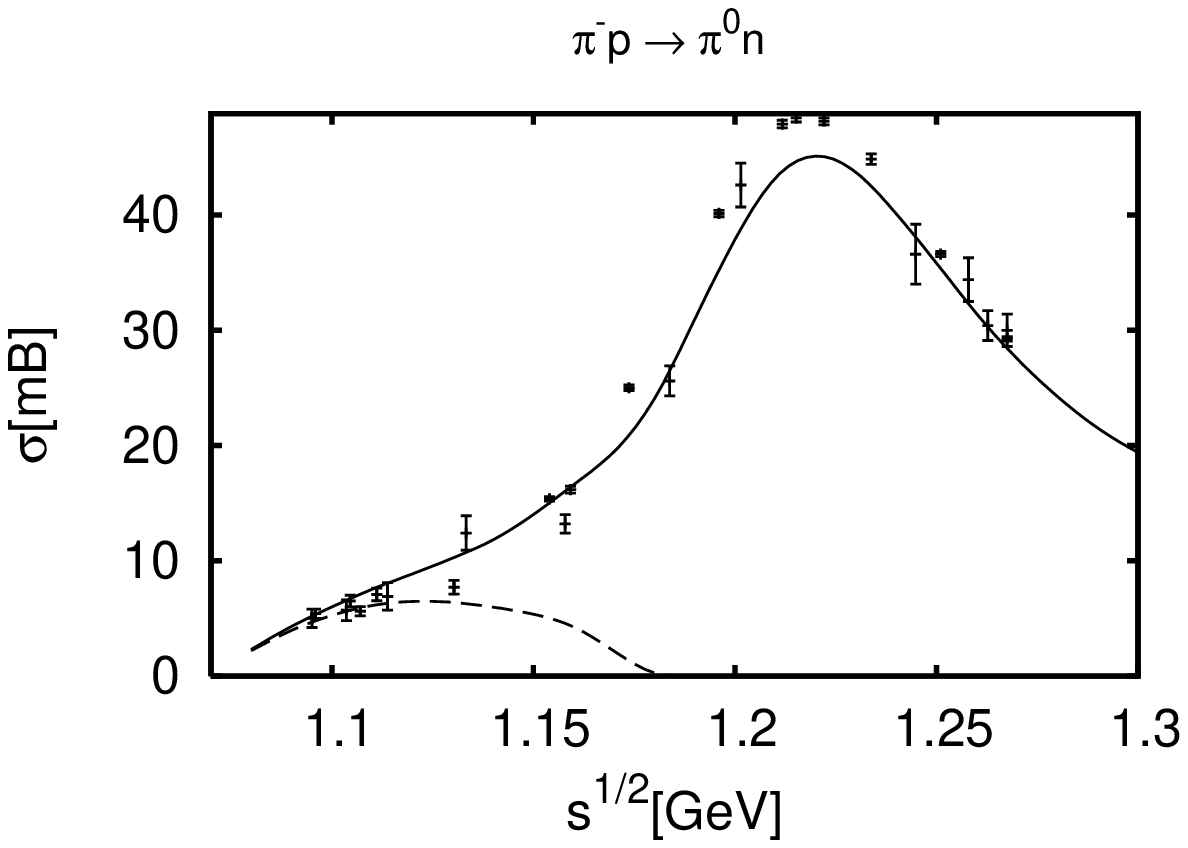}
\caption{Cross sections for \red{elastic scattering and charge exchange reactions} in pion-nucleon scattering. The solid lines show the full cross 
sections, whereas the dashed lines represent the background contributions. Data are taken from \cite{Carter:1971tj,Landoldt}.}
\label{backfig}
\end{figure}

All different events in pion nucleon scattering can be categorized into four different isospin channels
\begin{eqnarray}
\sigma_{\pi^{-} n \rightarrow \pi^{-} n}=\sigma_{\pi^{+} p \rightarrow \pi^{+} p} \; ,\label{c1}\\
\sigma_{\pi^{-} p \rightarrow \pi^{0} n}=\sigma_{\pi^{0} n \rightarrow \pi^{-} p}
=\sigma_{\pi^{+} n \rightarrow \pi^{0} p}=\sigma_{\pi^{0} p \rightarrow \pi^{+} n} \; , \label{c2}\\
\sigma_{\pi^{-} p \rightarrow \pi^{-}p}=\sigma_{\pi^{+} n \rightarrow \pi^{+} n} \; , \label{c3}\\
\sigma_{\pi^{0} n \rightarrow \pi^{0} n}=\sigma_{\pi^{0} p \rightarrow \pi^{0} p} \; . \label{c4}
\end{eqnarray}
The cross sections in the individual channels are either connected by time reversal or isospin symmetry. The first channel (eq. \ref{c1}) is a pure isospin $I=3/2$ scattering process, whereas the other three channels are mixtures of $I=1/2$ and $I=3/2$. The cross section for the $I=3/2$ channel $\sigma_{\pi \ N \rightarrow \Delta \rightarrow \pi \ N}$ is given explicitly in \cite{effeDoktor} based on the resonance analysis by Manley and Saleski \cite{ManleySaleski}. As can be seen in the upper left panel of \fig{\ref{backfig}}, the first channel is well described solely by the $\Delta$-resonance scattering; therefore, we neglect a background term in this channel. In the second and third channels (eq. \ref{c2} and \ref{c3}) there are good data sets for 
\begin{eqnarray*}
\pi^{-} p \rightarrow \pi^{-} \ p \\
\pi^{-} p \rightarrow \pi^{0} \ n.
\end{eqnarray*}
down to very low energies. Hence, we introduce a background term on top of our resonance contributions for a better description of those channels as shown in the remaining panels of \fig{\ref{backfig}}. The last channel (eq. \ref{c4}) is inaccessible for experiment, therefore we can not introduce any background term. 

\begin{figure}[ht!]
\centering
\includegraphics[angle=0,width=8cm]{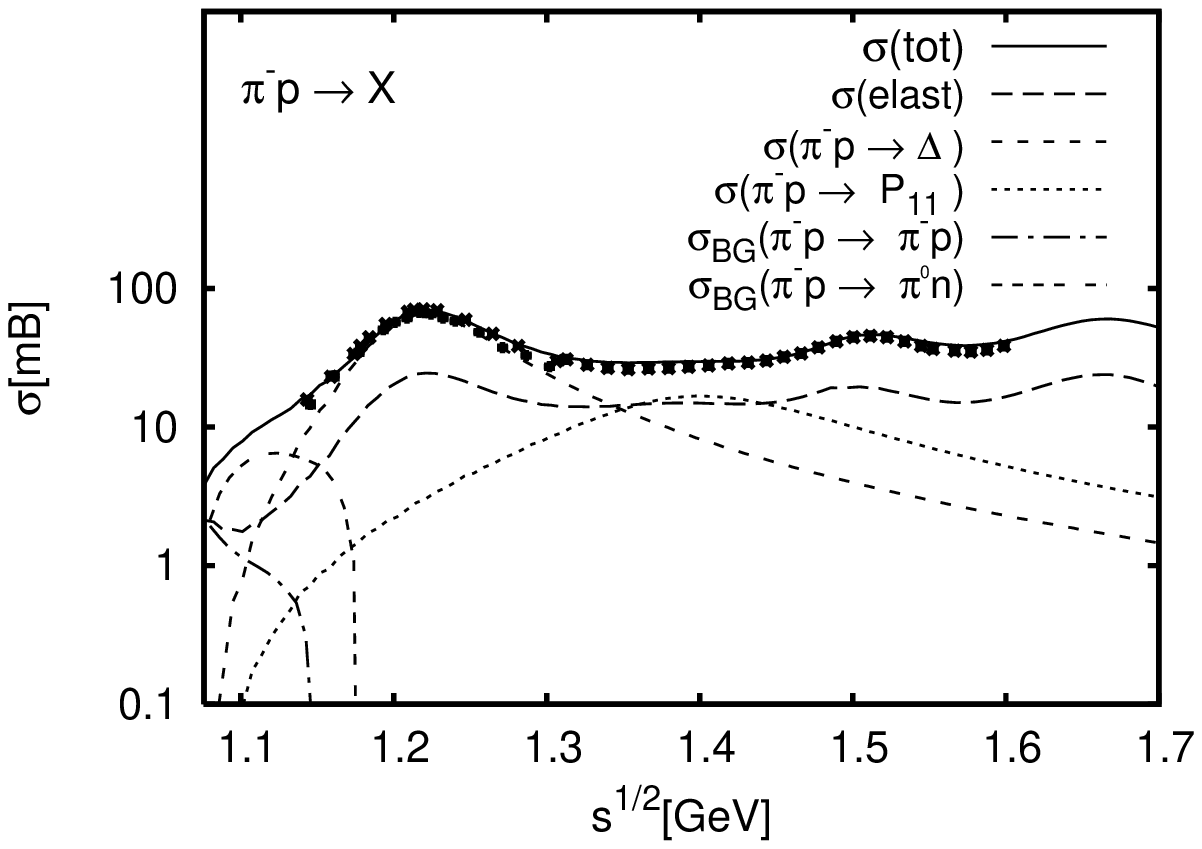}
\includegraphics[angle=0,width=8cm]{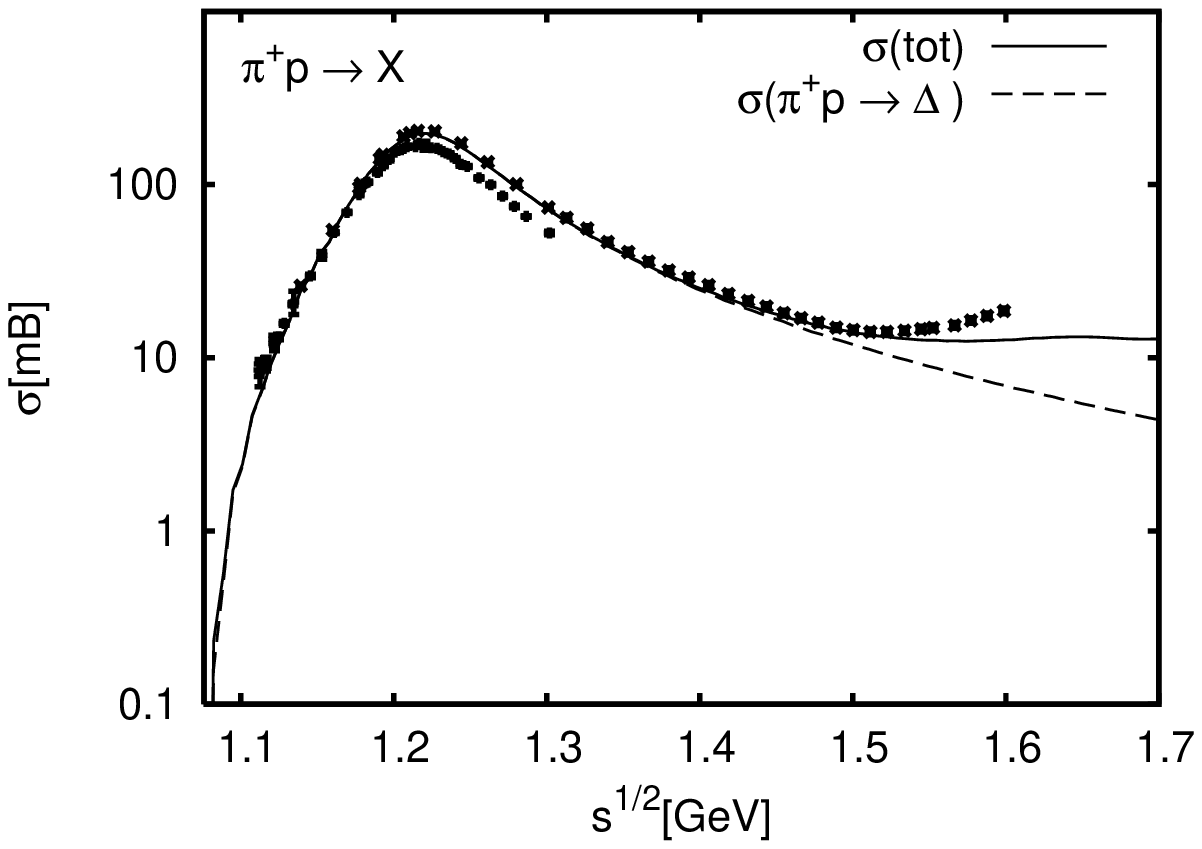}
\includegraphics[angle=0,width=8cm]{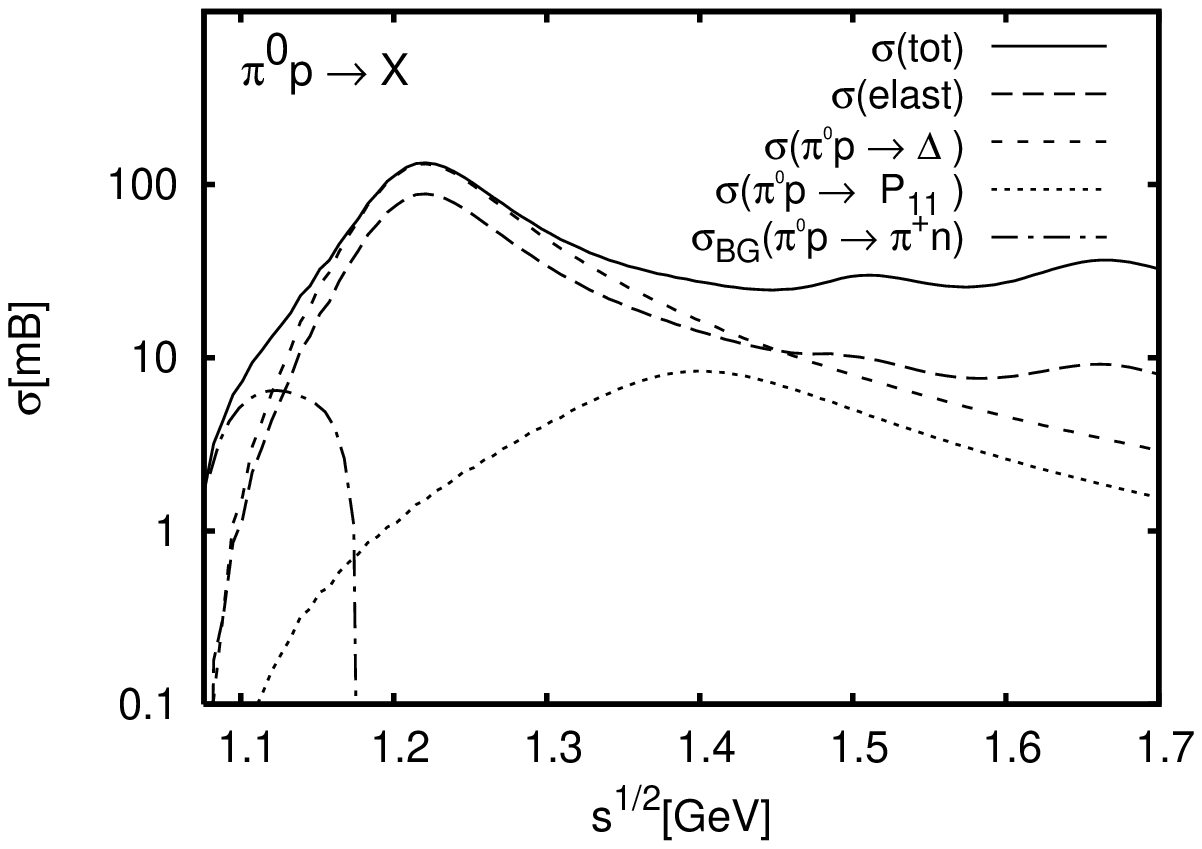}
\caption{Elementary cross sections for \red{elastic and inelastic} scattering of pions and protons. Note that the resonance production processes contribute to elastic and inelastic scattering processes. The data \cite{Carter:1971tj,Davidson:1972ky,Kriss:1999cv} are for $\sigma$(tot).}
\label{pionProton}
\end{figure}

Concerning total cross sections, the relevant contributions to the scattering of pions and protons are shown in \fig {\ref{pionProton}} for the different pion charge states. The cross sections on the neutron follow by isospin inversion. 

\begin{figure}
\centering
\includegraphics[angle=0,width=7.5cm]{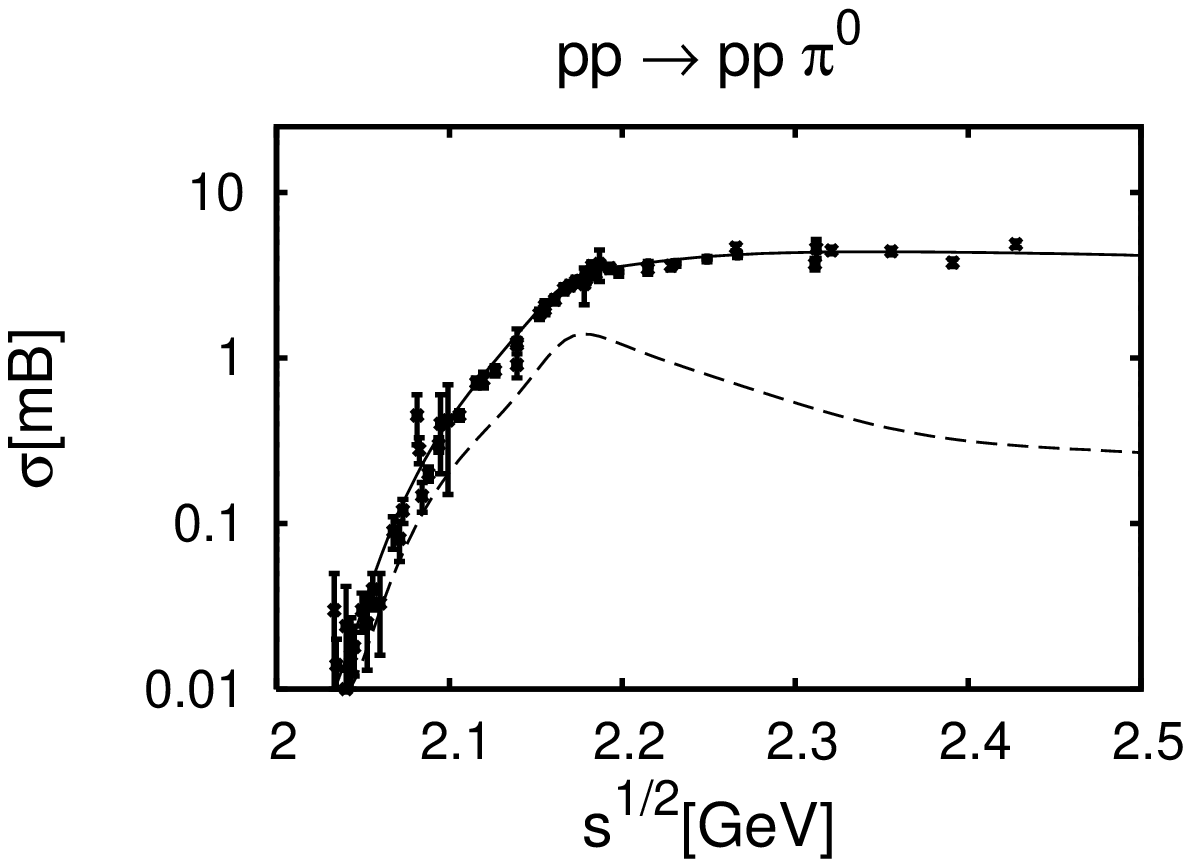}
\includegraphics[angle=0,width=7.5cm]{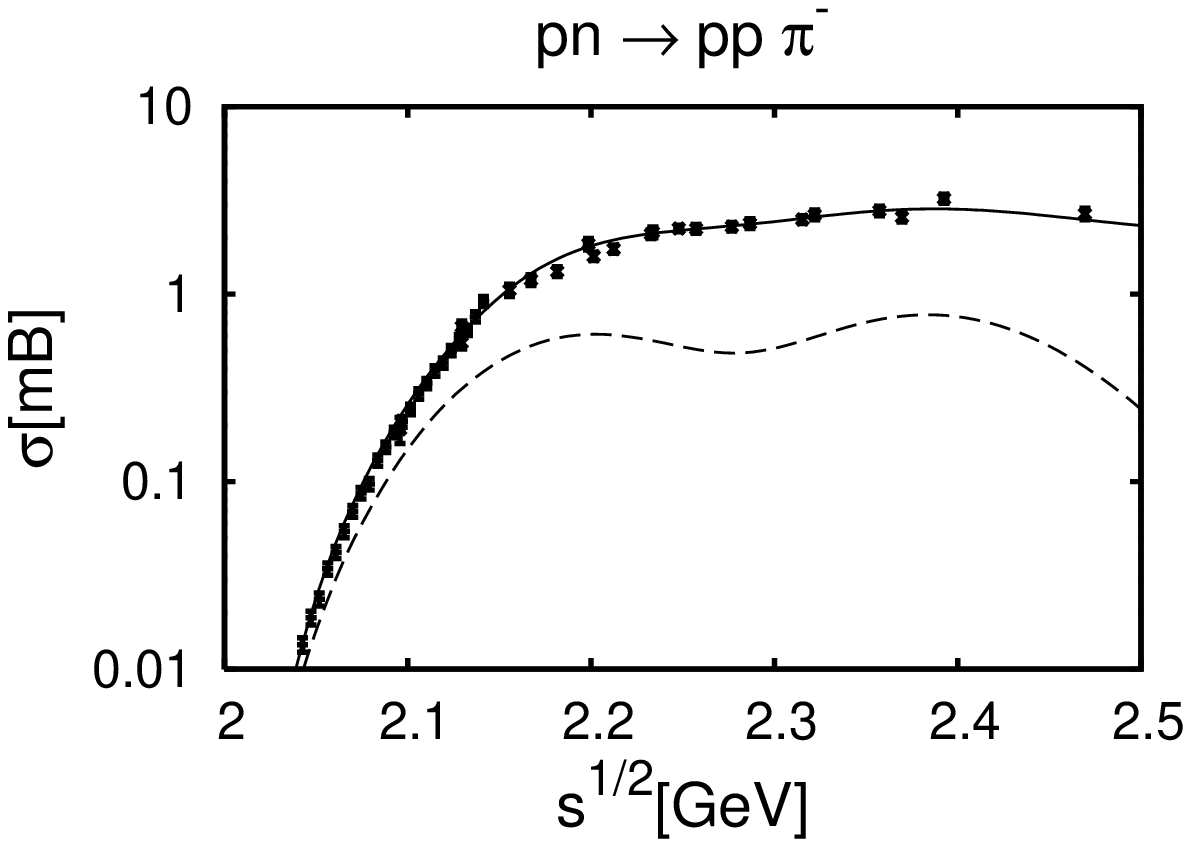}
\includegraphics[angle=0,width=7.5cm]{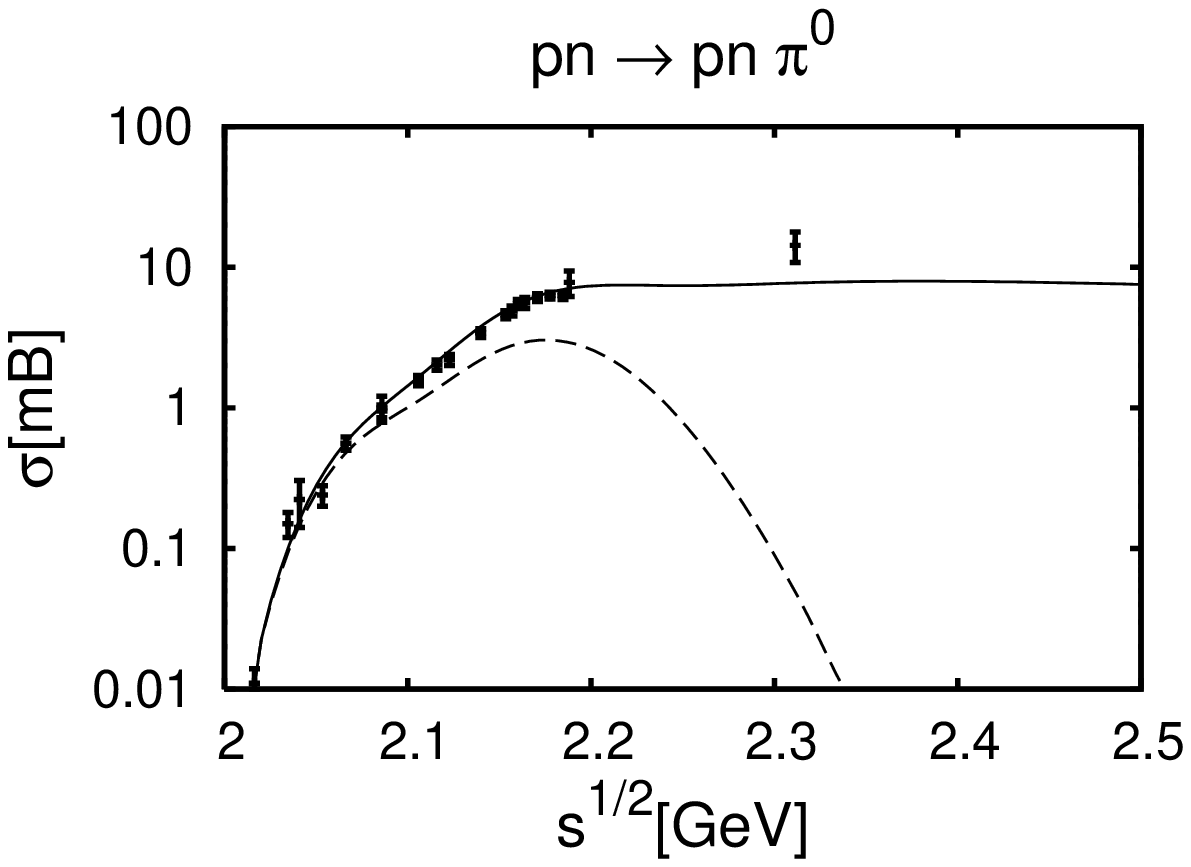}
\includegraphics[angle=0,width=7.5cm]{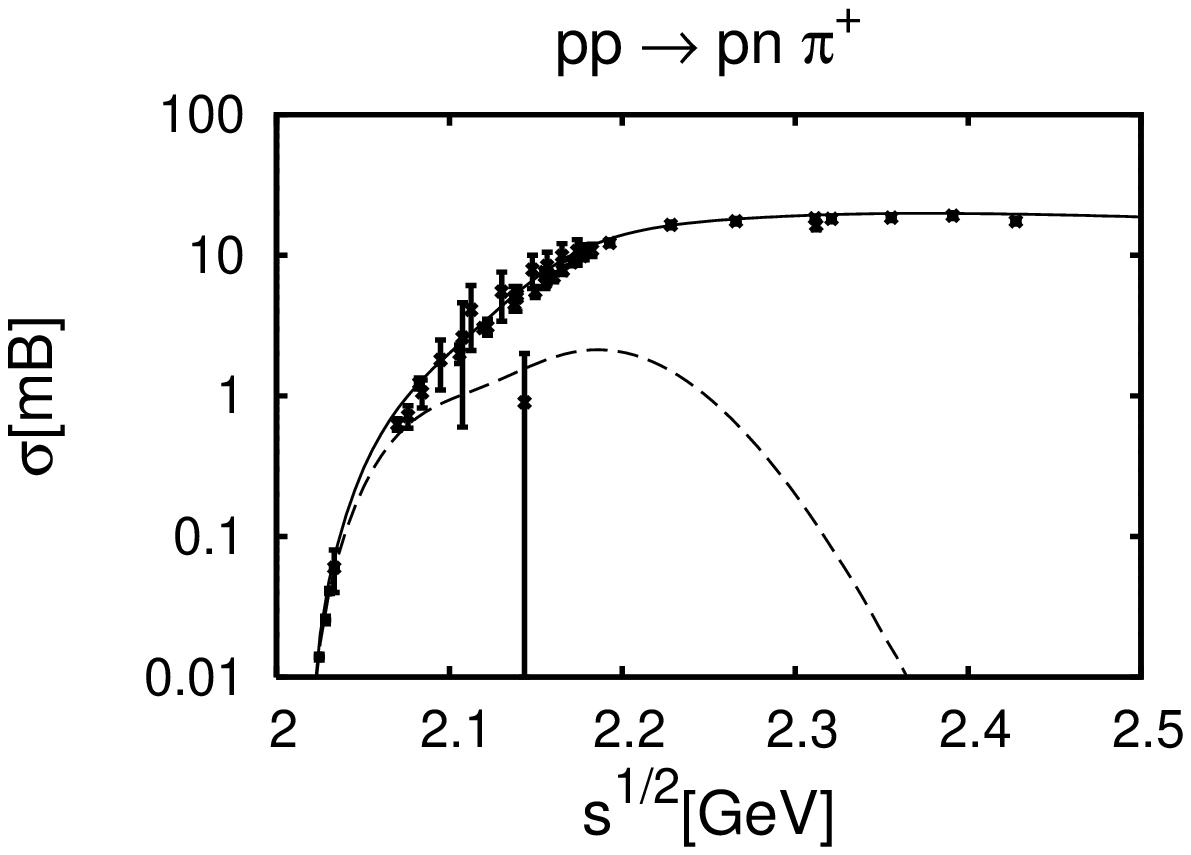}
\caption{Elementary cross sections for different $NN\rightarrow NN\pi$  isospin channels. The solid lines show the full cross section, whereas the dashed lines represent the non-resonant background contribution. The data are taken from \cite{Landoldt,Andreev:1988tv,Daum:2001yh,Hardie:1997mg,Tsuboyama:1988mq,Shimizu:1982dx,Bondar:1995zv}.}
\label{NN_NNPi}
\end{figure}
The process $ NN\rightarrow NN \pi$ has been extensively studied in several experiments over the last twenty years \cite{Landoldt,Andreev:1988tv,Daum:2001yh,Hardie:1997mg,Tsuboyama:1988mq,Shimizu:1982dx,Bondar:1995zv}. We can therefore construct well defined background cross sections on top of our resonance contributions for all possible isospin channels. In \fig{\ref{NN_NNPi}} we show the relevant cross sections. 

\begin{figure}
\centering
\includegraphics[angle=0,width=7.5cm]{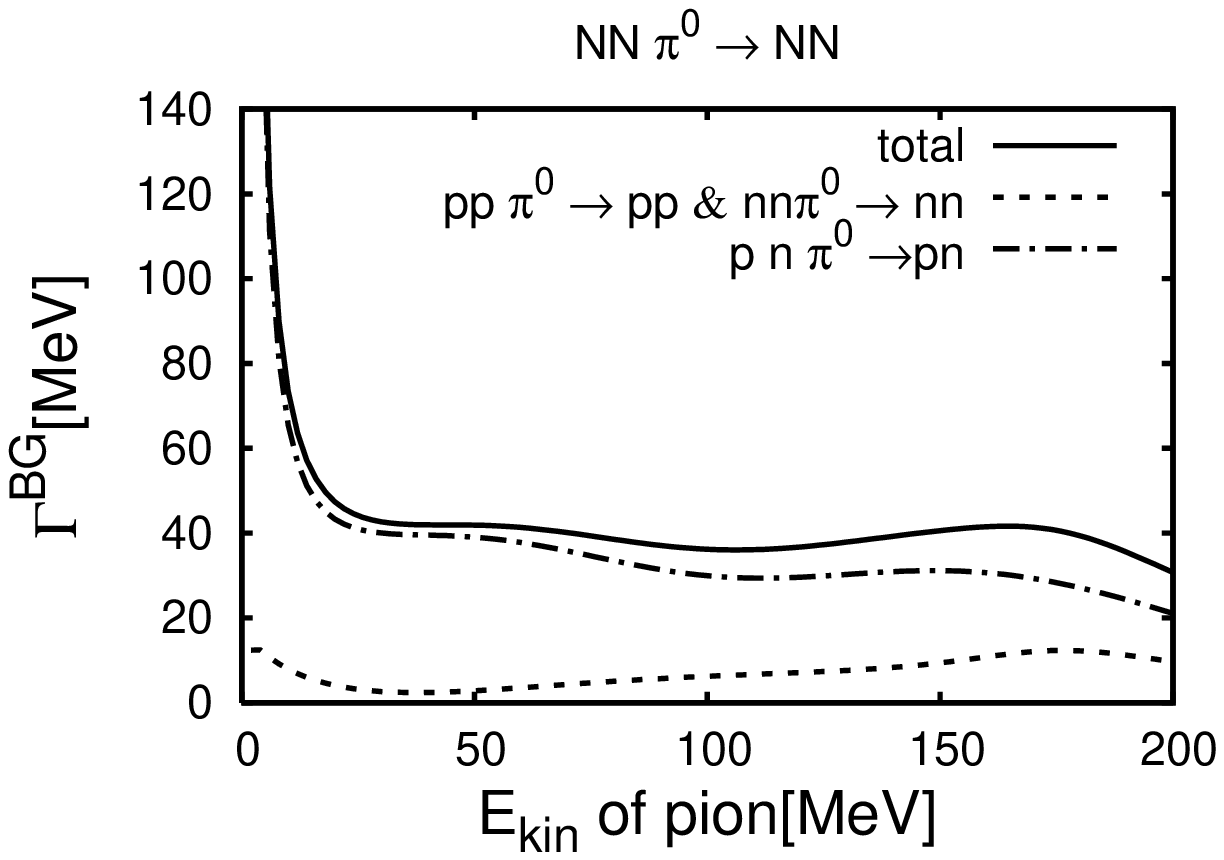}
\includegraphics[angle=0,width=7.5cm]{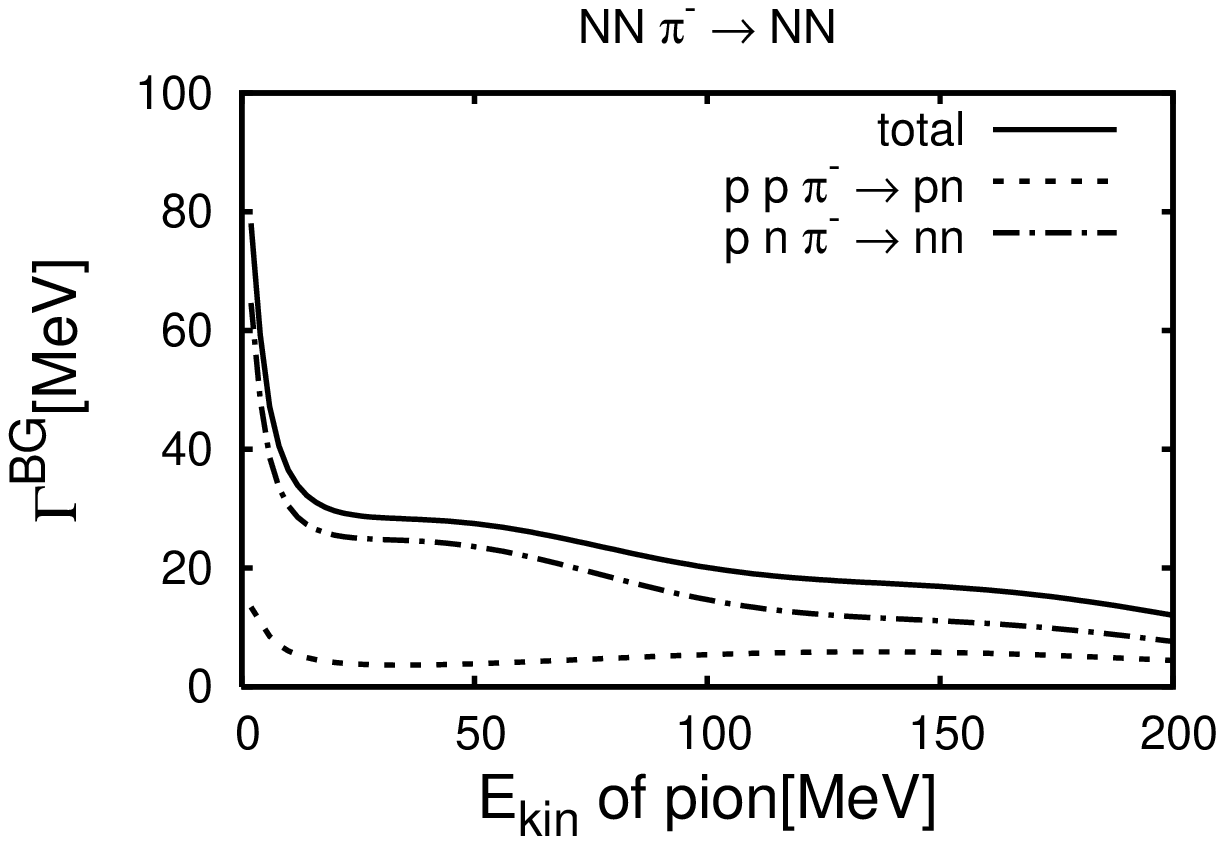}
\caption{The $NN\pi\rightarrow NN$ background absorption rate for symmetric nuclear matter at $\rho=\rho_{0}$ . In-medium-modifications are not included in these decay widths. The rate for $\pi^0$ is shown on the left and for $\pi^-$ on the right panel.}
\label{gamma_NNPi_NN}
\end{figure}

The $\pi NN\rightarrow NN$ process is described by the two-step process $\pi N \rightarrow R$ followed by $R N \rightarrow NN$, and a direct background contribution. One defines the pion absorption rate
\begin{eqnarray}
\Gamma_{N_{A}N_{B}\pi \rightarrow N_{a}N_{b} }&=&S_{ab} S_{AB}\frac{\left|\vec{p}_{ab}\right|}{4\pi\sqrt{s}}\frac{|\M|^{2} }{2 E_{\pi}} \frac{\rho_{N_{A}}}{2 E_{A}} \frac{\rho_{N_{B}}}{2 E_{B}} \label{DecRate} \; .
\label{gammaDef}
\end{eqnarray}
This rate depends on the densities $\rho_{N_{A}}$ and $\rho_{N_{B}}$ of the nucleons in the initial state. The matrix element $\M$ can be calculated by detailed balance, assuming that it only depends on the Mandelstam variable s. 
\[
|\M(s)|=|\M_{N_{a} N_{b} \rightarrow N_{A}  N_{B}  \pi}(s)|=|\M_{N_{A}  N_{B}  \pi  \rightarrow \ N_{a}  N_{b} }(s)|
\]
To obtain this matrix element \cor{we consider more closely} the $NN\rightarrow NN\pi$ process. The cross section for this process is given by
\begin{eqnarray}
\sigma_{\ N_{a}  N_{b} \rightarrow N_{A}  N_{B} \pi }
&=&S_{AB}  \int \frac{(2\pi)^{4}}{4\left|\vec{p}_{ab}\right|\sqrt{s}}   \delta^{4}\left(p_{a}+p_{b}-p_{A}-p_{B}-p_{\pi}\right) 
 |\M_{N_{a}  N_{b} \rightarrow N_{A}  N_{B}  \pi}|^{2} \\
&& \times \frac{d^{3}p_{A}}{(2\pi)^{3}\ 2 E_{A}}  \frac{d^{3}p_{B}}{(2\pi)^{3} 2 E_{B}}   \frac{d^{3}p_{\pi}}{(2\pi)^{3} 2 E_{\pi}} \nonumber \\
&\cong& S_{AB}  \frac{1}{(2\pi)^{3} 64\left|\vec{p}_{ab}\right|\sqrt{s}^{3}}   |\M|^{2} \int^{(m^2_{AB})_{max}}_{(m^2_{AB})_{min}} dm_{AB}^{2} \int^{(m^2_{A\pi})_{max}}_{(m^2_{A\pi})_{min}} dm_{A\pi}^{2} \label{piNN}
\end{eqnarray}
with
\begin{eqnarray*}
m_{xy}^{2}&=&(p_{x}+p_{y})^{2} \\
(m^2_{AB})_{min}&=&	(m_A+m_B)^2 \\
(m^2_{AB})_{max}&=&	(\sqrt{s}-m_\pi)^2 \\
(m^2_{A\pi})_{min/max}&=&	\left(E^{\star}_A+E^{\star}_\pi\right)^2-\left(\sqrt{(E_A^\star)^2-m_A^2}\pm \sqrt{(E_\pi^\star)^2-m_\pi^2}      \right)^2 \\
E_\pi^\star &=&\frac{s-m_{AB}^2-m_\pi^2}{2m_{AB}} \\
E_A^\star   &=&\frac{m^2_{AB}+m_{A}^2-m_B^2}{2m_{AB}}
\end{eqnarray*}
and
\begin{eqnarray*}
S_{AB}=
\left\lbrace
\begin{array}{cc}
\frac{1}{2} & \mbox{if particles A and B are identical} \nonumber \\
1           & \mbox{otherwise} \nonumber
\end{array} \right.  
\end{eqnarray*}
as symmetry factor; $\vec{p}_{ab}$ denotes the momentum of the particles $a$ and $b$ in their center of mass frame. Equation (\ref{piNN}) \cor{holds under the assumption that the matrix element is only dependent on the Mandelstam }$s$. So we get
\begin{eqnarray}
|\M(s)|^{2} 
=\left( S_{AB}  \frac{1}{64(2\pi)^{3} \left|\vec{p}_{ab}\right|\sqrt{s}^{3}} \right. 
\left. \int dm_{AB}^{2} dm_{A\pi}^{2} \right)^{-1} \sigma_{ N_{a}  N_{b} \rightarrow N_{A}  N_{B}  \pi }\; . \nonumber 
\end{eqnarray}
After inserting this result for $|\M(s)|^{2}$ into equation (\ref{gammaDef}) we find that $\Gamma_{N_{A}  N_{B}  \pi  \rightarrow  N_{a}  N_{b}}$ depends linearly on $ \sigma_{ N_{a}  N_{b} \rightarrow N_{A}  N_{B} \pi  }$. This cross section $\sigma_{ N_{a}  N_{b} \rightarrow N_{A}  N_{B}  \pi  }$ is, according to equation (\ref{backDef}), a sum \cor{of background and resonance contributions}. Therefore, $\Gamma_{N_{A}  N_{B}  \pi  \rightarrow  N_{a}  N_{b}}$ can also be split \cor{into} a resonance and a background contribution.
\[
\Gamma_{N_{A}  N_{B}  \pi  \rightarrow  N_{a}  N_{b}}= \Gamma^{\BG}_{N_{A}  N_{B}  \pi  \rightarrow  N_{a} N_{b}}+\Gamma^{\mbox{\begin{scriptsize}resonance contribution\end{scriptsize}}}_{N_{A}  N_{B}  \pi  \rightarrow  N_{a} N_{b}}
\]
with 
\begin{eqnarray*}
\Gamma^{\BG}_{N_{A}  N_{B}  \pi  \rightarrow  N_{a} N_{b}}&\sim & \sigma^{\BG}_{N  N \rightarrow N  N  \pi}\\[0.2cm]
\Gamma^{\mbox{\begin{scriptsize}resonance contribution\end{scriptsize}}}_{N_{A}  N_{B}  \pi  \rightarrow  N_{a}  N_{b}} &\sim &  \sigma^{\mbox{\begin{scriptsize}resonance contribution\end{scriptsize}}}_{ N  N \rightarrow N  N  \pi} \; .
\end{eqnarray*}
More details concerning this absorption rate can be found in \cite{Effenberger:1996rc} and \cite{diplom}. The resonance absorption part is included in the collision term for the resonances, which are propagated explicitly. The background absorption rate $\Gamma^{\BG}_{N_{A}  N_{B}  \pi  \rightarrow  N_{a} N_{b}}$ is shown in \fig{\ref{gamma_NNPi_NN}} for symmetric nuclear matter at $\rho=\rho_0$ with $\rho_0=0.168\ \mathrm{fm}^{-3}$ being normal nuclear matter density. For positive pions we get the same results as for negative ones due to isospin symmetry. Notice that in-medium modifications are not accounted for so far.

\subsection{Medium modifications}\label{medMod}
As medium modifications we implement, besides effects like Pauli-blocking, Fermi-motion of the nucleons and Coulomb forces, also different types of hadronic potentials for the particles and density dependent modifications of the decay widths. We emphasize that the imaginary parts of the potentials are in general not included since their effects are already accounted for by the collision term. Only in the case of the $\Delta$ resonance we follow a prescription which includes imaginary contributions of the optical potential as explained below. 

In position-space the nucleons are initialized according to Wood-Saxon-Distributions. For the momenta we utilize a local Thomas-Fermi-Approximation and express the local Fermi-momenta for \cor{the neutrons ($p^n_{F}$) and the protons ($p^p_{F}$)} as \[p_{F}^{n,p}(\vec{r})=\left(3\pi^2\rho^{n,p}(\vec{r})\right)^{\frac{1}{3}}\; ,\] where $\rho^{n,p}(\vec{r})$ denote the densities of the nucleons. The initial phase-space distribution of the nucleons is therefore given by
 \[
 f_{\mathrm{init}}^{n,p}(\vec{r},\vec{p})=N\,\Theta\left(p^{n,p}_F(\vec{r})-\left|\vec{p}\right|\right) .
\]
with $N$ being a normalization factor. Due to the low particle multiplicities, \red{the probability that a final state is Pauli-blocked $P_{PB}$} was approximated using the analytic ground-state expression
\[
P_{PB}^{n,p}(\vec{r},\vec{p})=\Theta\left(p_F^{n,p}(\vec{r})-\left|\vec{p}\right|\right) 
\]
instead of calculating the Pauli-blocking probability based on $f_1^{\mathrm{nucleon}}$ as necessary in calculations of, e.g., heavy-ion-collisions.

\subsubsection{Baryon potentials}
\begin{figure}
\begin{center}
\includegraphics[]{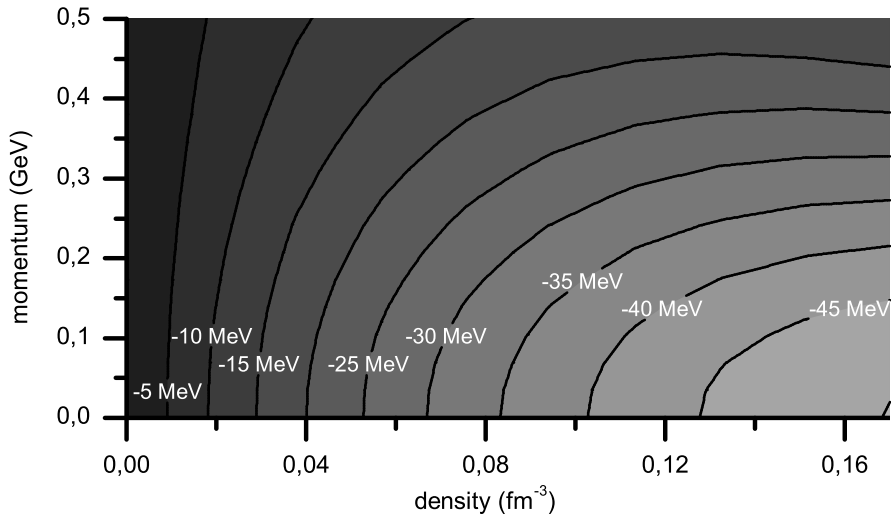}
\end{center}
\caption{The hadronic potential $A_{0}$ of the $\Delta$ in the LRF as a function of momentum and density in units of MeV.}
\label{delPot}
\end{figure}
\begin{figure}
\centering
	\includegraphics[]{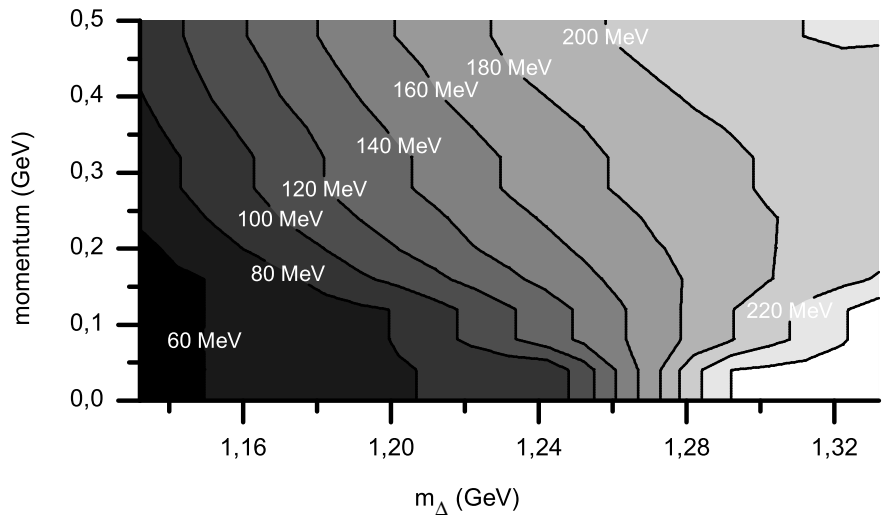}

\caption{The width $\Gamma_{\mathrm{tot}}$ of the $\Delta$ resonance at $\rho=\rho_0$ in units of MeV \red{as a function of momentum and invariant mass $m_\Delta$.} The total width $\Gamma_{\mathrm{tot}}$ is the sum of the in-medium contributions and the Pauli-blocked free width $\overset{\sim}{\Gamma}$ . }
\label{delWidth}
\end{figure}

The potentials are introduced as \cor{time-like} components of vector potentials in the local rest-frame\,(LRF)~\cite{Teis:1996kx}. For our purposes the most important mean field potentials are those acting on the nucleon, the $\Delta$ resonance and the $\pi$ meson. The nucleon potential is described by a momentum-dependent mean-field potential (for explicit details and parameters see~\cite{Teis:1996kx}).\cor{ Phenomenology} tells us that the $\Delta$ potential has a depth of about $-30 \MeV$ at $\rho_{0}$\,\cite{ericsonWeise,Peters:1998mb}. 
Comparing to a momentum independent nucleon potential, which is approximately $-50 \MeV$ \cor{deep}, the $\Delta$ potential is, \cor{therefore,} taken to be 
\begin{eqnarray*}
	A_0^{\Delta}(\vec{p},\vec{r})=\frac{2}{3}\ A_0^{\mathrm{nucleon}}(\vec{p},\vec{r}).
	\label{DeltaPotential}
\end{eqnarray*}
Here we \cor{assume} the same momentum dependence for the nucleon and the $\Delta$ potentials. The potential acting on the $\Delta$ is shown in \fig{\ref{delPot}} as function of momentum and density in the local rest-frame.
\subsubsection{Modifications of the $\Delta$ width}
The $\Delta$ resonance is explicitly propagated in the model. Besides explicit $N\Delta \rightarrow N \Delta$ and $NN \rightarrow N \Delta$ collisions and the vacuum decay channel $\Delta\rightarrow \pi N$, the model implements also an absorption probability for this resonance to account for $N\Delta \rightarrow N N$  and $N N \Delta \rightarrow N N N$ processes~\cite{effeDoktor}. This probability is based on the corresponding contributions to the width of the $\Delta$, derived in a microscopic model by Oset et al.~\cite{osetSpreading}.  
In \fig{\ref{delWidth}} we show the \cor{total} width 
\[
\Gamma_{\mathrm{tot}}^\Delta=\Gamma^\Delta_{\mathrm{med,abs}}+\Gamma_{N\Delta \longrightarrow N \Delta}+\overset{\sim}{\Gamma} \,^{\Delta}_{vac}
\]
of the $\Delta$ at normal nuclear matter density as implemented in the model. $\overset{\sim}{\Gamma} \,^{\Delta}_{vac}$ denotes the vacuum decay channel\cor{, including Pauli blocking,} and 
\[
\Gamma^{\Delta}_{\mathrm{med,abs}}=\Gamma_{N\Delta \longrightarrow N N}+\Gamma_{N N \Delta \longrightarrow N N N}
\]
is the absorptive contribution due to interactions with the nuclear medium \cor{as given by the model}. 
\subsubsection{Pion potential}
\begin{figure}[b]
	\centering
	\centering \includegraphics[]{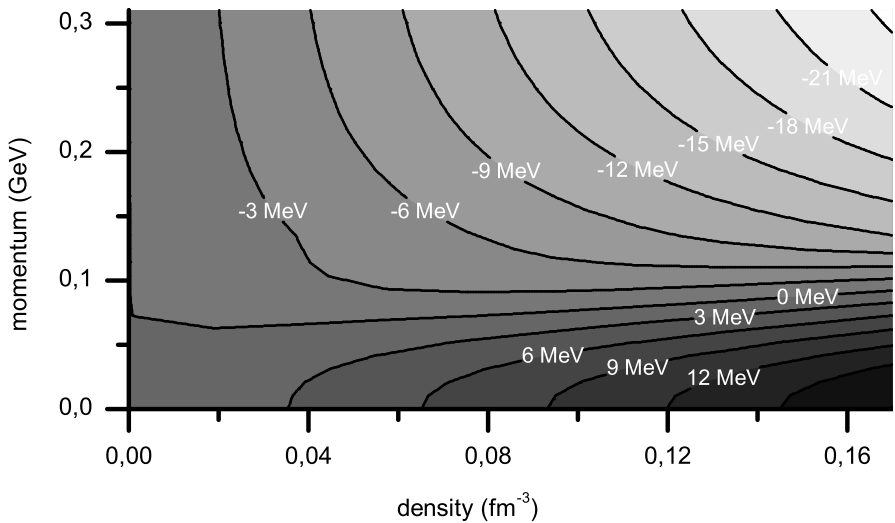}
	\caption{Hadronic potential $A_0$ of the pion as function of nucleon density and pion momentum in symmetric nuclear matter in units of MeV.}
	\label{potPlot}
\end{figure}

For the first time we have also included a realistic potential in the Hamiltonian of the BUU equation for the $\pi$ meson at very low energies. Below $80 \MeV$ pion momentum a model by Nieves et al.~\cite{osetlow} \cor{was used.}\del{ In this work an optical potential was derived based upon an microscopic model for pion nucleon scattering.} In general, the pion spectral function shows at low energies three peak structures - the so called pion, nucleon-hole and $\Delta$-hole branches. At \cor{momenta smaller than 150 MeV} the pion is most probably found in the pion branch \cor{in which} the dispersion relation does not differ much \cor{from} the free one of the pion. Therefore we evaluate the optical potential at the vacuum position of the dispersion relation. 

In latest experiments at GSI~\cite{Geissel:2002ur} and MAMI~\cite{Kohl:2001fx} new results on the pionic self energy at threshold were obtained. The results hint at a value of about $A^0_{\pi}(p=0)\simeq 25 \MeV$. The potential by Nieves et al.~\cite{osetlow} amounts to a somewhat lower value of about $18 \MeV$ in symmetric nuclear matter at normal matter density. It is important to realize that \red{such a repulsive} potential limits the applicability of any semi-classical model to pion kinetic energies $\gtrsim 20 \MeV$ in the central density region. 

Above $140 \MeV$ a first-order $\Delta$-hole result was \cor{used}. In this energy regime the potential acting on the pion is defined by averaging over the spectral strength in the pion and the $\Delta$-hole branches. To ensure a continuous derivative of the Hamiltonian, both models were matched in the region of $p=80-140 \MeV$. 
The potential for the pion is shown in \fig{\ref{potPlot}}. It is repulsive in the low-momentum regime, for higher momenta it turns attractive. The repulsion is caused by a only weakly momentum dependent $S$-wave term, while the attraction is due to a momentum dependent $P$-wave term, which vanishes at zero momentum and gains strength with increasing momentum. At constant density the potential is, therefore, continously decreasing with momentum in the range considered here.

\section{Results of the simulations}
\subsection{The mean free path of pions in nuclear matter}\label{meanFree}
Here, we discuss simple properties of the pion in the medium and compare to other theoretical calculations. In particular we investigate the total width \[\Gamma_{\mathrm{tot}}=\ln\left(-\frac{d(\ln\left(N_{\pi}\right))}{dt}\right)\]
and mean free path $\lambda$. These \cor{quantities} are connected via the relation $\lambda=v/\Gamma_{\mathrm{tot}}$, where $v$ denotes the velocity of the pions. In our model the loss of pions per time-step $\Delta t$ is given by 
\begin{eqnarray*}
\frac{1}{N_{\pi}} \frac{\Delta N_{\pi}(E,\rho_{n},\rho_{p})}{\Delta t}=- \sum_{N=n,p} \int d^3p_N e^{-\sigma_N(p_{\pi},p_N) \rho_N v(p_{\pi},\rho_p,\rho_n)\Delta t}  \\
- \sum_{(N,M)\in\left\lbrace (n,p),(n,n),(p,p)\right\rbrace} \int d^3p_N \int d^3p_M  e^{-\Gamma^{\BG}(p_{\pi},p_N,p_M,\rho_N,\rho_M) \Delta t} = -e^{-\Gamma_{tot} \Delta t}
\end{eqnarray*}
where $v$ denotes the pion velocity in the medium, $\sigma_N$ - the  $\pi N$ scattering cross section and $\Gamma^{\BG}$ is the absorptive three-body width of the pion and \cor{$\Gamma_{tot}$ the total width}. For the cross sections and the width we include the medium modifications discussed earlier. The integrals above are performed over the Fermi spheres of neutrons and protons.

In practice, we obtain $\Gamma_{\mathrm{tot}}$ by performing a Monte-Carlo simulation with pions and nucleons initialized in a box with continuous boundary conditions within our BUU framework, including all the medium modifications and necessary collision rates. 

The momentum $p_{\pi}$ is calculated in the medium from the dispersion relation
\begin{eqnarray*}
E_{\mathrm{total}}&=&\sqrt{p_{\pi}^{2}+m_{\pi}^2}+A_0^{\pi}(p)+V_{C}
\end{eqnarray*}
with the hadronic potential $A_0^{\pi}$ and the Coulomb potential $V_C$. Finally, $\Gamma_{\mathrm{tot}}$ and $\lambda$ can be easily extracted by observing the time evolution of the pions. All results are shown as functions of the experimental observable 
\[
E_{\mathrm{kin}}^{\mathrm{Vacuum}}= E_{\mathrm{total}}-m_{\pi}
\]
\cor{in the classically allowed region $E_{\mathrm{kin}}^{\mathrm{Vacuum}}\geq A_0^{\pi}(p)+V_{C}$.}

\subsubsection{Isospin dependence}
To begin with, we study the case in which we omit all potentials. In the GiBUU-simulation we use the same mass of $m_{\pi}=138 \MeV$ for all pions regardless of the charge. For both the positive and the negative pions the same data are used as input for the vacuum cross sections. Due to isospin symmetry the positive and negative charge states have the same properties in symmetric matter if electromagnetic forces are neglected.
\begin{figure}
\centering
\includegraphics[width=8cm]{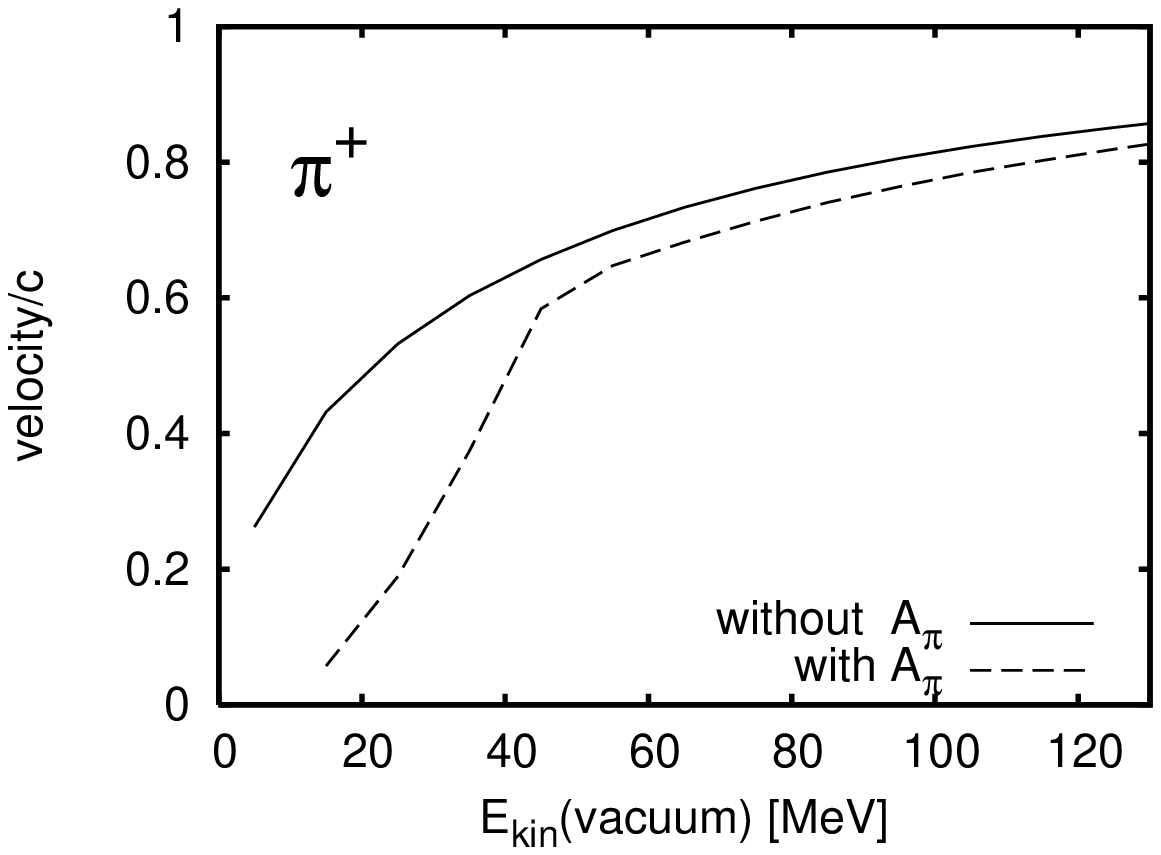}
\includegraphics[width=8cm]{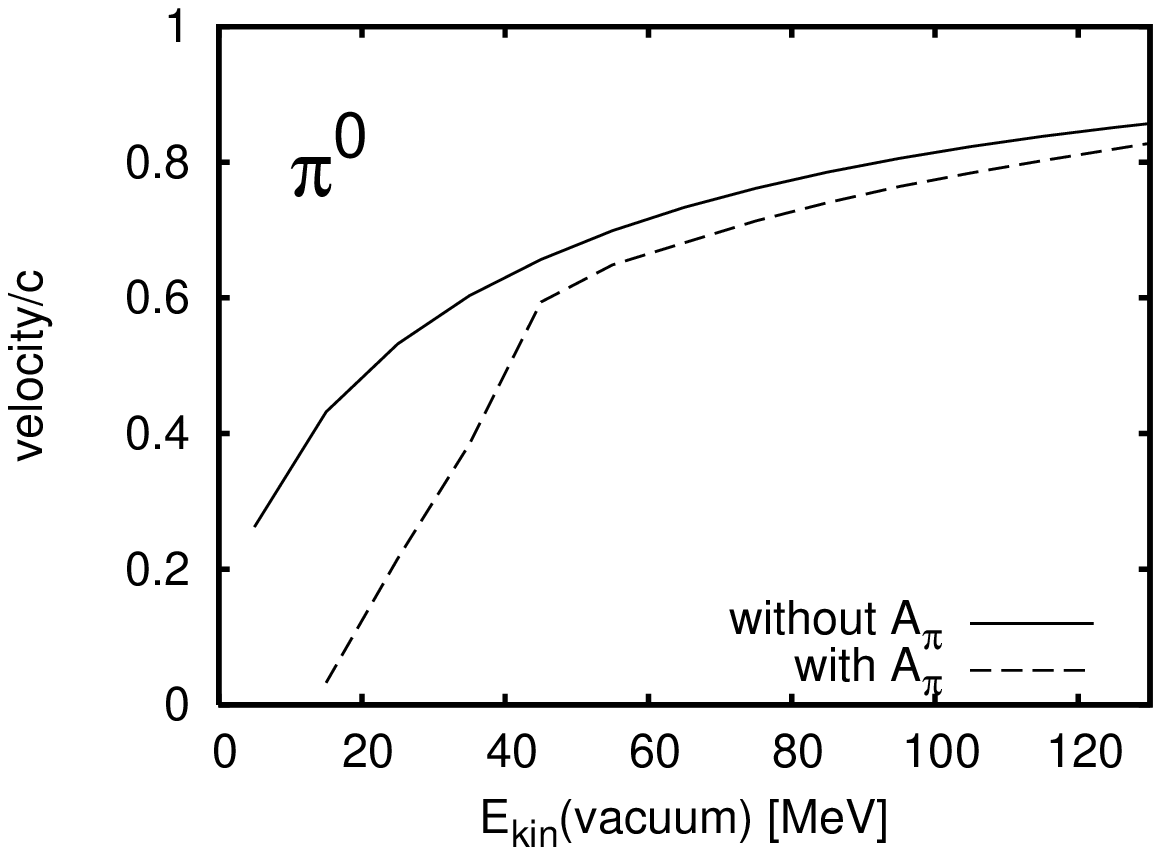}\\
\includegraphics[width=8cm]{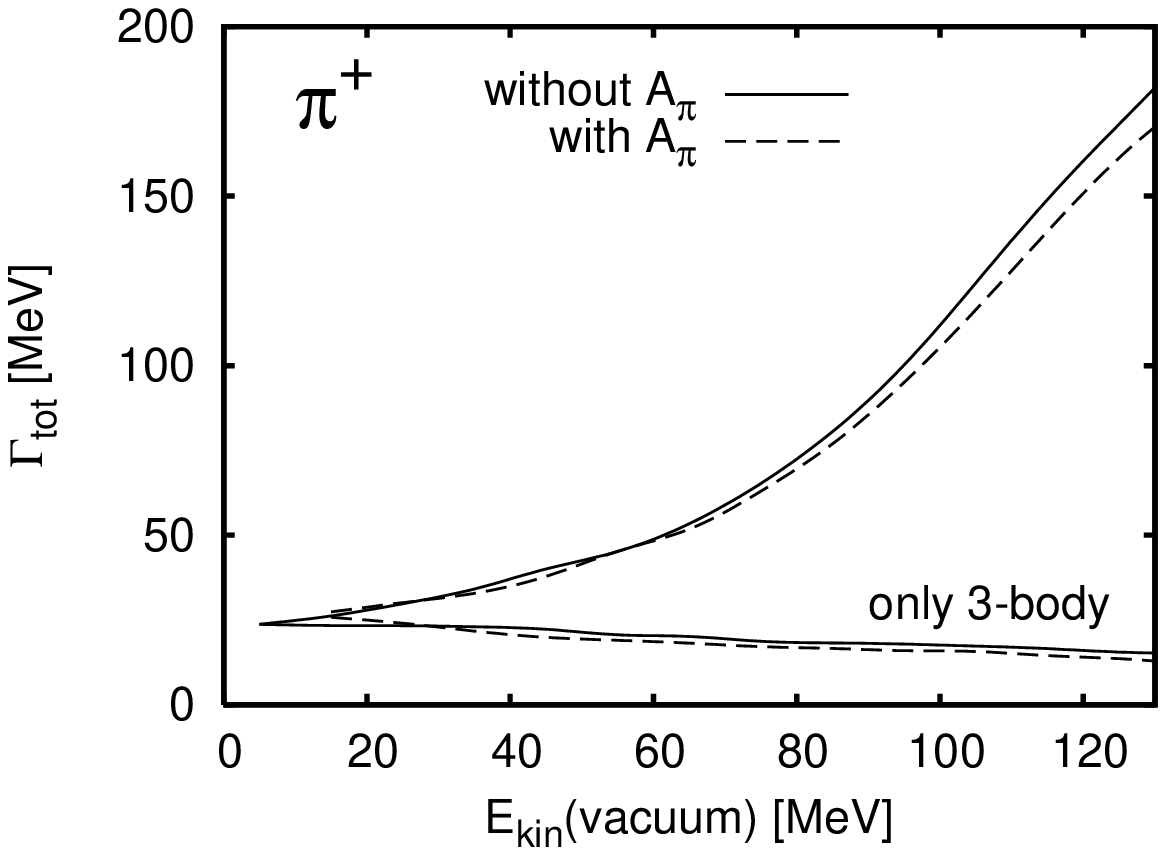}
\includegraphics[width=8cm]{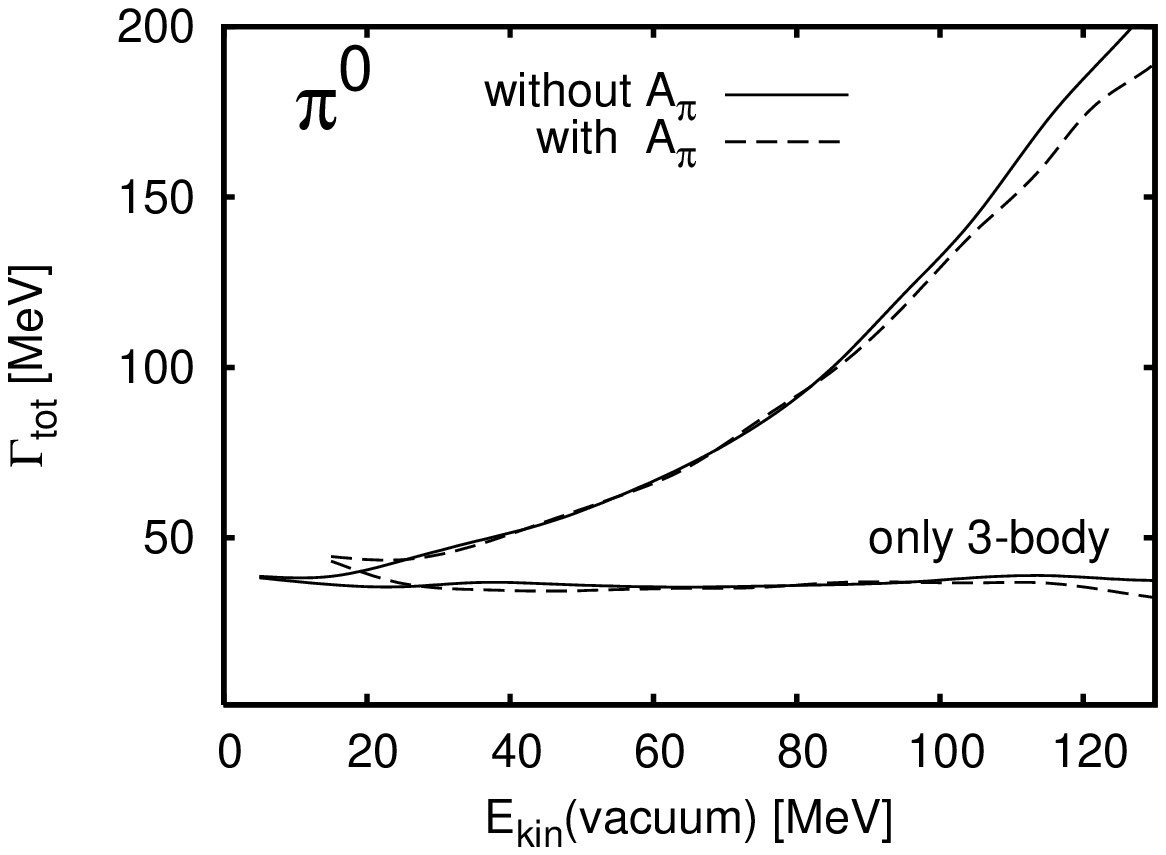}\\
\caption{BUU results without electro-magnetic forces inside symmetric nuclear matter at $\rho_0=0.168 fm^{-3}$. The results are shown with (dashed) and without (solid line) hadronic pion potential. The upper panels show the velocity of the pions in nuclear matter, the lower ones visualize the full width of the pion. The lower curves in the lower panel denoted "only 3-body" do not include two-body processes or resonance production.\del{Hence they show the importance of the included $NN\pi\rightarrow NN$ process.}}
\label{noCoulomb}
\end{figure}
In \fig{\ref{noCoulomb}} we observe that the positive and neutral pions do actually have the same velocities, but not the same width in nuclear matter. The difference is quite small except for very low energies. The reason for this can be found in the elementary absorption and scattering processes. As mentioned before, we have used vacuum data to pin down the elementary reaction rates. Therefore the  absorptive width of the charged pions is fixed by the elementary reactions  
\begin{eqnarray*}
 p\ p &\rightarrow &\ p \ n \ \pi^{+} \\
 p\ n &\rightarrow &\ p \ p \ \pi^{-}
\end{eqnarray*}
while the absorptive width of the $\pi^{0}$ is determined by
\begin{eqnarray*}
 p\ p &\rightarrow &\ p \ p \ \pi^{0} \\
 p\ n &\rightarrow &\ p \ n \ \pi^{0}
\end{eqnarray*}
through detailed balance. Fig. \ref{gamma_NNPi_NN} shows that the three-body absorption rate for the neutral pions at low energies is higher than for the charged ones. This explains the larger width of the $\pi^{0}$ near threshold.


\subsubsection{Collisional width in the medium}
Using the neutral pion as an example, we first discuss the lower right panel of \fig{\ref{noCoulomb}}. At low energies there is practically no difference in the width between the simulation with (dashed lines) and without (solid lines) pion potential. Due to the absence of resonance contributions in this energy regime, the width of the pion is dominated by its absorption via the $N N \pi \to N N $ background. At higher energies a small difference in the two curves can be observed. Here the decay width is somewhat smaller if the pion potential is included.

This difference can be understood by investigating the resonance production process $N \pi \rightarrow \Delta$ in detail \cite{diplom}. At fixed pion energy, the attractive pion potential leads to a higher momentum of the pion and, therefore, also to a higher momentum of the produced $\Delta$, resulting in a decrease of its mass. 
In the considered energy regime this mass is smaller than the pole mass. 
As a consequence, the production of the $\Delta$ resonance is less probable and the width of the pion decreases. 

\subsubsection{Velocity of the $\pi$ mesons in medium.}
The velocity of the pions is given by Hamilton's equation
\begin{eqnarray}
\frac{\partial r_{i}}{\partial t}&=&\frac{\partial H}{\partial p_{i}}=\frac{\partial\left( \sqrt{p^{2}+m^{2}}+A_{0}^{\pi}+V_{C}\right)}{\partial p_{i}} 
=\frac{p_{i}}{\sqrt{p^{2}+m^{2}}}+\frac{\partial A_{0}^{\pi}}{\partial p_{i}}.
\label{veloEQ}
\end{eqnarray}
The hadronic potential is repulsive for low values of $E_{tot}$ and of the same order as $E_{\mathrm{kin}}$. Therefore, in the first term of equation (\ref{veloEQ}) the pion momentum is considerably smaller and, as shown in the upper panels of \fig{\ref{noCoulomb}}, the velocity is strongly modified at low values of $E_{kin}$. The second term $\partial A_{0}^{\pi}/\partial p$ is always negative. It is large in absolute magnitude for low energies and small for energies greater than $80 \MeV$ and leads to an overall reduction of the velocities.  

\subsubsection{Results for the mean free path.}

\cor{After having extracted} the width of the pion in the nuclear matter restframe directly from our numerical simulation the mean free path is \cor{obtained} by $\lambda=v/\Gamma$. Therefore we now\del{need to} consider the two effects discussed earlier in order to understand the changes in the mean free path: modification of the decay width $\Gamma$ and modification of the velocity $v$ due to the potentials.
\begin{figure}
\centering
\includegraphics[width=8cm]{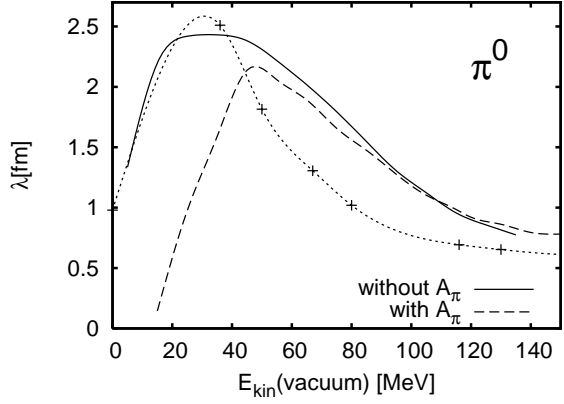}
\caption{Mean free path of a neutral pion in symmetric nuclear matter at $\rho_0=0.168 fm^{-3}$. We show the result without (solid line) and with (dashed line) hadronic potential for the pion; all other standard medium modifications are included. For comparison, the result of Mehrem et al. \cite{MehremRadi} is also shown (crosses linked by a dotted interpolating line), which will be discussed later.}
\label{MeanFreePlot}
\end{figure}
The effect of including the hadronic potential becomes visible in \fig{\ref{MeanFreePlot}}, especially at very low energies. The mean free path drops rapidly at low energies compared to the simulation without hadronic potential. This sharp decrease of the mean free path as a function of asymptotic kinetic energy stems from the velocity decrease at low energies when including the \cor{repulsive} hadronic potential. At larger values of the kinetic energy the effect of \del{decreasing}width and \del{decreasing}velocity just compensate each other. 

\begin{figure}
\centering
\includegraphics[width=8cm]{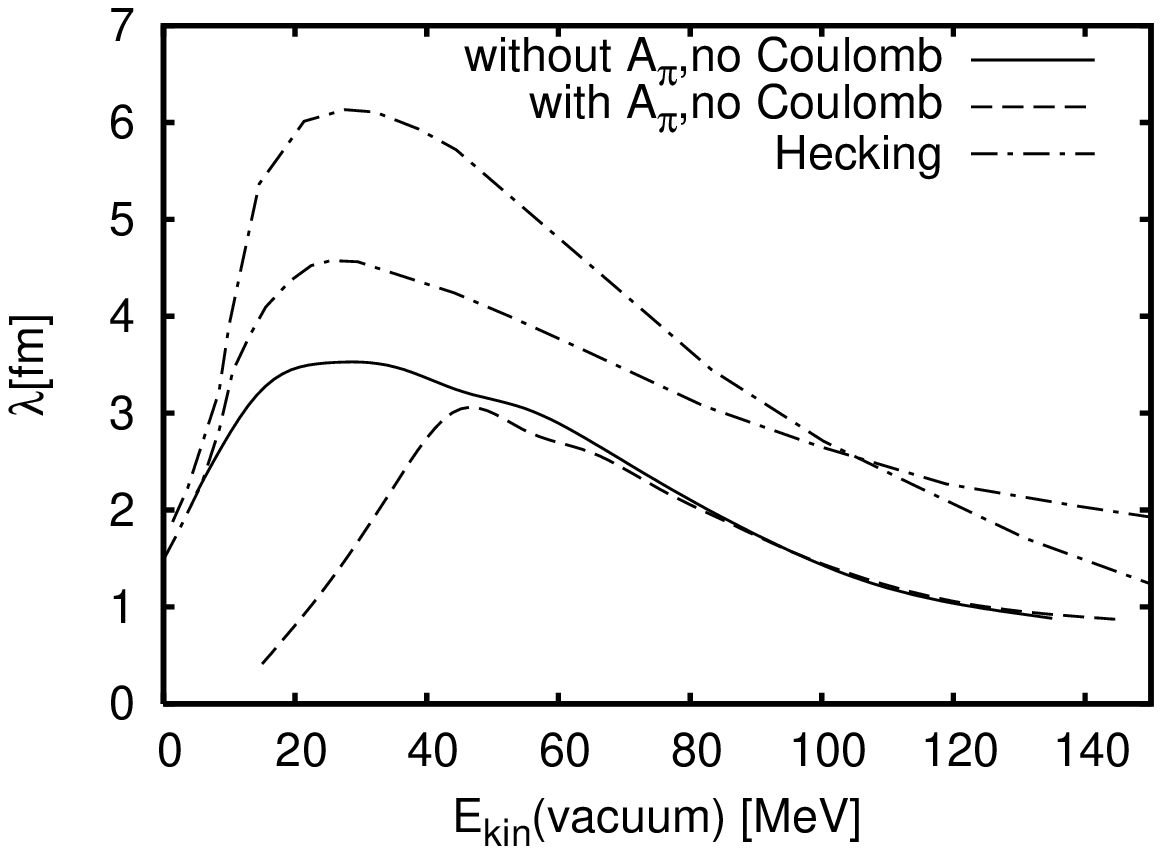}
\includegraphics[width=8cm]{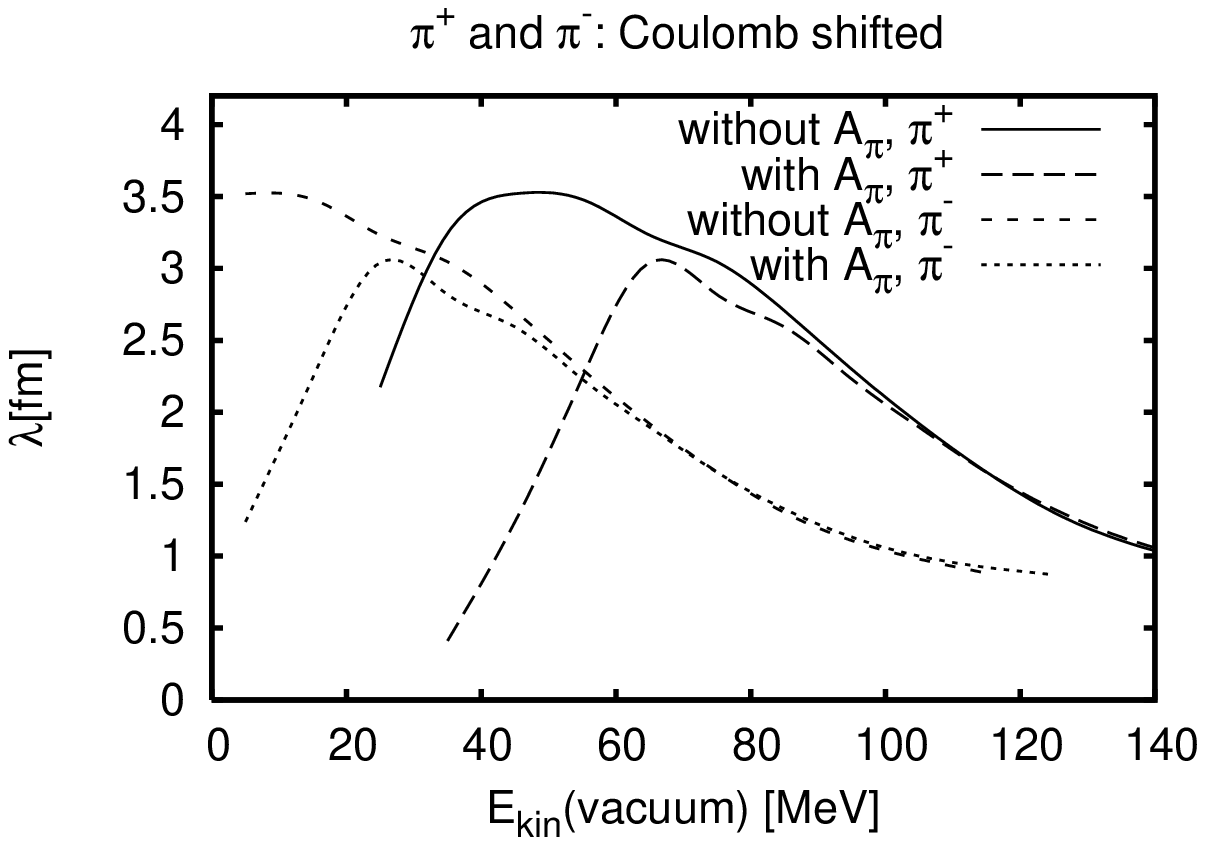}
\caption{Influence of the electro-magnetic forces on the mean free path of the charged pions  ($\rho=0.168 fm^{-3},\ \rho_{p}=\rho_{n}=\frac{\rho}{2}$). On the left panel we show results of our model without Coulomb potential - in this case the charged pions have identical mean free paths. The curves presented with dashed dotted lines are charge-averaged results of P. Hecking \cite{Hecking}, which we will discuss later. The panel on the right shows the mean free path of the charged mesons, including Coulomb effects. We explicitly investigate the result with and without hadronic potential and Coulomb potential for the pion, all other standard medium modifications are included.}
\label{CoulVergleich}
\end{figure}
For charged pions, electromagnetic forces play an important role, especially at low energies. Since in nuclear matter calculations we \cor{cannot} introduce any Coulomb potential based on the actual density, we assume a reasonable constant Coulomb potential \red{of $V_{C}=\pm 20 \MeV$ depending on the charge of the pion}. 
The results for the mean free path are shown in \fig{\ref{CoulVergleich}}. The most important effect of the hadronic potential is the large decrease of the mean free path at very low energies, qualitatively similar to that already discussed for the $\pi^0$ above.

\begin{figure}[t]
\centering
\includegraphics[width=10cm]{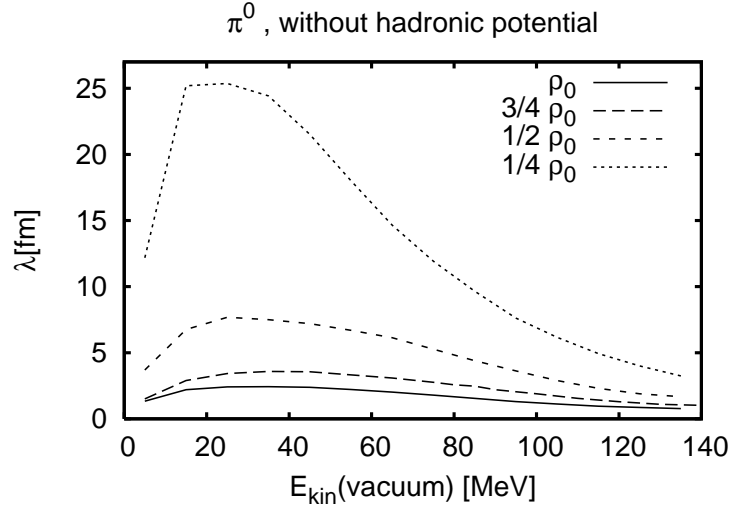}
\caption{The mean free path of a $\pi^0$ investigated at different densities of the nuclear matter. The hadronic potential for the pion is not included.}
\label{densDependence}
\end{figure}

We have also studied the density dependence of the mean free path, choosing the $\pi^0$ as a showcase. The results are shown in \fig{\ref{densDependence}}. It is important to note that the density dependence is highly nonlinear, contrary to the low-density limit. This non-linearity is generated by the $NN\pi\rightarrow NN$ process, which goes to first order quadratically with density, and by the implicit density dependence in the medium modifications.




\subsubsection{The vacuum approximation.}\label{vacApproxSec}

We now compare our results to the eikonal vacuum approximation \cite{ericsonWeise,Cassing},commonly found in the literature. In this approximation the mean free path is given by
\begin{eqnarray}
\lambda=\frac{1}{2\ \Im(p)}=\frac{1}{\rho\ \sigma_{Vac}} \label{MeanFreeDef}\; .
\end{eqnarray}
$\Im(p)$ denotes the imaginary part of the momentum. 
Using the total cross sections shown in \fig{\ref{pionProton}}, we can now evaluate the mean free path. We assume $\sigma_{\pi^{-} n\to X}= \sigma_{\pi^{+} p\to X}$ and $\sigma_{\pi^{+} n\to X}= \sigma_{\pi^{-} p\to X}$ because there exist no data for $\pi$ scattering off neutrons.

\begin{figure}
 \centering
 \includegraphics[width=8.5cm]{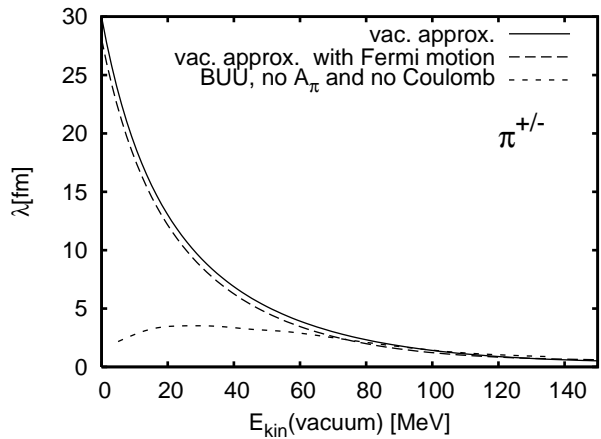}
 \caption{Mean free path of the charged pions in the naive vacuum approximation. The inclusion of Fermi motion does not have a sizeable impact. A BUU calculation without Coulomb forces and without hadronic potential for the pion is shown for comparison.}
 \label{FermiMotion}
 \end{figure}

In the case of symmetric matter, we calculate $\sigma_{Vac}$ for the $\pi^{+}$ and $\pi^{-}$ by averaging over the neutron and proton contribution to the cross section:
\begin{eqnarray*}
\sigma_{\mathrm{Vac}}^{\pm}&=&\frac{1}{2}\left(\sigma_{\pi^{\pm} n\to X}+\sigma_{\pi^{\pm} p\to X} \right)
=\frac{1}{2}\left(\sigma_{\pi^{+} p\to X}+\sigma_{\pi^{-} p\to X} \right)\; .
\end{eqnarray*}
For charged pions the mean free path in this approximation is shown in \fig{\ref{FermiMotion}}. Comparing it to a  full BUU calculation, which considers the Pauli blocking of the final states, as well as the Fermi motion of the initial states and three-body interactions, we see a dramatic discrepancy at small energies.

For completeness we have also investigated the influence of Fermi-motion and calculated the mean free path by integrating over the interactions in the Fermi sea:
\begin{eqnarray}
\lambda=\frac{1}{4\ \iiint\limits_{p_F} \sigma_{Vac}(\sqrt{s}) \frac{d^{3}p}{\left(2\pi\right)^{3}}}\label{FermiMeanFree} \; .
\end{eqnarray}
The result can also be seen in \fig{\ref{FermiMotion}}. The modifications of the mean free path due to the Fermi motion in the vacuum approximation are very small.

The vacuum approximation does not consider any kind of \cor{many}-body interactions at low energy. \cor{Only the two body $\pi N$ interaction is allowed. }In our model, detailed balance dictated us to introduce a $NN\pi\to NN$ process to account for the $NN\to NN \pi$ channel. In \fig{\ref{noCoulomb}} we see the large contribution of this process to the width at low energies. Such a three body  $NN\pi$ interaction is not included in the vacuum approximation. At higher energies, especially above $120 \MeV$ the vacuum approximation coincides with our results, since there the dominant process is the two body process $N\pi\rightarrow \Delta$.

We conclude that the naive vacuum approximation is qualitatively and quantitatively not reliable in the energy regime of $E_{\mathrm{kin}} \lesssim 70 \MeV$ where multi-body collisions, potential effects and Pauli-blocking are important.

\subsubsection{Comparison to optical model results.}

In \cite{Hecking} Hecking published calculations of the pion mean-free-path \cor{based on} two different types of\cor{ phenomenological optical} potentials $V_{\mathrm{opt}}$. The first parameter set for an optical potential is taken out of Stricker's analysis \cite{stricker}, the second one is calculated by Chai and Riska \cite{ChaiRiska}\del{ using a different parameterization of the optical potential}. Starting from equation (\ref{MeanFreeDef}) one can calculate the imaginary part of the momentum using the dispersion relation $p^{2}=E^{2}- \left(m^{2}+2E\; V_{\mathrm{opt}}(p,E)\right)$ approximating $\Pi=2E\ V_{\mathrm{opt}}$ by ignoring terms of the order $O\left(V_{\mathrm{opt}}^2/m^2\right)$. Instead of solving the real and imaginary parts of the dispersion relation, Hecking approximates the real part of the dispersion relation by the vacuum solution $\Re\left[p^{2}\right]=E^{2}-m^{2}$ and uses this approximation in the equation for the imaginary part $\Im\left[p^{2}\right]=-\Im\left[2E\ V_{\mathrm{opt}}(p,E=\sqrt{p^{2}+m^{2}})\right]$. This last equation now defines also $\Im\left[p\right]$, which can be used in equation (\ref{MeanFreeDef}) to obtain the mean free path. To minimize Coulomb effects, he defines an averaged mean free path
\[
\lambda_{\mathrm{avg}}=\frac{1}{2}\left(\lambda_{\pi^{+}}+\lambda_{\pi^{-}}\right) \; .
\]
In order to compare to his results we have performed a calculation without Coulomb forces. 
In \fig{\ref{CoulVergleich}}, Hecking's results are compared to the full BUU calculations. As a most prominent feature one observes that our results are much lower than Hecking's over the full energy range. The work by  Mehrem et al.~\cite{MehremRadi} explains this discrepancy.
Solving the full dispersion relation with an optical potential calculated by J. A. Carr \footnote{The parameters are given in~\cite{MehremRadi}}, it reports a mean free path of the pion which is \cor{smaller than} Hecking's result. In this work the large difference is attributed to Hecking's approximations described above.  

As a benchmark for our model, the result of Mehrem et al.~\cite{MehremRadi} for the mean free path is shown in \fig{\ref{MeanFreePlot}} in comparison to our result obtained with the BUU simulation~\footnote{The real part of $V_{\mathrm{opt}}$, which is used in Mehrem's work, is very strong at low energies with $V_{\mathrm{opt}}\approx 48 \MeV$ at $E_{\mathrm{kin}}^{\mathrm{vac}}=0$. A direct comparison by including this real part in our semi-classical model is, therefore, not possible.}. Including the hadronic potential in our model we notice that the mean free path decreases considerably at low pion kinetic energies. There the hadronic potential becomes repulsive with $V\gtrsim E_{kin}$ and the semi-classical model breaks down, as mentioned earlier. In contrast to this, quantum mechanics allows for tunneling, i. e. propagation, into such classically forbidden regions. 


A proper discussion of the mean free path is obviously important in the analysis of experiments with final state pions being produced inside the nuclear medium. When working in a quantum mechanical framework solving the full dispersion relation, as done e.g. by Mehrem et al., is very important.  Ad hoc assumptions have a large effect on the pion mean free path and this has a dramatic effect on any final state analysis.

Since the mean free path is not directly observable, it is \cor{ultimately} an open question whether transport gives a reasonable mean free path for the pion. This can only be answered by experiment. A test for our model assumptions will\cor{, therefore,} be absorption cross sections which \cor{we address} in the next paragraph.
\subsection{The scattering of pions off nuclei}\label{scattering}
\begin{figure}[h!]
\centering
\includegraphics[]{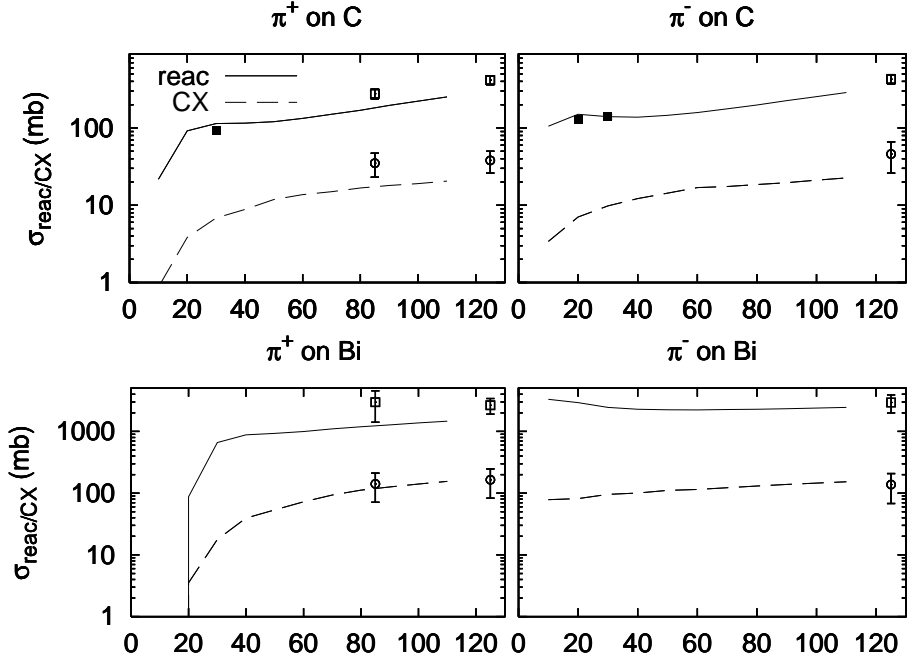}
\caption{Reaction ("reac"; solid lines) and charge exchange ("CX"; dashed lines) cross sections for $^{12}C$ and $^{209}Bi$. The data points are taken from \cite{ashery} (open circles: $\sigma_{CX}$, open squares: $\sigma_{\mathrm{reac}}$) and  \cite{Friedman:1991it} (full quares: $\sigma_{\mathrm{reac}}$).}
\label{reacData}
\end{figure}

\red{Low-energy pion scattering experiments have been studied extensively with elementary targets (e.g.~\cite{Carter:1971tj,Davidson:1972ky,Kriss:1999cv,Sadler:2004yq}), however, there exist only a few data points for pions scattering off complex nuclei~\cite{Carroll:1976hj,Clough:1974qt,Wilkin:1973xd,ashery,Friedman:1991it,nakai,byfield}. 

The total cross section~\cite{Carroll:1976hj,Clough:1974qt,Wilkin:1973xd} includes the coherent  contribution which would have to be 
substracted, before comparing them to our results (for a discussion on this issue see e.g.~\cite{ashery}). We study here solely 
reaction, charge exchange (CX) and absorption cross sections. 
In the considered energy regime, reaction and CX cross section measurements are rare~\cite{ashery,Friedman:1991it}. In 
\fig{\ref{reacData}} we present our results for reaction and CX cross sections for $^{12}C$ and $^{209}Bi$. The overall agreement 
with the few existing data points is satisfactory.} \red{Fortunately, the experimental situation for absorption cross 
sections~\cite{ashery,nakai,byfield} is more promising. Therefore, this observable can be used to evaluate the quality of our model.} 
\begin{figure}[b]
\centering
\includegraphics[width=7cm]{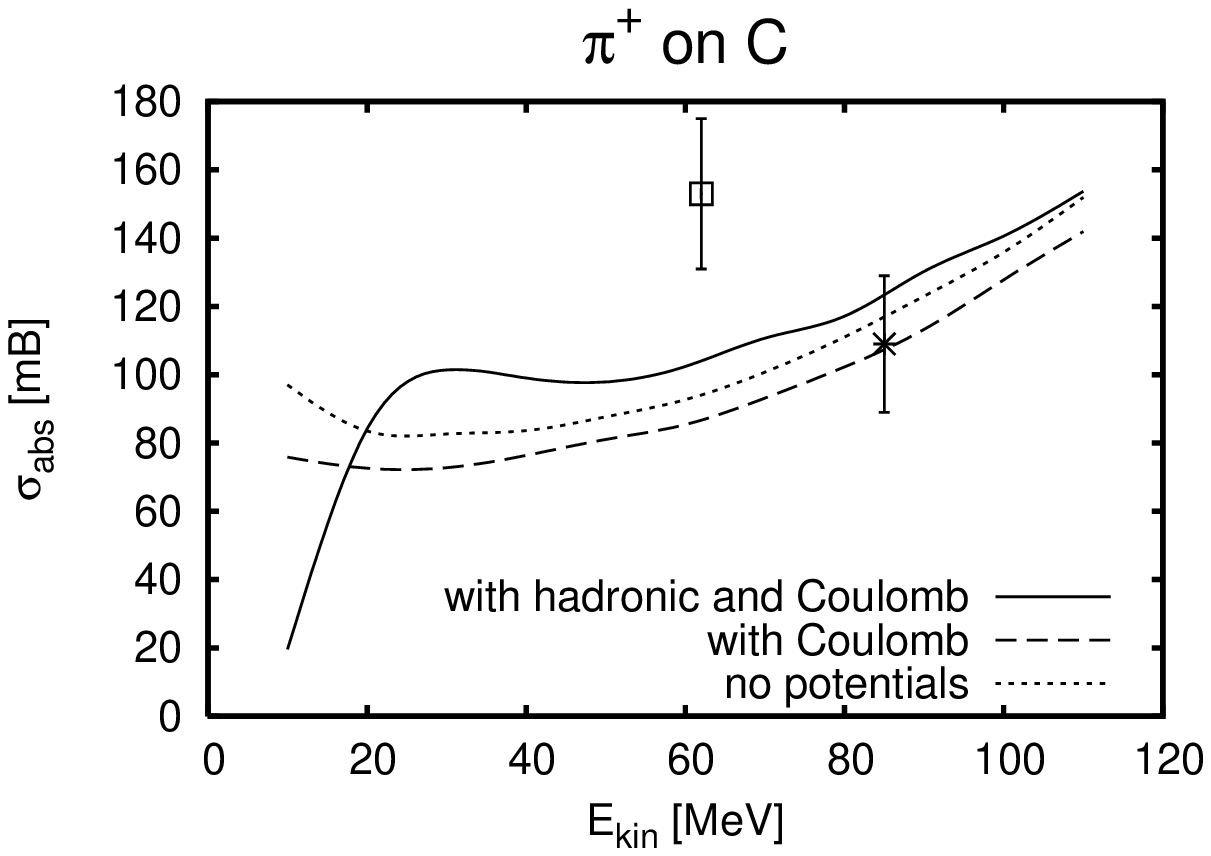}
\includegraphics[width=7cm]{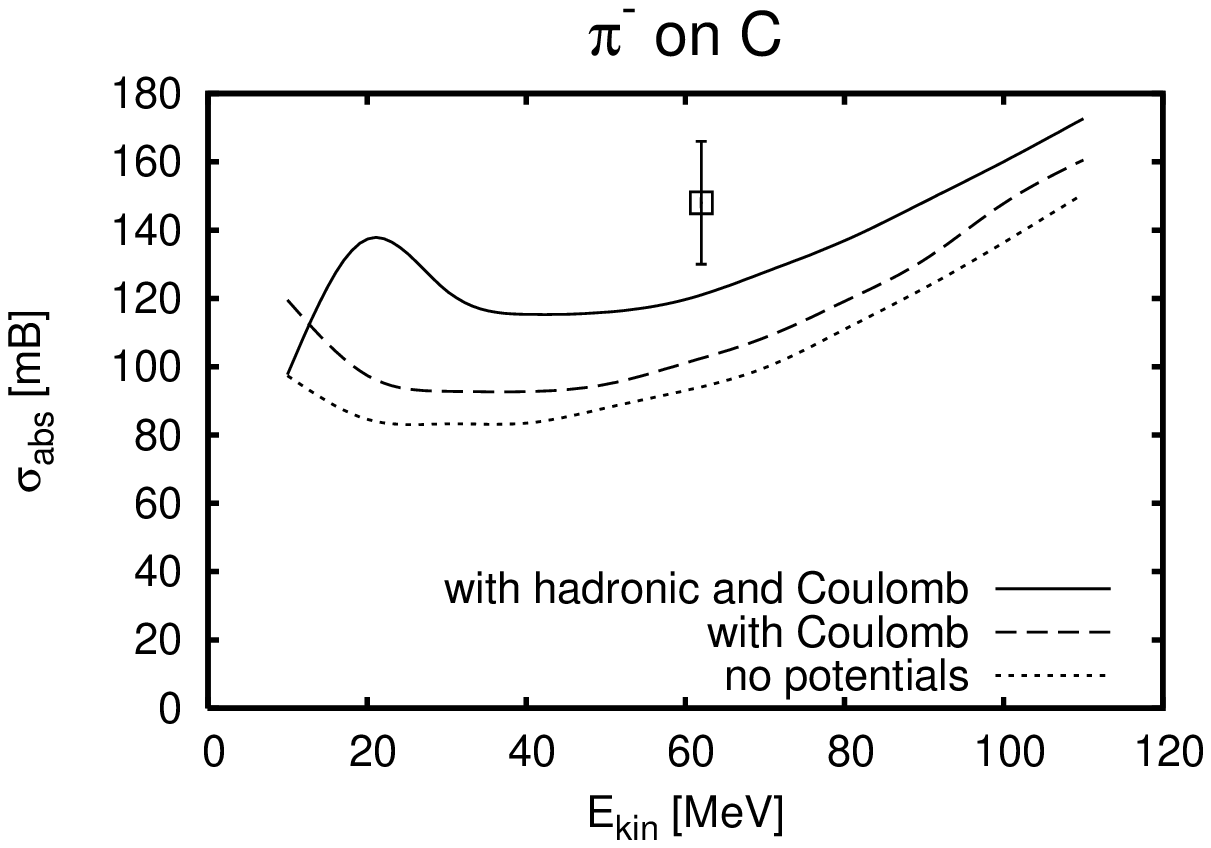}
\includegraphics[width=7cm]{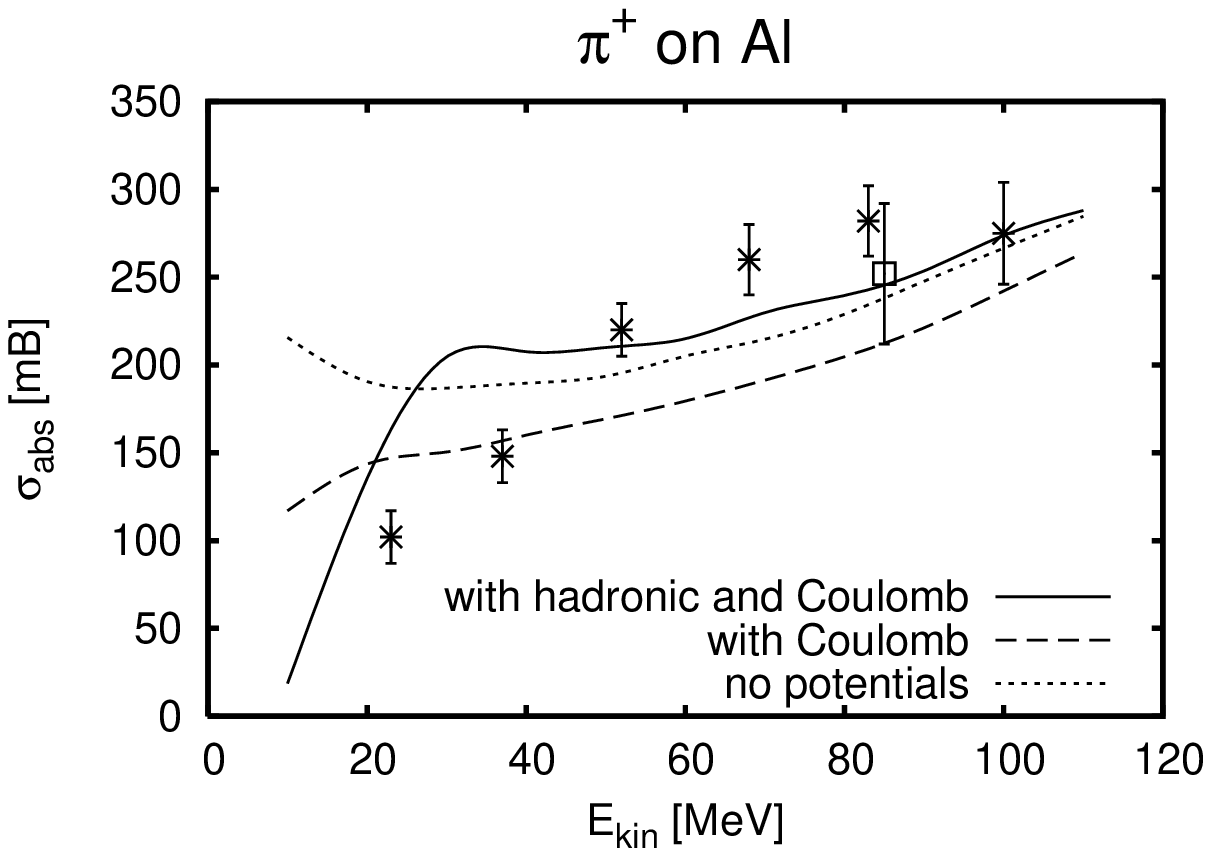}
\includegraphics[width=7cm]{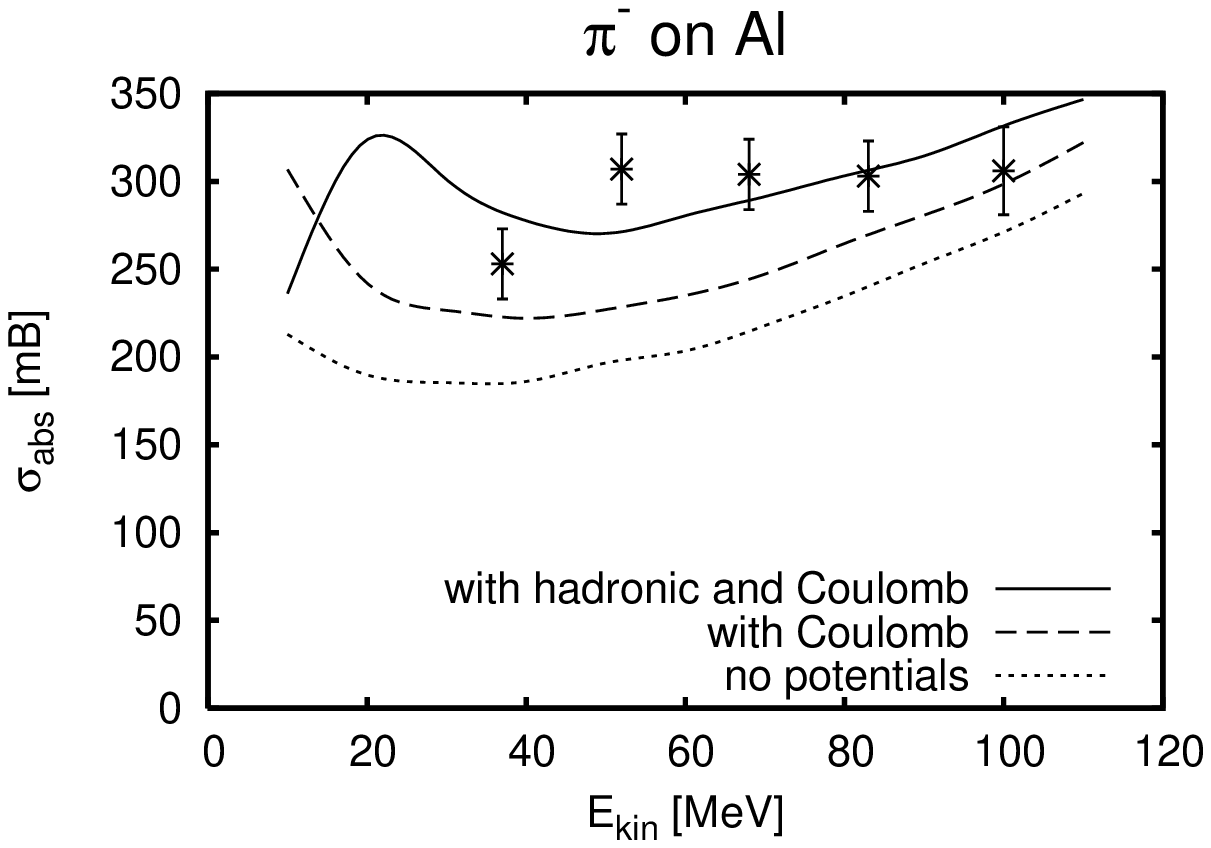}
\caption{Absorption on light nuclei depending on the choice of potentials for the pion. The data points are taken from \cite{ashery,nakai,byfield}. In all plots the standard hadronic potentials for the nucleon and $\Delta$ resonance are used. \del{We explicitly vary the inclusion of Coulomb potential and hadronic potential for the pion.}}
\label{absPlot1}
\end{figure}

In a quantum mechanical approach one cannot calculate absorption cross sections \red{in the general case. Only at very low 
energies, where the quasielastic contribution to the total reaction cross section is negligible, such an absorption cross section can be obtained(e.g.~\cite{osetlow}).} One thus relies on eikonal approximations to split the reaction cross section into a quasi-elastic and absorptive part. On the contrary, in the BUU simulation the calculation of an absorption cross section is straightforward due to the semi-classical treatment.

\begin{figure}[t]
\centering
\includegraphics[width=7cm]{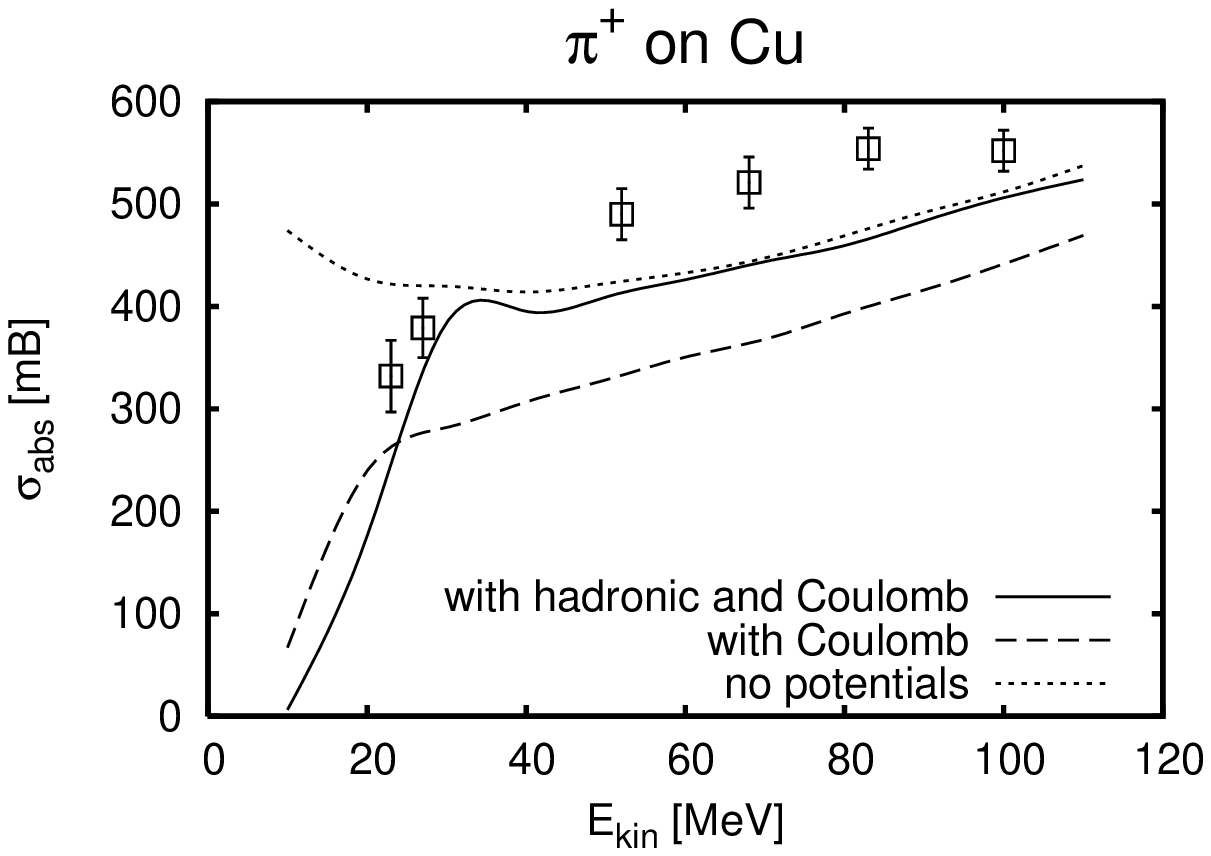}
\includegraphics[width=7cm]{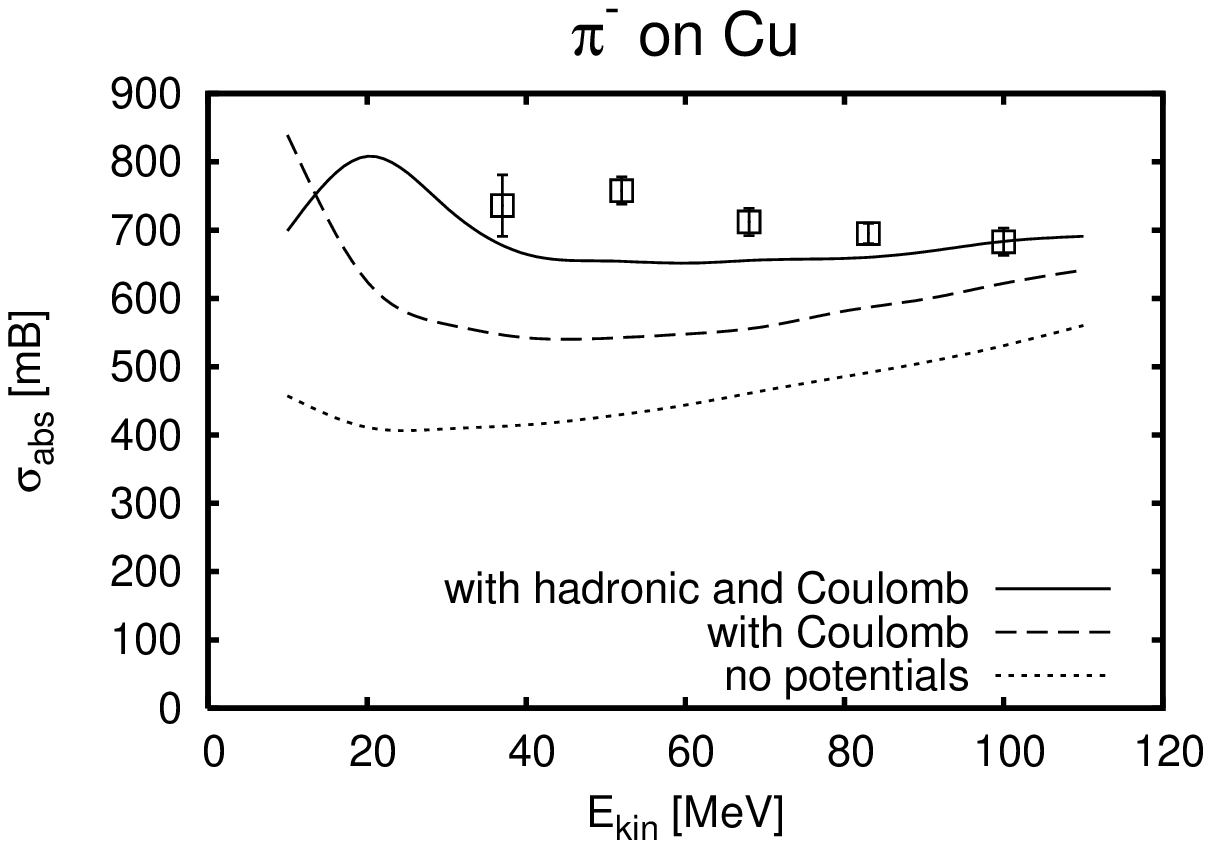}
\includegraphics[width=7cm]{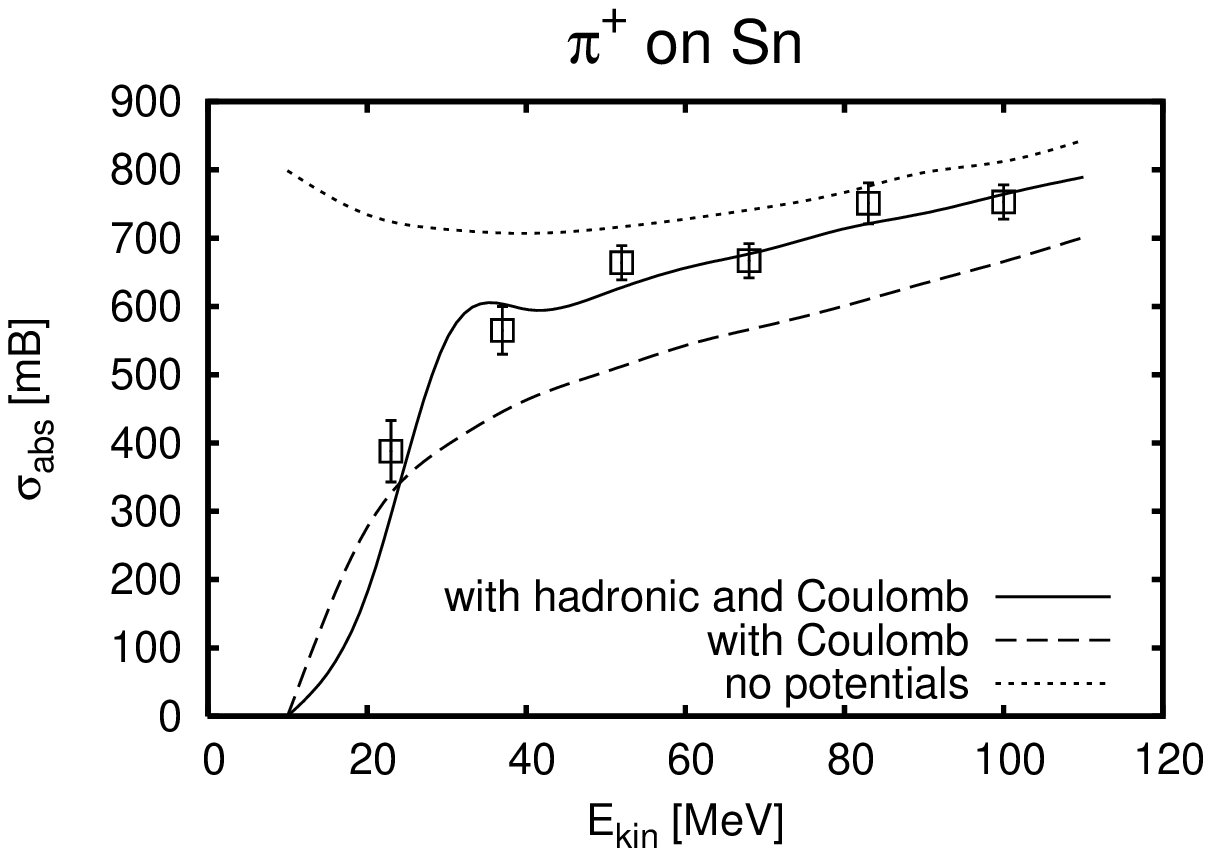}
\includegraphics[width=7cm]{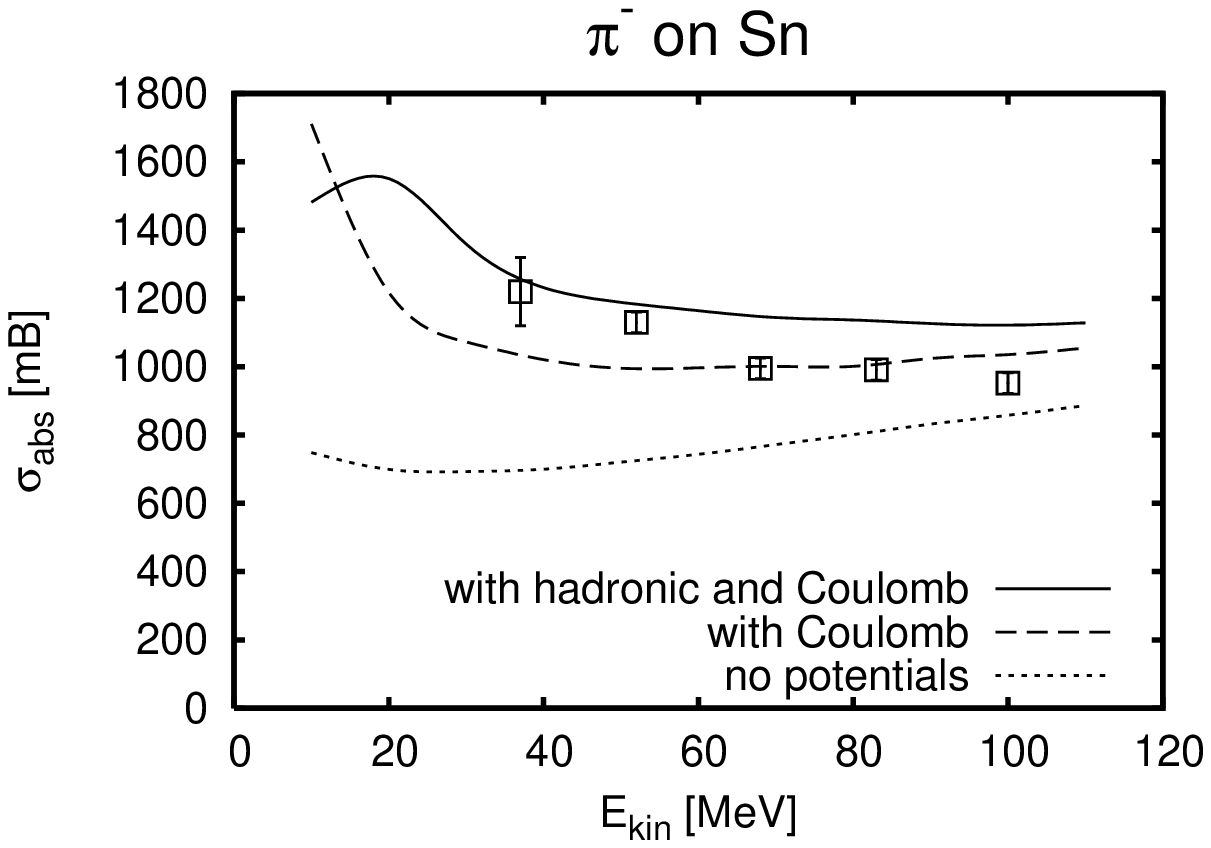}
\includegraphics[width=7cm]{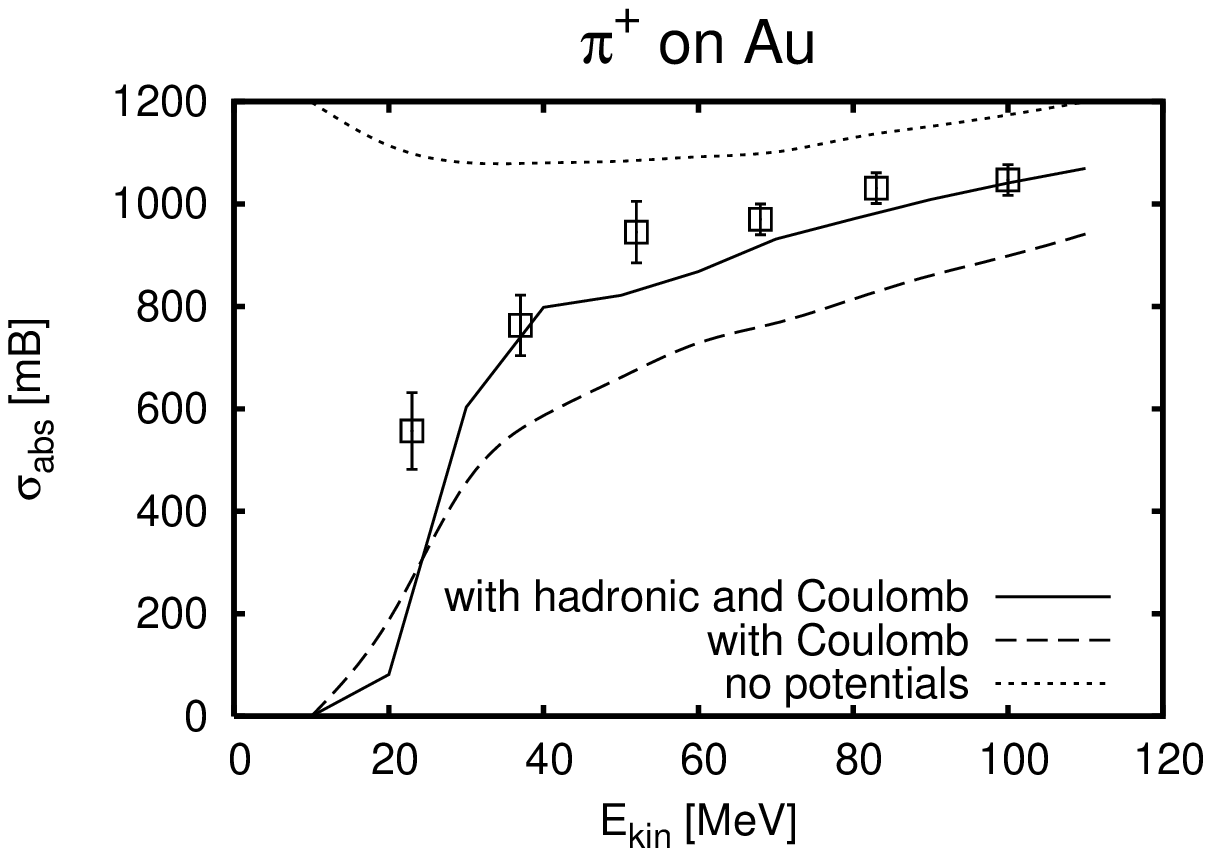}
\includegraphics[width=7cm]{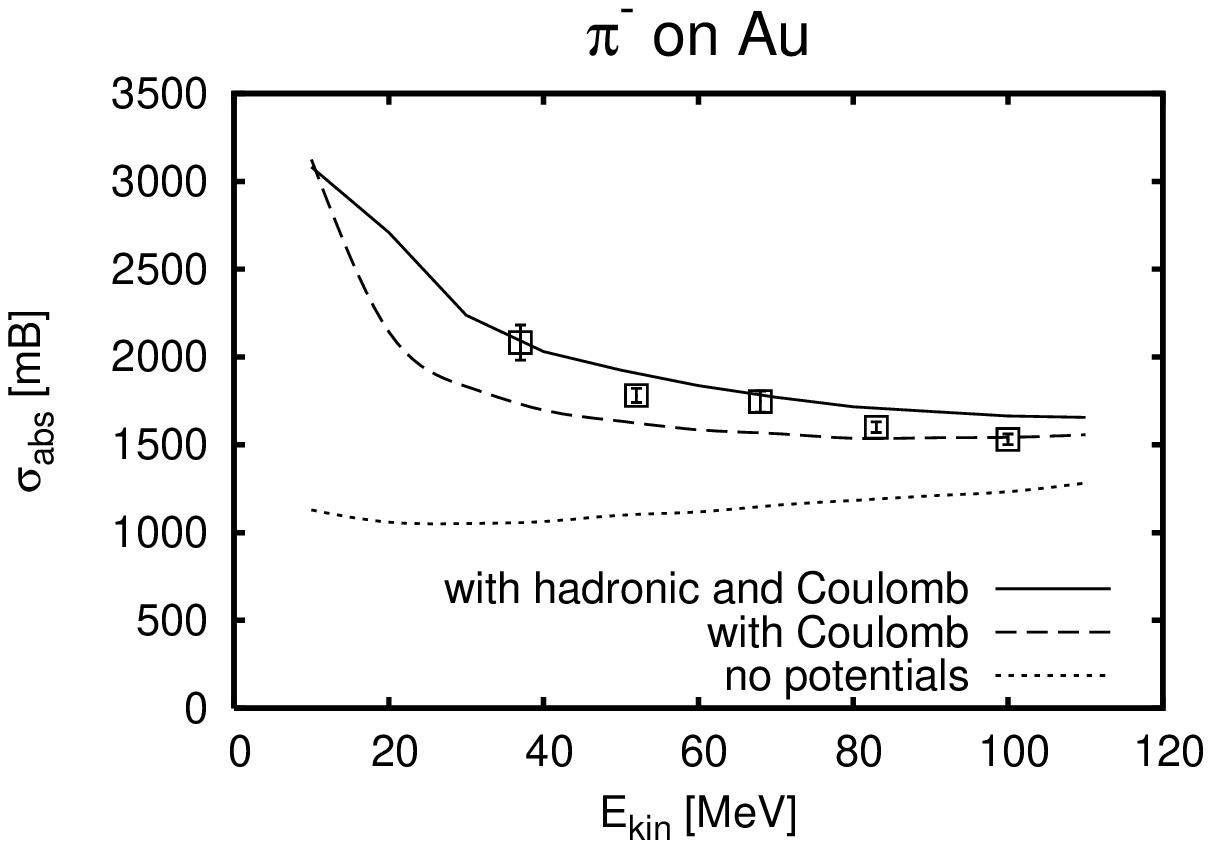}
\caption{Absorption on heavier nuclei depending on the isospin of potentials for the pion. Data points are taken from \cite{nakai}. In all plots the standard hadronic potentials for the nucleon and $\Delta$ resonance are used. We explicitly vary the inclusion of Coulomb potential and hadronic potential for the pion.}
\label{absPlot2}
\end{figure}

In figures \ref{absPlot1} and \ref{absPlot2} we show calculations for different nuclei and different bombarding energies of the 
pions. Comparing the curves obtained without any potential (dotted) to those with the Coulomb potential included (dashed), we see 
that the Coulomb potential alone has only a small influence on light nuclei (fig.~\ref{absPlot1}), but is very important at low 
energies for heavy nuclei (fig.~\ref{absPlot2}). Its long range leads to a sizeable deformation of the trajectories already long 
before the pions reach the nucleus. Therefore the negative pions can interact with the nucleus even if they have large impact 
parameter while the positive pions are deflected. In presence of the Coulomb potential we see a reduction of the cross section for 
the $\pi^{+}$ and a large increase of the cross section for the $\pi^{-}$. \red{This agrees to the findings of Nieves et 
al.~\cite{osetPionicAtoms}, who pointed out the relevance of the Coulomb potential in their quantum mechanical calculation of 
absorption and reaction cross sections.}

When one includes the hadronic potential for the pion, another overall effect sets in. Once the pion enters the nucleus it is influenced by the short-range hadronic potential, which amounts to $-40 \MeV$ at high momenta and to $+20 \MeV$ at low momenta, as well as the Coulomb potential which amounts to roughly $\pm 10 \MeV$ in a medium size nucleus, and to roughly $\pm 20 \MeV$ \cor{in the case of Pb}. At very low energies the two potentials nearly compensate for the negative pion, while they add up to a strongly repulsive potential in the case of a $\pi^{+}$.

In \fig{\ref{absPlot1}} we see a fair agreement on C and Al with the data, when the hadronic potential is included. Addressing heavier nuclei, fig. \ref{absPlot2} shows the absorption cross sections for \cor{Cu, Sn and Au}. The overall agreement to data is very good if the hadronic potential is included, especially for \cor{Sn and Au}. The curves which include a hadronic potential for the pion show a prominent kink structure at roughly $30 \MeV$ ($20 \MeV$) for the positive (negative) pion. This kink is caused by the repulsive character of the pion potential at low energies. On one hand, the mean free path decreases rapidly at very low energies (compare \fig{\ref{MeanFreePlot}}) - this causes the absorption cross section to rise. On the other hand, the repulsive potential pushes the pions out of the nucleus or even reflects them. Below the kink the repulsive feature is more prominent; above the kink the decreasing mean free path is more important. 

As a conclusion, we find that it is critical to include Coulomb corrections.  On top, the absorption cross sections are sensitive to the hadronic potential of the pion, i.e. to the real part of the self energy in the medium. As we have already seen in \fig{\ref{CoulVergleich}}, the mean free path is quite insensitive to the hadronic potential except at very low energies. We thus conclude that the modification of the trajectories of the pion is the main effect of the hadronic potential. In its repulsive regime the hadronic potential pushes the pion outwards and the overall path of the pion inside the nucleus becomes shorter. Therefore the probability of absorption is decreased. The attractive behavior at larger energies causes the opposite effect. 

The overall agreement to the data \red{is satisfactory in spite of some discrepancies, especially in Al and 
Cu.} Considering the fact that the pions have very large wave lengths at such 
low energies, the success of \cor{the}\del{our} semi-classical BUU model is quite astonishing. Due to the large wave length one expects also many-body correlations and quantum interference effects to be important. \cor{Many-body effects are partially included via the mean fields acting on pions and baryons and the modification of the $\Delta$ width. Besides this we included only $1\leftrightarrow2$, $2\leftrightarrow2$ and $2\leftrightarrow3$ body processes in the collision term. We take the success as an evidence that no higher order correlations than the latter ones are necessary to describe pion absorption.}



\subsection{Photon-induced $\pi \pi$ production in nuclei}\label{piPiPhoto}
\subsubsection{Overview}
\begin{figure}[th]
\begin{center}
\includegraphics[width=7cm]{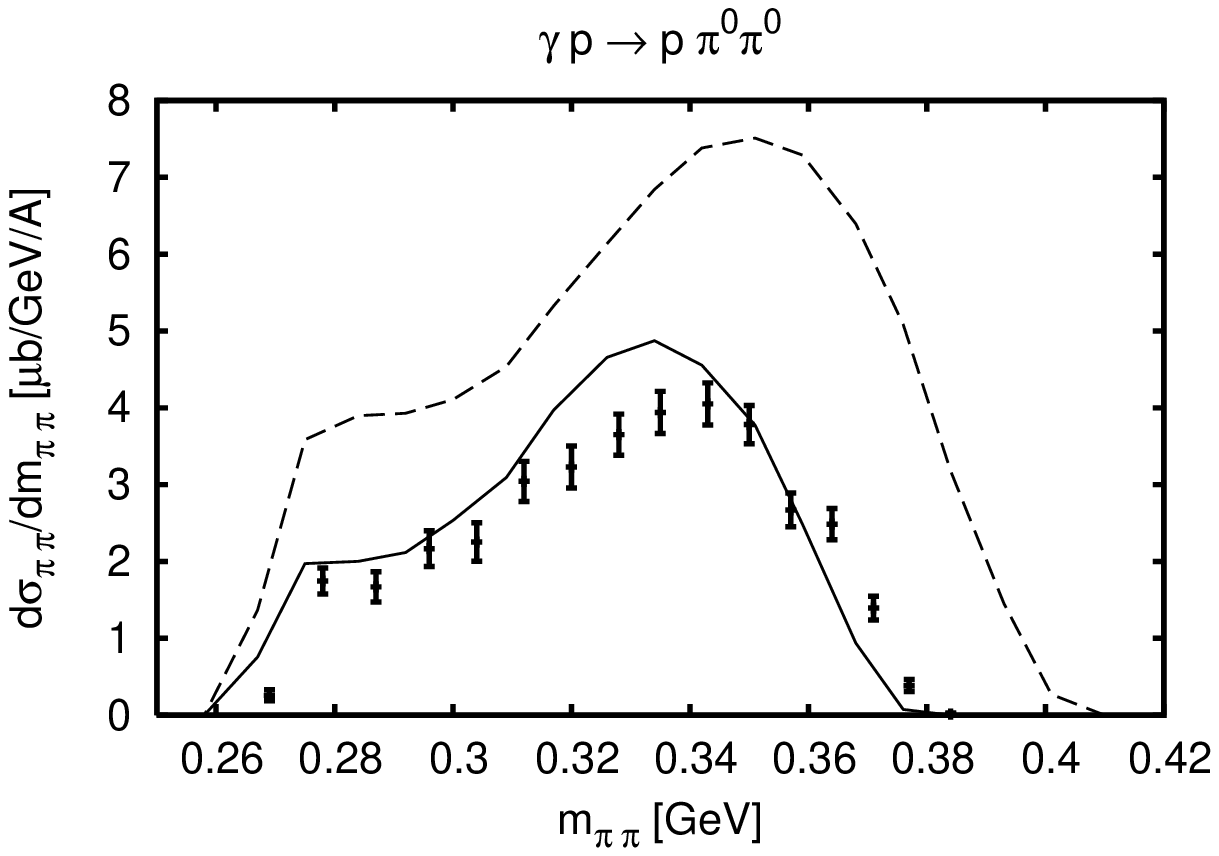}
\includegraphics[width=7cm]{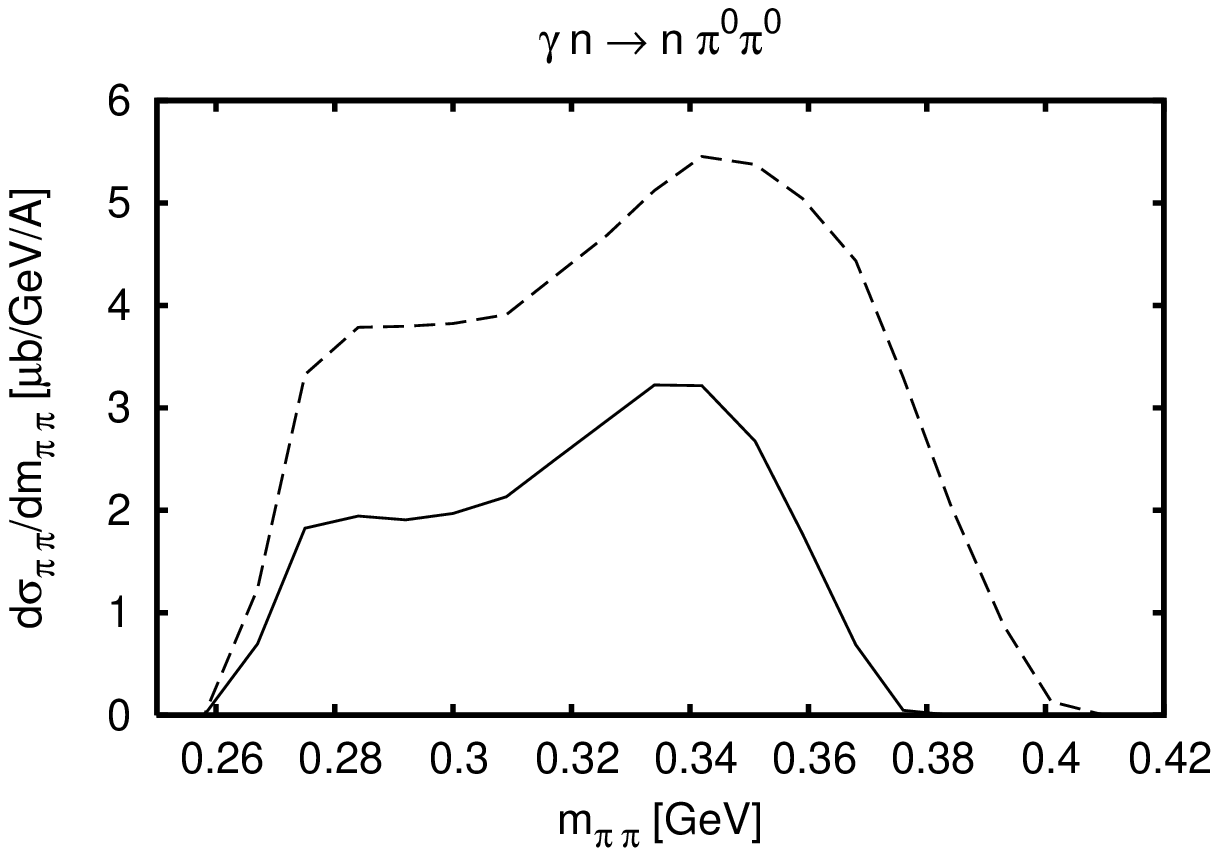}
\includegraphics[width=7cm]{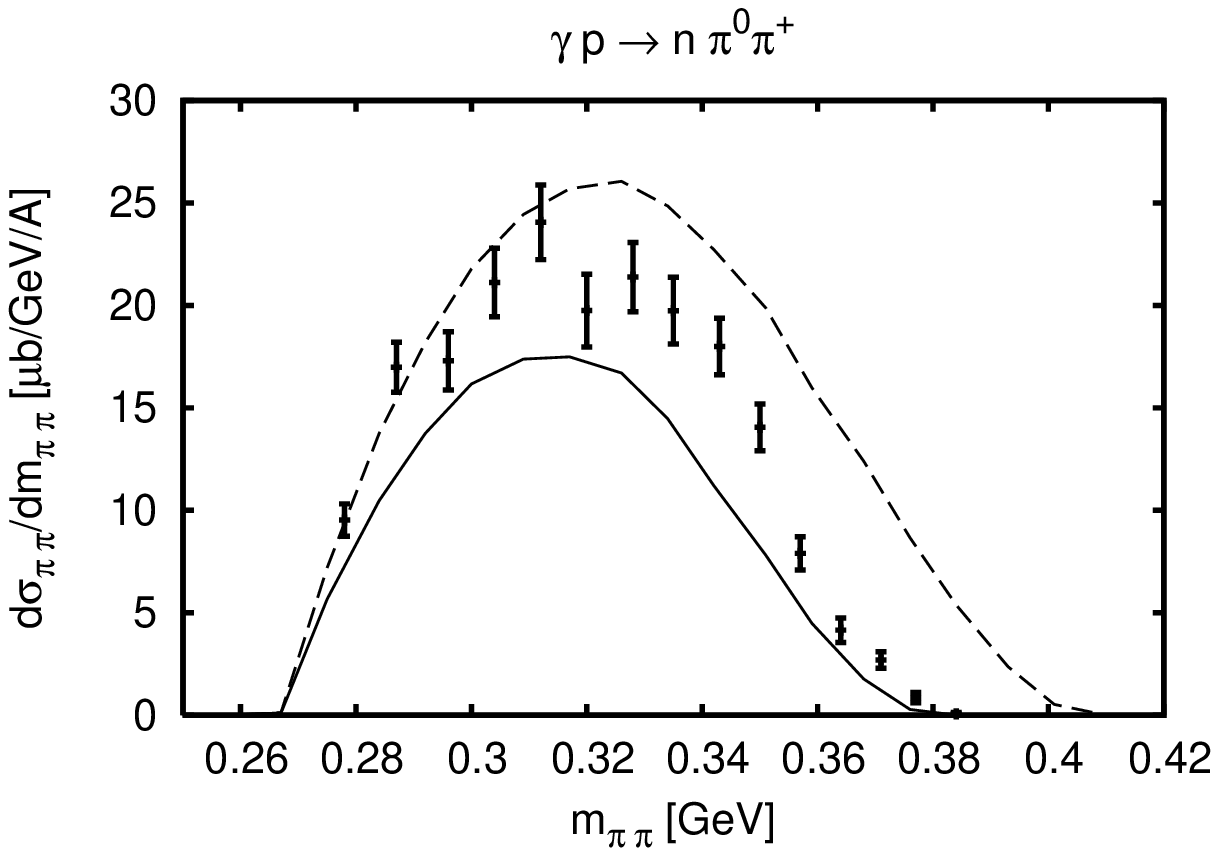}
\includegraphics[width=7cm]{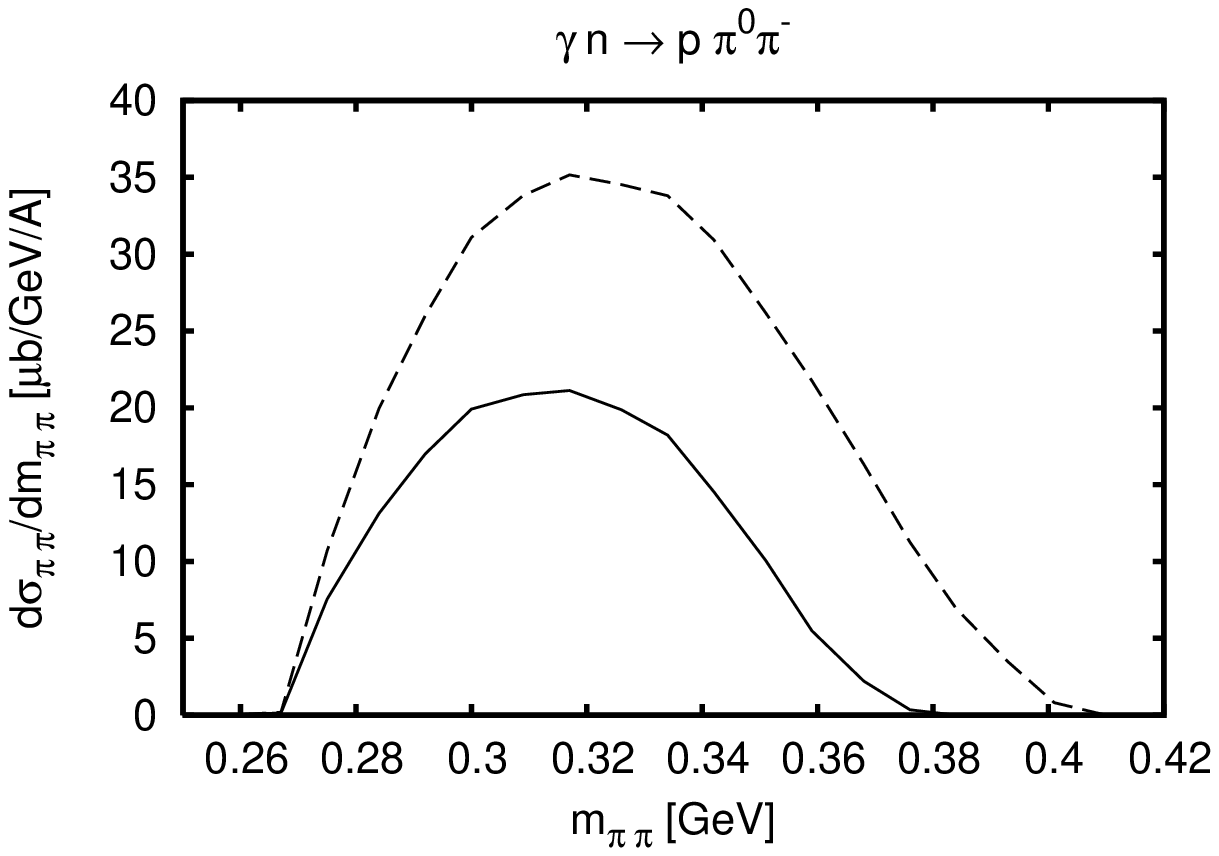}
\includegraphics[width=7cm]{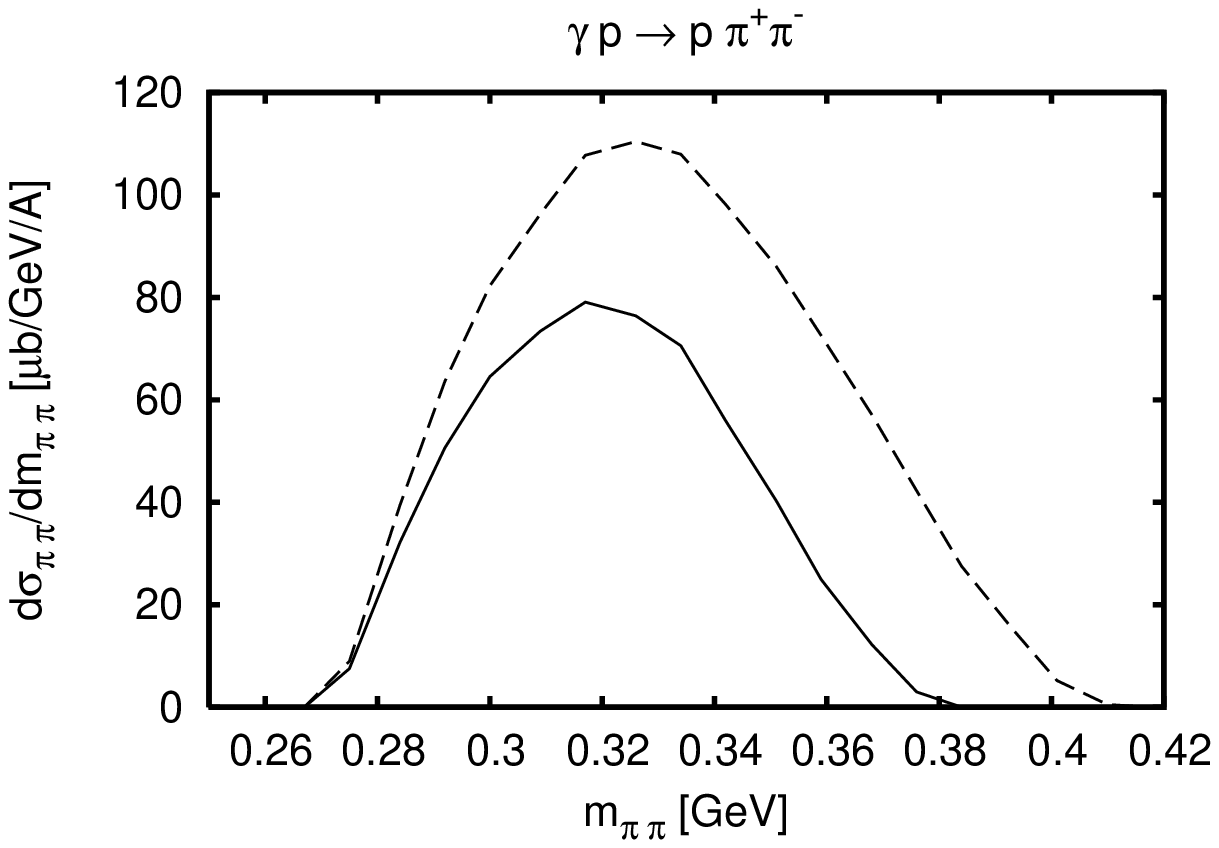}
\includegraphics[width=7cm]{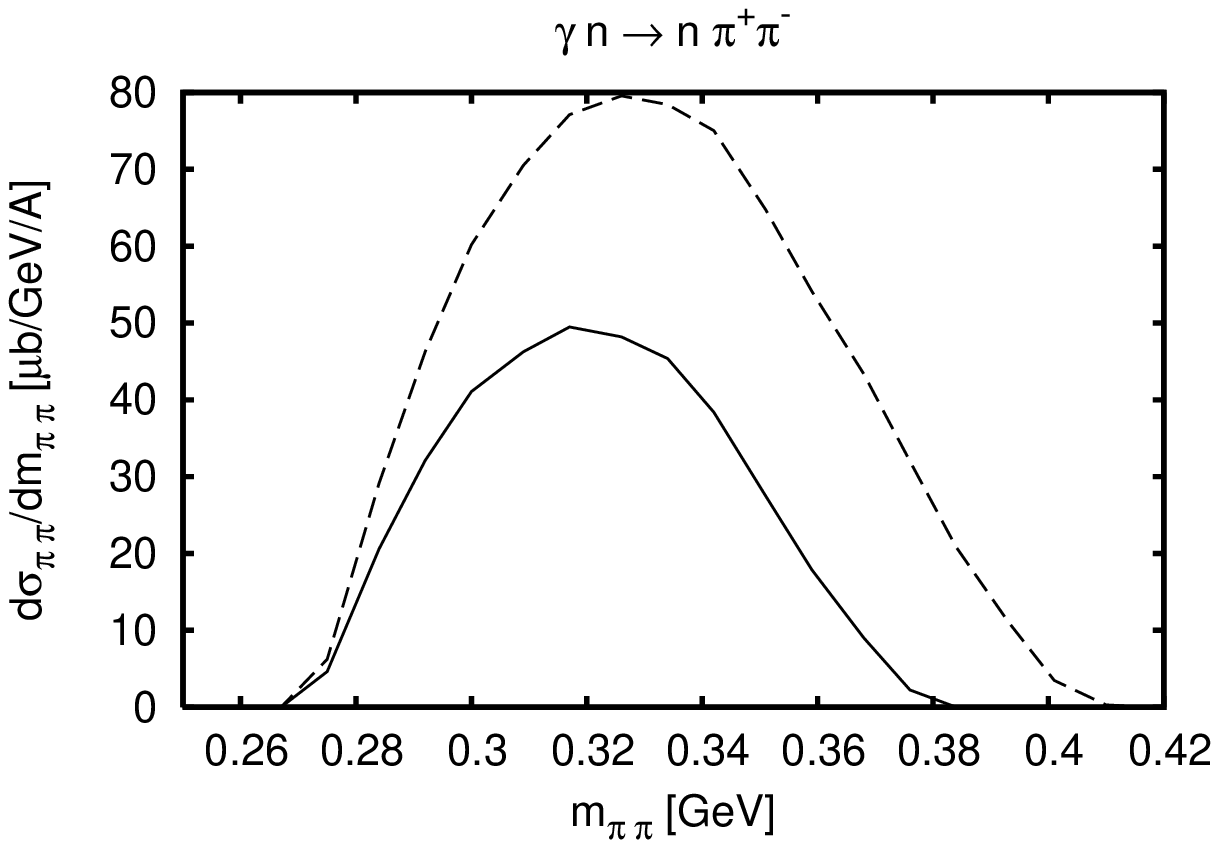}
\caption{Two pion invariant mass distributions for \pn and \pc photoproduction off the proton and neutron obtained from the model of \cite{twoPi23,twoPi24}. We show results on two different energy bins, $E_{\gamma}=0.4-0.5 \GeV$ (upper curve) and $E_{\gamma}=0.4-0.46 \GeV$ (lower curve). Data for $E_{\gamma}=0.4-0.46 \GeV$ are taken from \cite{Messch}.}
\label{twoPi_elementar}
\end{center}
\end{figure}
\begin{figure}
\begin{center}
\includegraphics[width=0.6\textwidth]{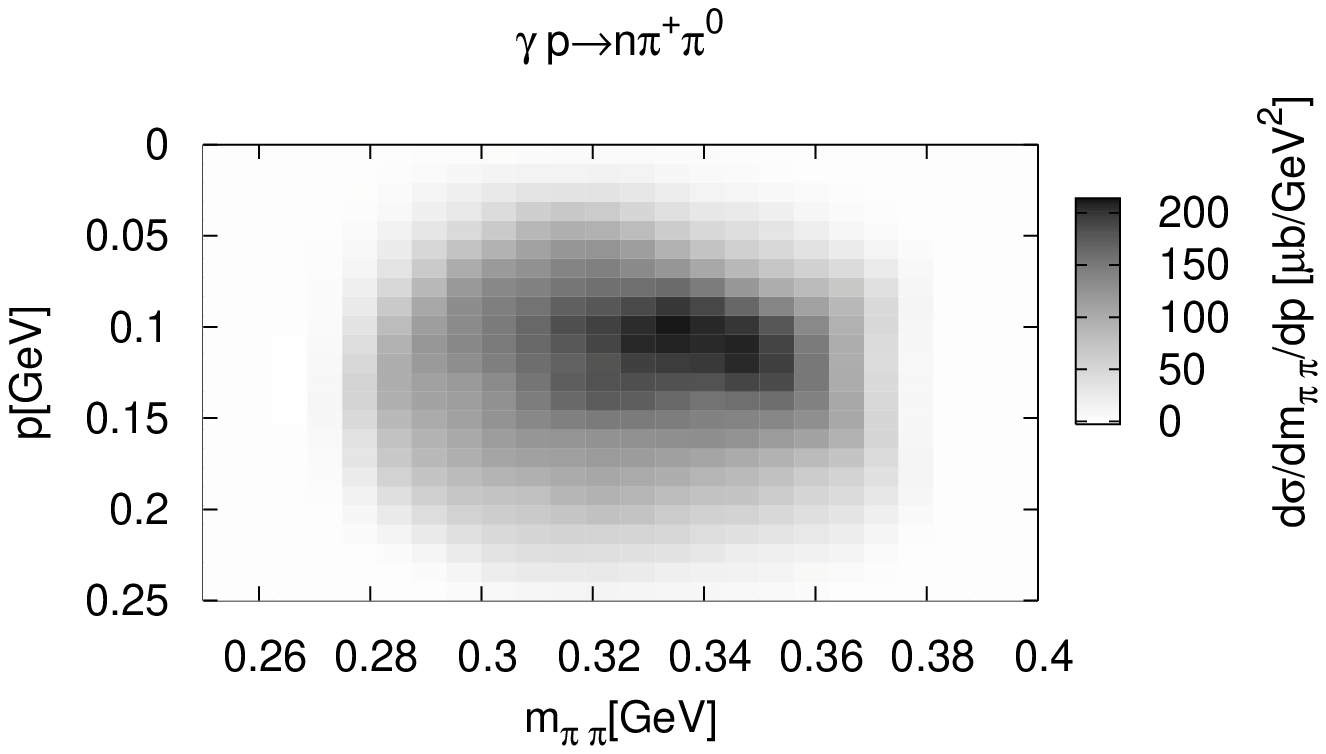} \\
\includegraphics[width=0.6\textwidth]{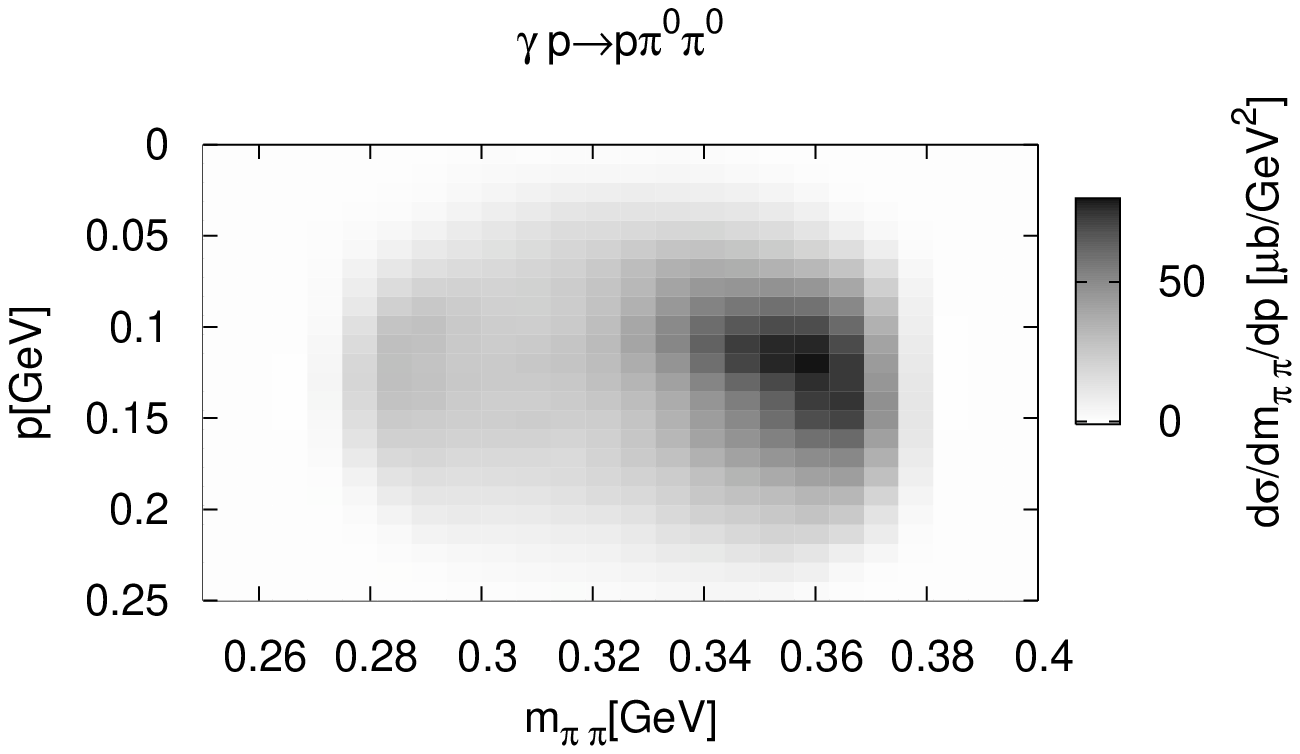}
\caption{The cross section $\frac{d\sigma}{dm_{\pi\pi} dp}$ for $\pi^+\pi^0$ and $\pi^0\pi^0$ photoproduction off the proton at $E_{\gamma}=0.45 \GeV$. The mass of the pion-pair is denoted $m_{\pi\pi}$ whereas $p$ denotes the absolute momentum of one of the pions.}
\label{dsigmadmdp}
\end{center}
\end{figure}
We now discuss the consequences of the pion mean free path on recent double pion production data off nuclei, which have been interpreted as a source of information about the spectral density in the $\sigma$ channel in nuclear matter~\cite{Messch,Roca:2002vd}. 

Already in~\cite{Muhlich:2004zj} we \cor{have} presented\del{ in detail} results on the double-pion photoproduction off nuclei using the transport approach. There the treatment of the pion final state differed from the treatment presented in the previous sections. In order to compare to a work by Roca et al.~\cite{Roca:2002vd}, we implemented there the same absorption probability for the pion as in their work. Therefore we did not propagate the resonances \cor{explicitly}, but rather used experimental data to describe the quasi-elastic $N\pi \rightarrow N \pi$ scattering. \cor{We also used} contributions to the imaginary part of the optical potential of \cite{osetlow} to model pion absorption. The aim there was to indicate the relevant role of conventional final-state interaction~(FSI) effects in two-pion photoproduction in nuclei. 

The authors of \cite{Roca:2002vd} achieved quite an impressive agreement with the experimental data, when studying double pion production in a many-body approach \cor{which allowed for pion-pion correlations}. However their treatment of the pion FSI was based upon a purely absorptive Glauber damping-factor calculated along straight line trajectories.

In the previous section we \cor{have shown} that our transport approach is still \cor{reliable}\del{meaningful} in the regime of long pionic wave lengths.\del{ Even if the semi-classical model is at its limits of applicability.} Since there is no fully quantum description for the incoherent $2\pi$ reaction available, we utilize the semi-classical final state description to simulate the $\pi \pi$-photo production experiment performed by the TAPS collaboration  \cite{Messch,SchadmandHabil,Schadmand:2005ji,Schadmand:2005xy}.
Originally \cor{this experiment} was intended to concentrate on the lower $400 \MeV\leq E_{\gamma} \leq 460 \MeV$ bin, focusing \del{to put the scope}on the threshold behavior of the $\pi \pi$ channel. Lately the experiment has evaluated a larger energy bin for the photon energy\cor{ in order to enhance statistics. Therefore,} also results for the larger bin with $400 \MeV\leq E_{\gamma} \leq 500 \MeV$ are presented here.
\subsubsection{Model for the elementary reaction}
For the elementary two pion production process on the nucleon we exploit the model of \cite{twoPi23,twoPi24}. This model provides a 
reliable input for the mass distributions of the pions in the elementary process. The total production cross sections are chosen 
according to the model in those channels where the agreement to data is satisfactory ($\gamma p\to n \pi^+\pi^0$,$\gamma n\to n 
\pi^+\pi^-$). In the remaining ones the model predicts lower total cross sections than those observed in experiment at the energy of 
interest for us. \cor{Hence, we use directly the data measured} by the TAPS and the DAPHNE collaborations 
\cite{Zabrodin:1997xd,Braghieri:1994rf,Kleber:2000qs,Wolf:2000qt,Kotulla:2003cx} to \cor{normalize the calculated cross sections, 
while we take the decay mass distributions from theory}\del{decrease the insecurity in the initial state cross sections}. The 
implemented elementary invariant-mass distributions are presented in \fig{\ref{twoPi_elementar}} for both a neutron and a proton 
target. \red{Presently, only data for invariant mass distributions on a proton target are available. The solid curves in 
}\red{\fig{\ref{twoPi_elementar}} 
represent the results of the elementary model~\cite{twoPi23,twoPi24} for $\pi\pi$ production for $E_\gamma=400-460 \MeV$ compared to 
the experimental data. The model describes the shape of the data quite well; for the semi-charged channel the cross section is 
somewhat underestimated. The dashed curves in \fig{\ref{twoPi_elementar}} show the predictions for $E_\gamma=400-500 \MeV$.} 

To illustrate the momentum distribution of the pions, we concentrate on a $\gamma p$ reaction with a fixed photon energy of $450 \MeV$. This distribution is shown in \fig{\ref{dsigmadmdp}} as function of both mass of the pion-pair and momentum of one of the pions. Most of the pions are produced with momenta around $120 \MeV$, corresponding to $E_{\mathrm{kin}}\approx 44 \MeV$, i.e. just at the lower limit of the applicability of our semi-classical model at $\rho_0$. \red{Note that the momenta of the produced pions range between $0-250 \MeV$. In nuclei, the FSI is therefore dominated by the $\Delta$ resonance, $S$-wave scattering and the $NN\pi\to NN$ process.}
\subsubsection{Results and discussion}
\begin{figure}
\begin{center}
\includegraphics[width=7cm]{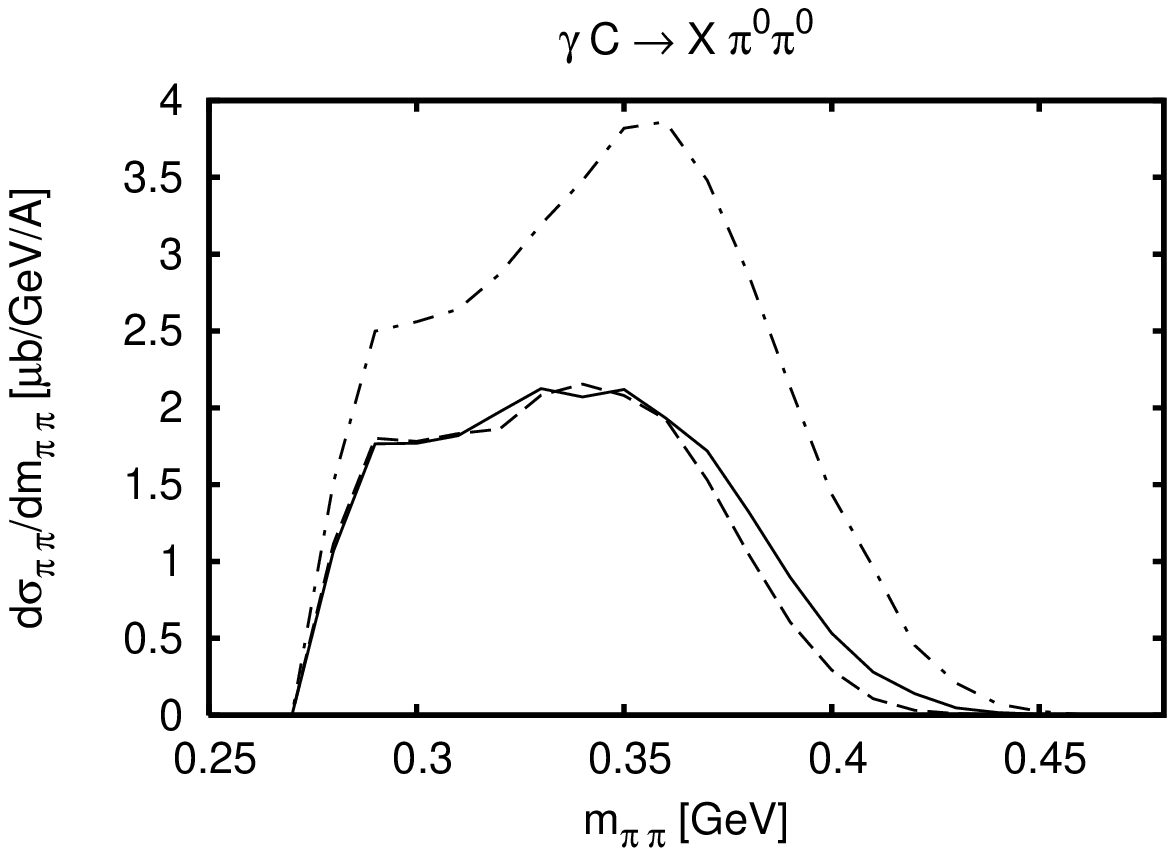}
\includegraphics[width=7cm]{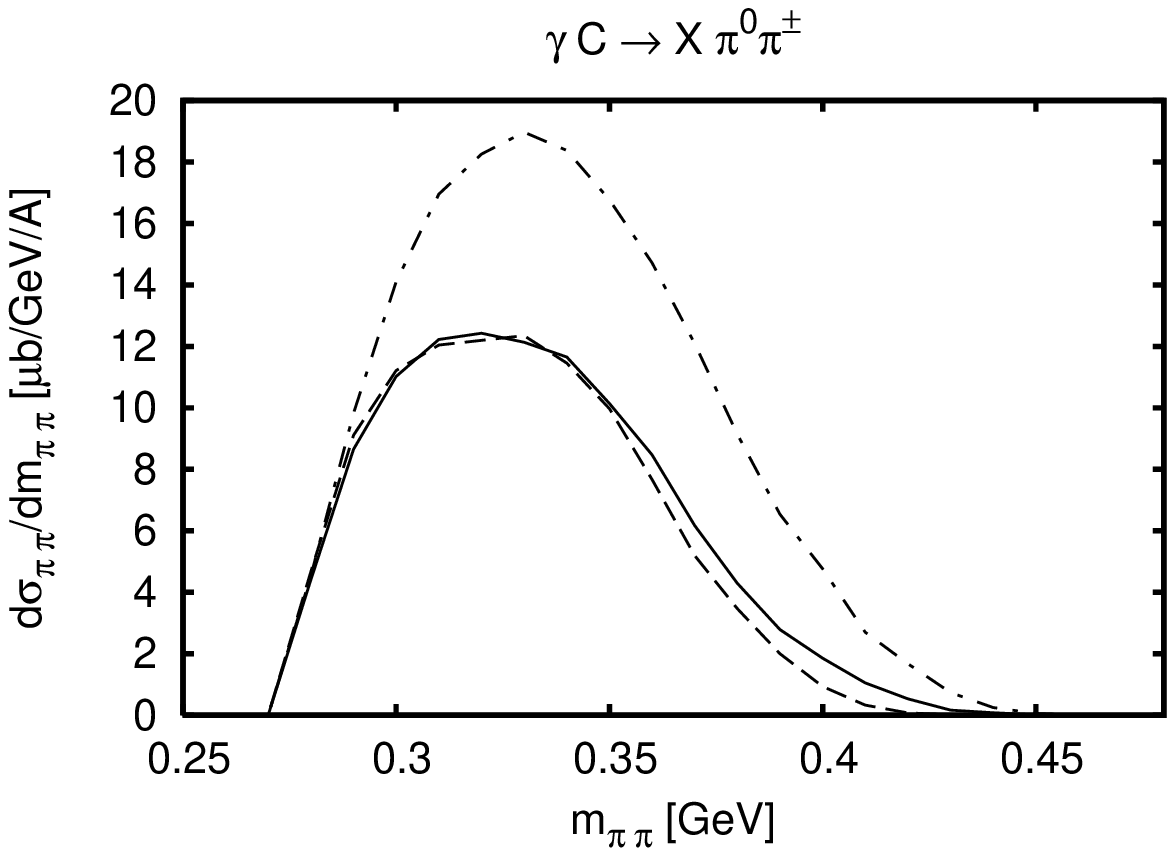}\\
\includegraphics[width=7cm]{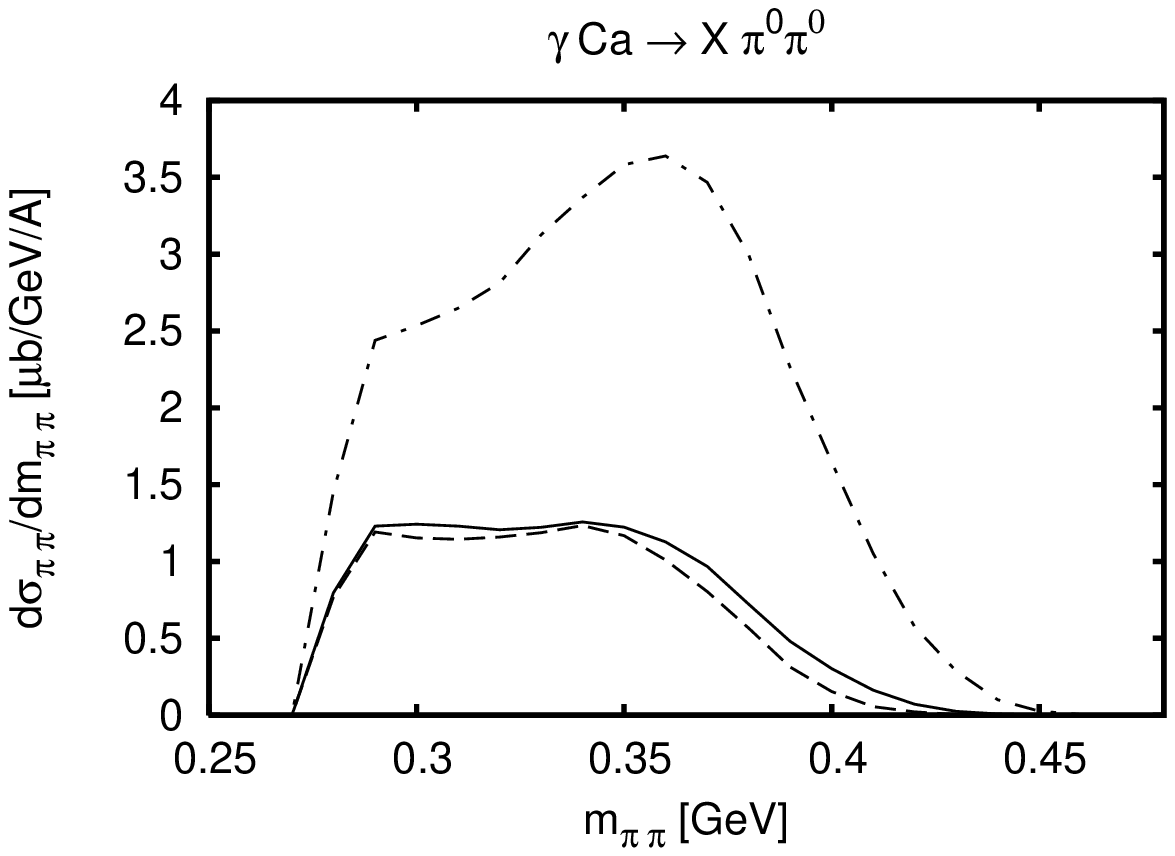}
\includegraphics[width=7cm]{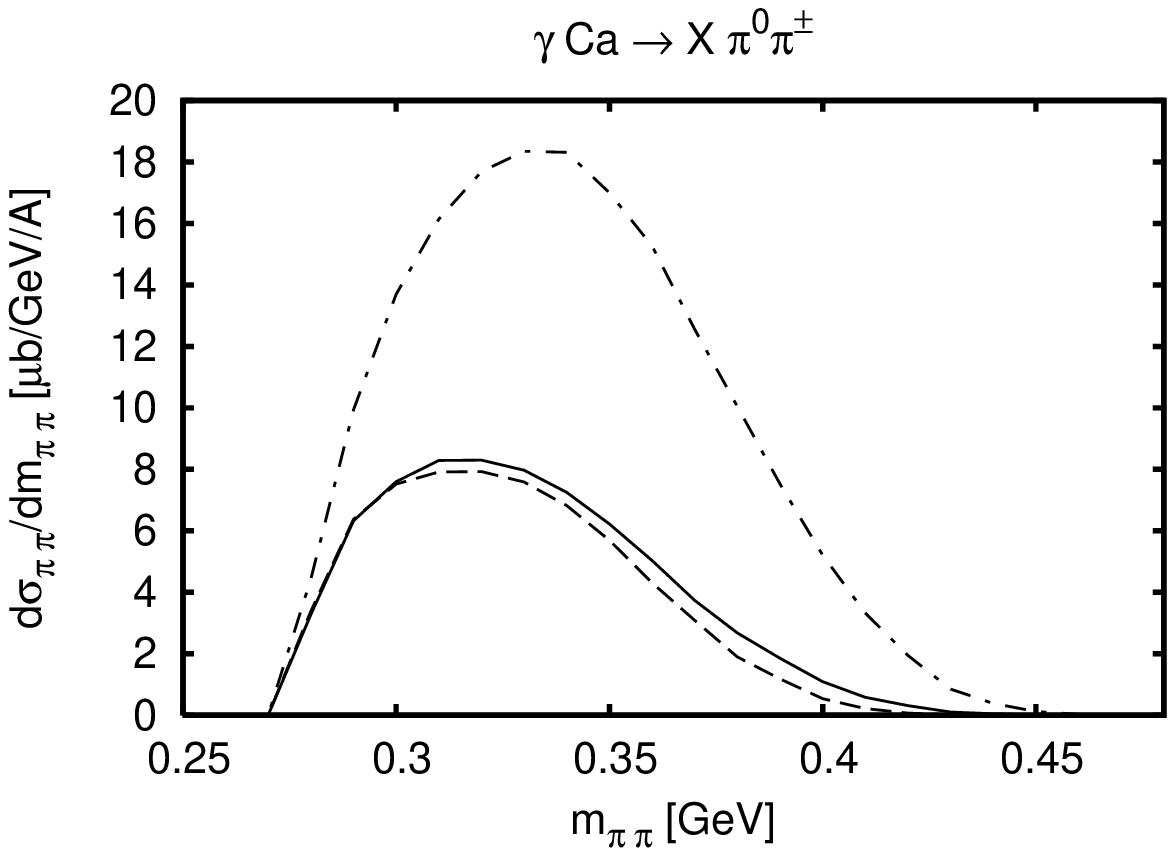}\\
\includegraphics[width=7cm]{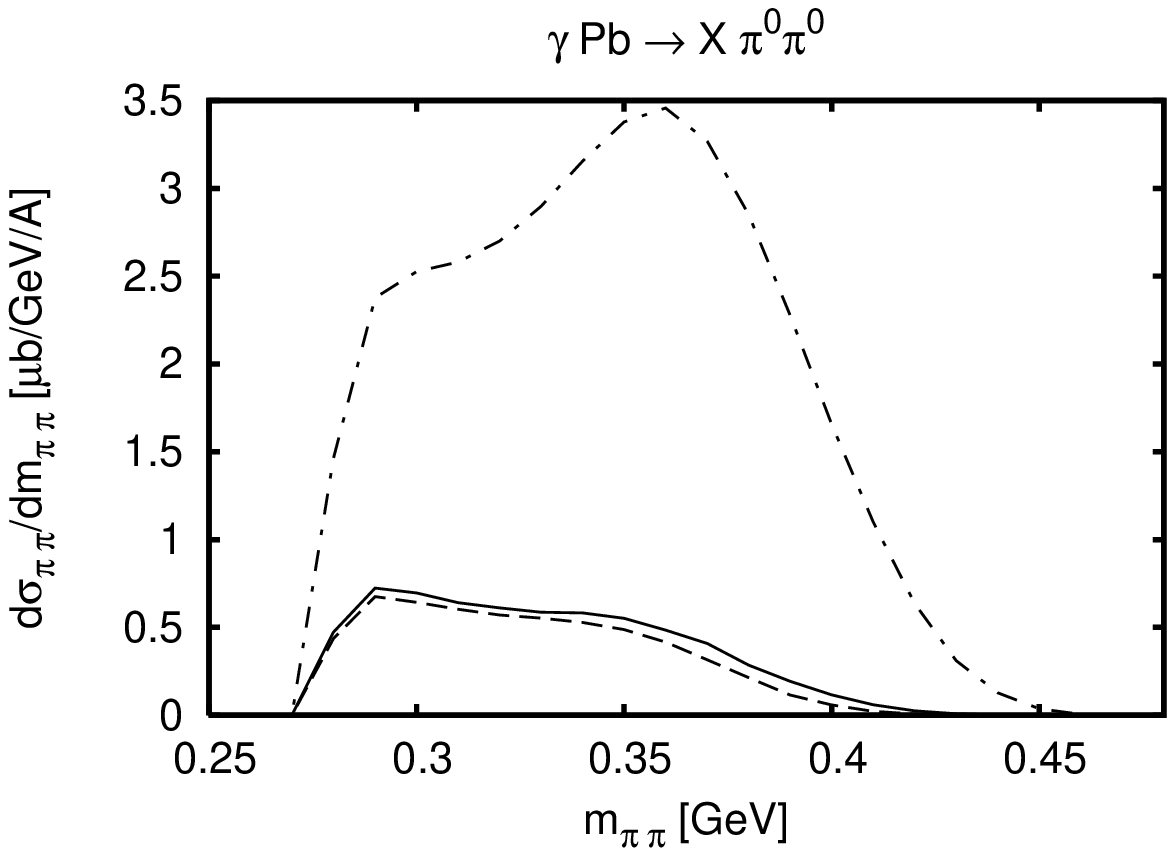}
\includegraphics[width=7cm]{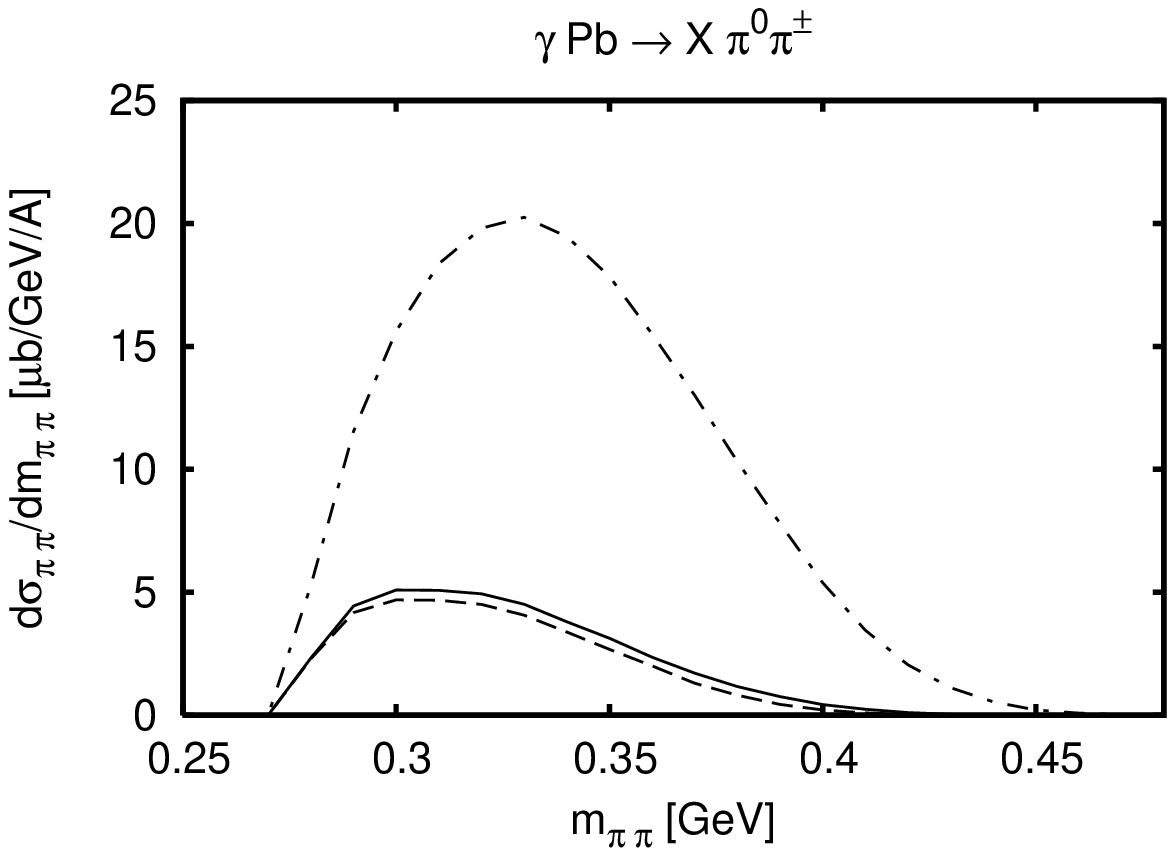}
\caption{Two pion invariant mass distributions for \pn and \pc photoproduction off $^{12}$C, $^{40}$Ca and $^{208}$Pb for $E_{\gamma}=0.4-0.46 \GeV$. The uncharged \pn channel is plotted on the left, the charged channel \pc is presented on the right hand side. We show results without final state interactions (dashed dotted) and results with FSI and respectively with (dashed) and without (solid) hadronic potential for the pion. }
\label{twoPi_kern}
\end{center}
\end{figure}
\begin{figure}
\begin{center}
\includegraphics[width=7cm]{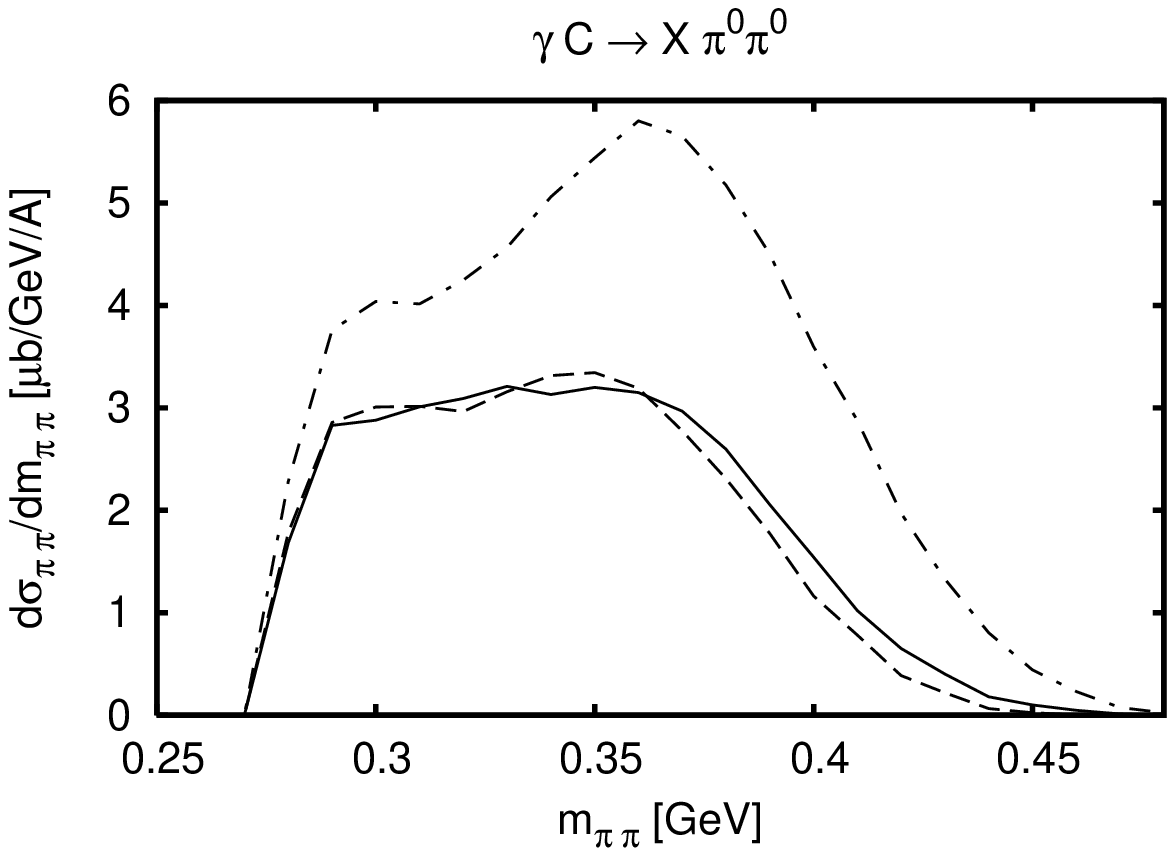}
\includegraphics[width=7cm]{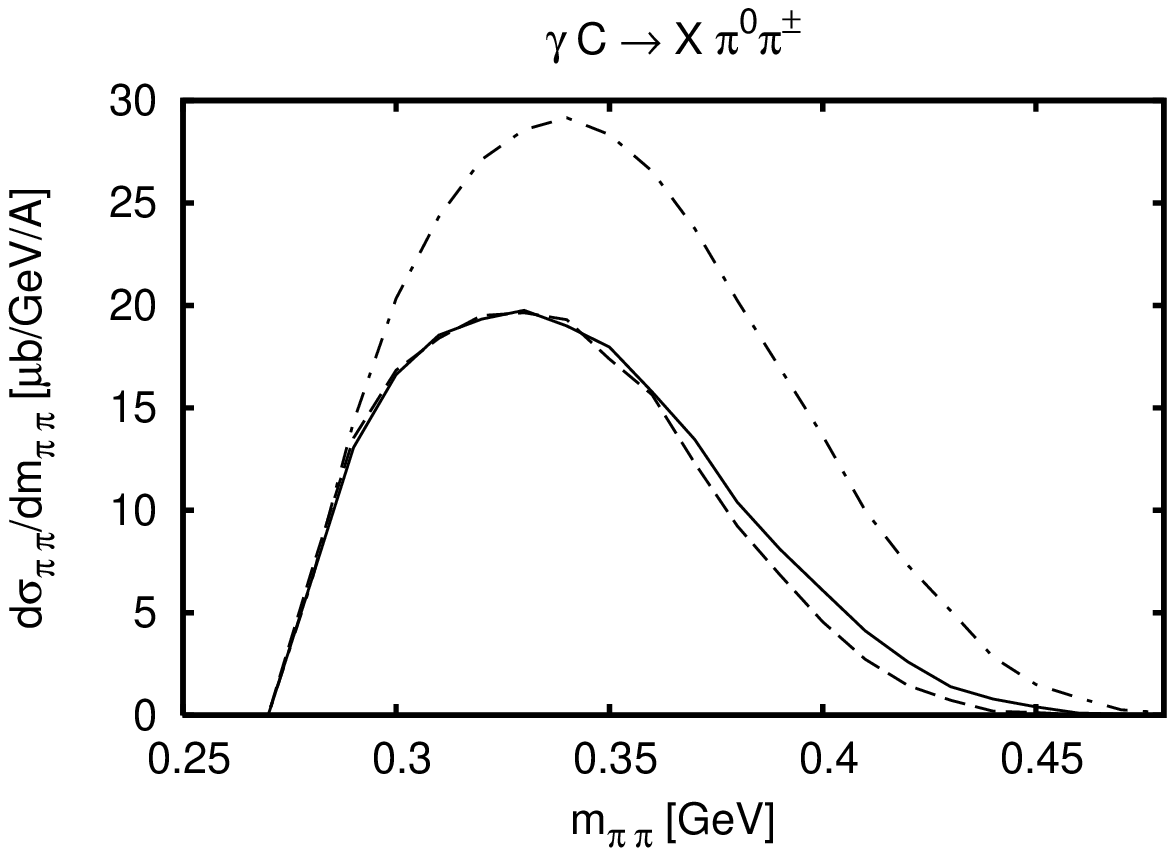}\\
\includegraphics[width=7cm]{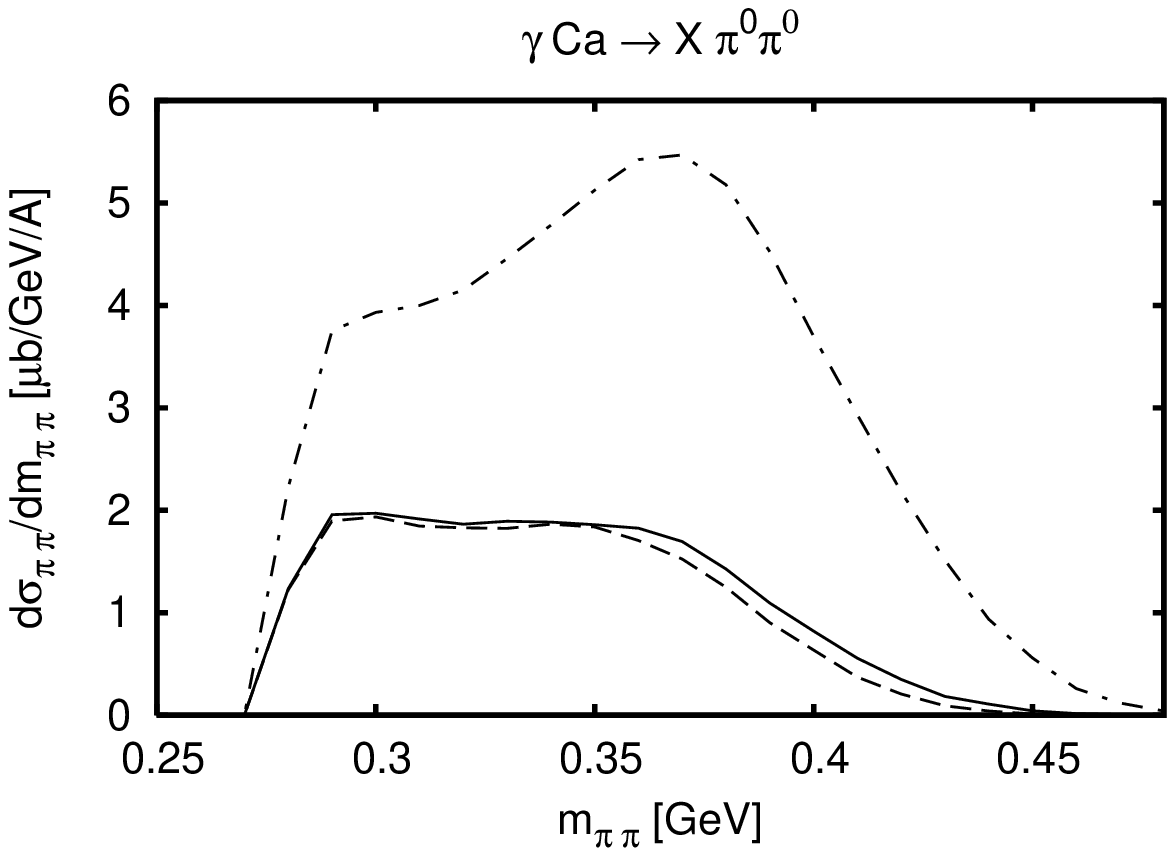}
\includegraphics[width=7cm]{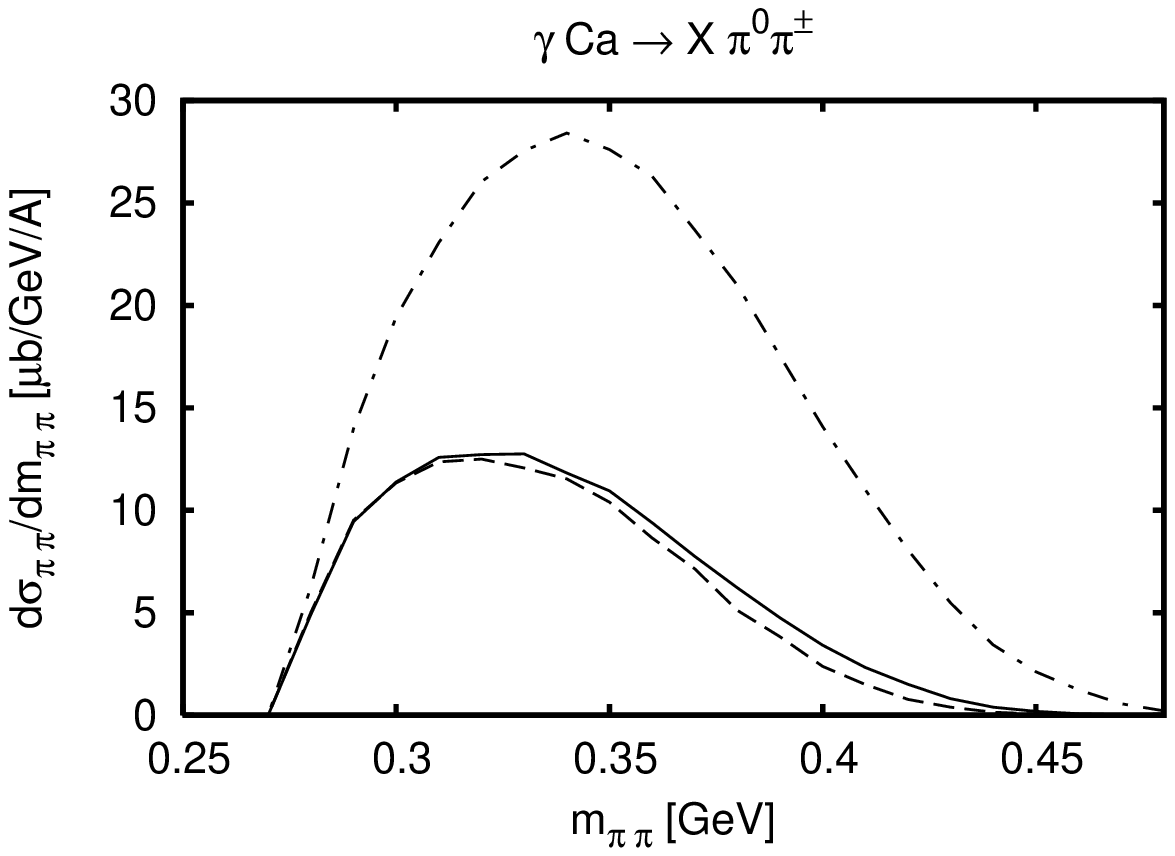}\\
\includegraphics[width=7cm]{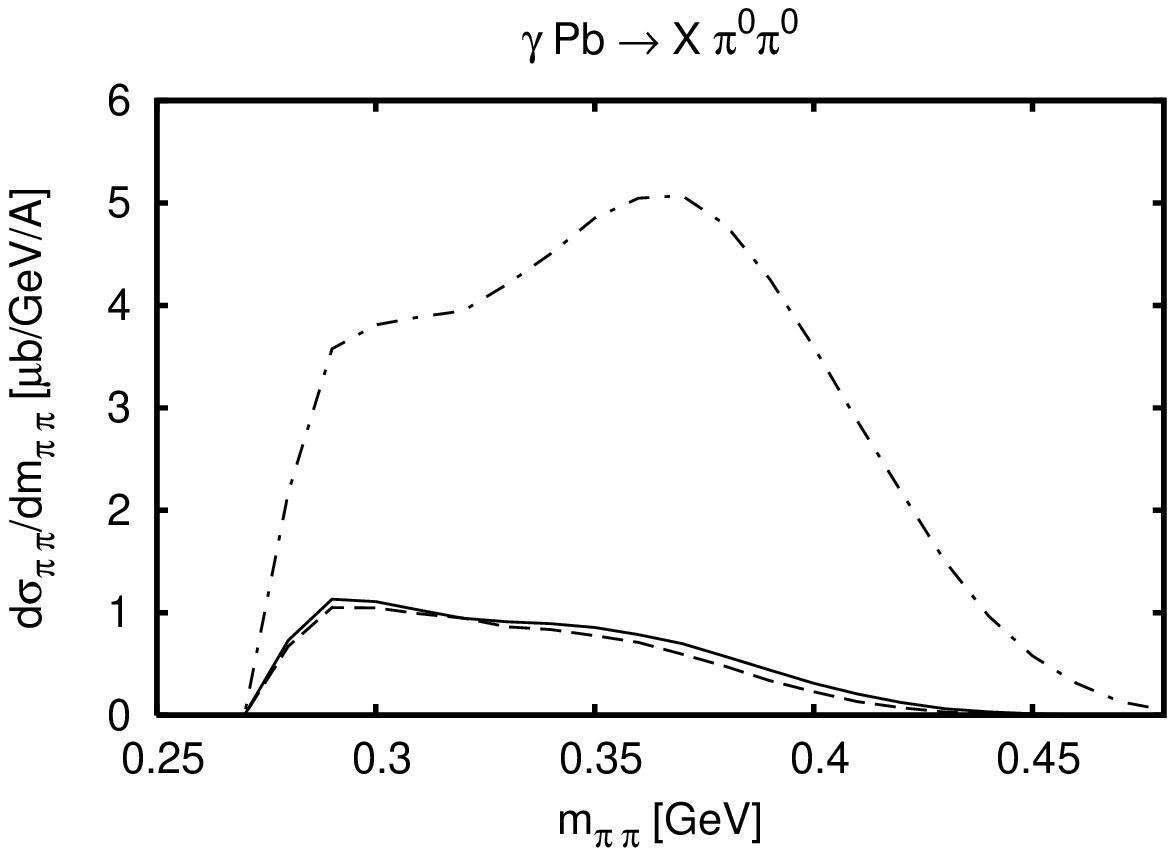}
\includegraphics[width=7cm]{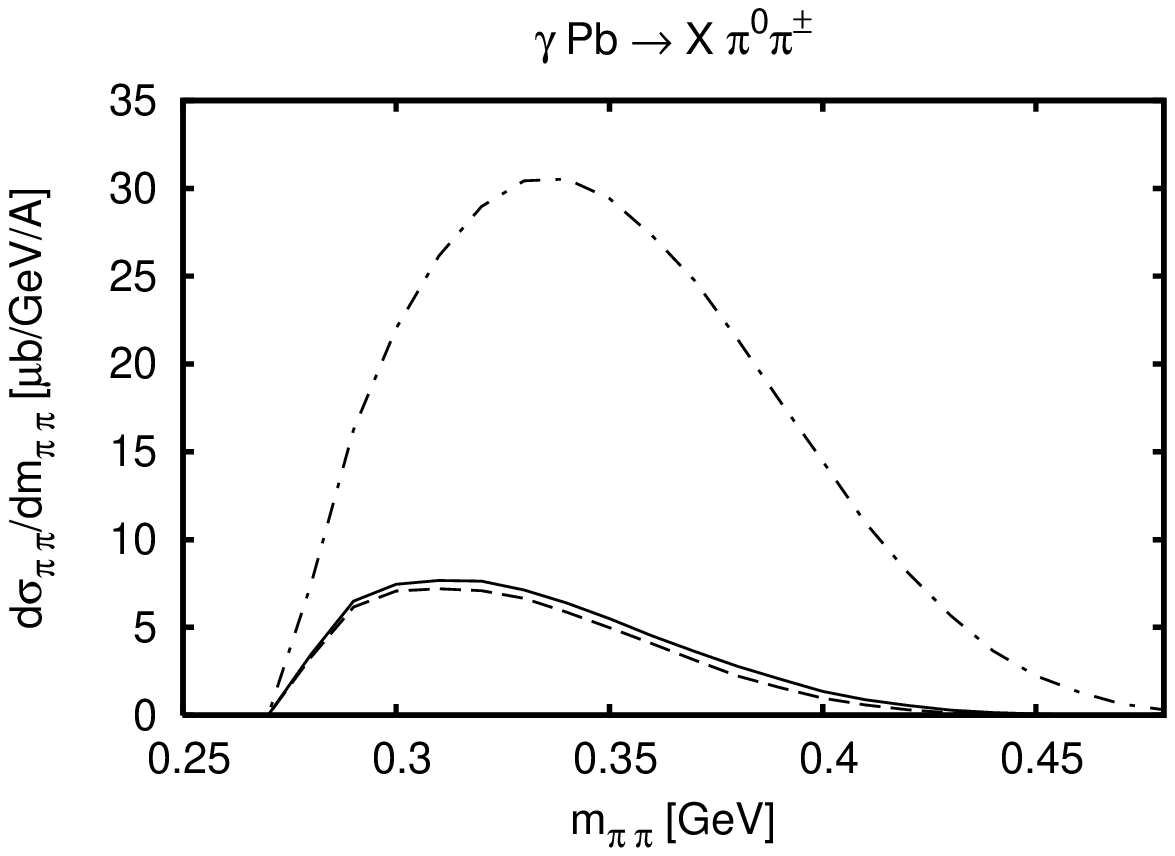}
\caption{Two pion invariant mass distributions for \pn and \pc photoproduction off $^{12}$C, $^{40}$Ca and $^{208}$Pb for $E_{\gamma}=0.4-0.5 \GeV$. The uncharged \pn channel is plotted on the left, the charged channel \pc is presented on the right hand side. We show results without final state interactions (dashed dotted) and results with FSI and respectively with (dashed) and without (solid) hadronic potential for the pion. }
\label{twoPi_kern_500}
\end{center}
\end{figure}
Our results for $\pi\pi$ photoproduction off \cor{C, Ca and Pb }nuclei are presented in \fig{\ref{twoPi_kern}} and 
\fig{\ref{twoPi_kern_500}}\red{; the two shown energy bins are the ones which are presently 
considered in the ongoing analysis by the TAPS group~\cite{Schadmand:2005xy,Schadmand:2005ji,Messch}}. As already discussed in 
\cite{Muhlich:2004zj}, we observe that absorption, 
elastic scattering and charge exchange processes cause a considerable change of the spectra with the peak of the mass distribution moving to lower masses due to rescattering.

\begin{figure}
\begin{center}
\includegraphics[width=7cm]{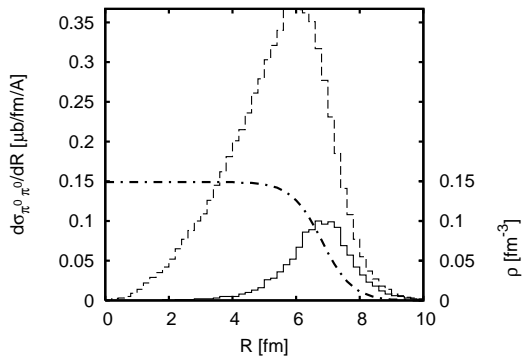}
\caption{Calculation without hadronic potential for the pion for $E_{\gamma}=500\; \MeV$ off Pb. The cross section for $\pi^0 \pi^0$ production is shown with FSI (solid steps) and without FSI (dashed steps) as function of the original production point $R$ of the pion pair. For illustration we show the density $\rho$ of the Pb nucleus (dashed dotted curve).}
\label{two_pi_radius}
\end{center}
\end{figure}

The hadronic potential for the pion causes only a minor effect, so the observed reaction is rather insensitive to this aspect of the model. This can be understood by analysing the production points of those pion pairs which are not absorbed and which are, finally, observed. In \fig{\ref{two_pi_radius}} the cross section for $\pi^0\pi^0$ production off Pb at $500 \MeV$ is shown as function of the production point $R$ of the pair. Without FSI the distribution $d\sigma_{\pi^0\pi^0}/dR$ is proportional to $\rho(R) R^2$; including FSI the distribution is shifted and centered around $6.8\,\mathrm{fm}$ which corresponds to roughly $\rho=0.076\, \mathrm{fm}^{-3}$. Hence, the potential has a minor effect on the observed pions due to the low density at the initial production point. 

Since the mean free path increases at lower momentum, a cut that keeps only low-momentum pions could reduce the sensitivity of the whole production process to the $\pi$-nucleus FSI. On the other hand, this would complicate the theoretical treatment of the process considerably, since quantum mechanical effects would become even more important.

First experimental results for this process have been presented in \cite{Messch}. Recently, those data have been reanalyzed by the TAPS collaboration. The preliminary results of this analysis can be found in \cite{Schadmand:2005xy,Schadmand:2005ji}, and a comparison of the former results to this new analysis is given in \cite{SchadmandHabil}. The old analysis is being re-evaluated, especially in the semi-charged channel. Therefore, we do not present these data sets here. At the moment, new sets of data, taken with a $4\pi$ setup with Crystal Ball and TAPS at MAMI, are being analyzed. With these a meaningful comparison with experiment will be possible.

\begin{figure}
\begin{center}
\includegraphics[width=7cm]{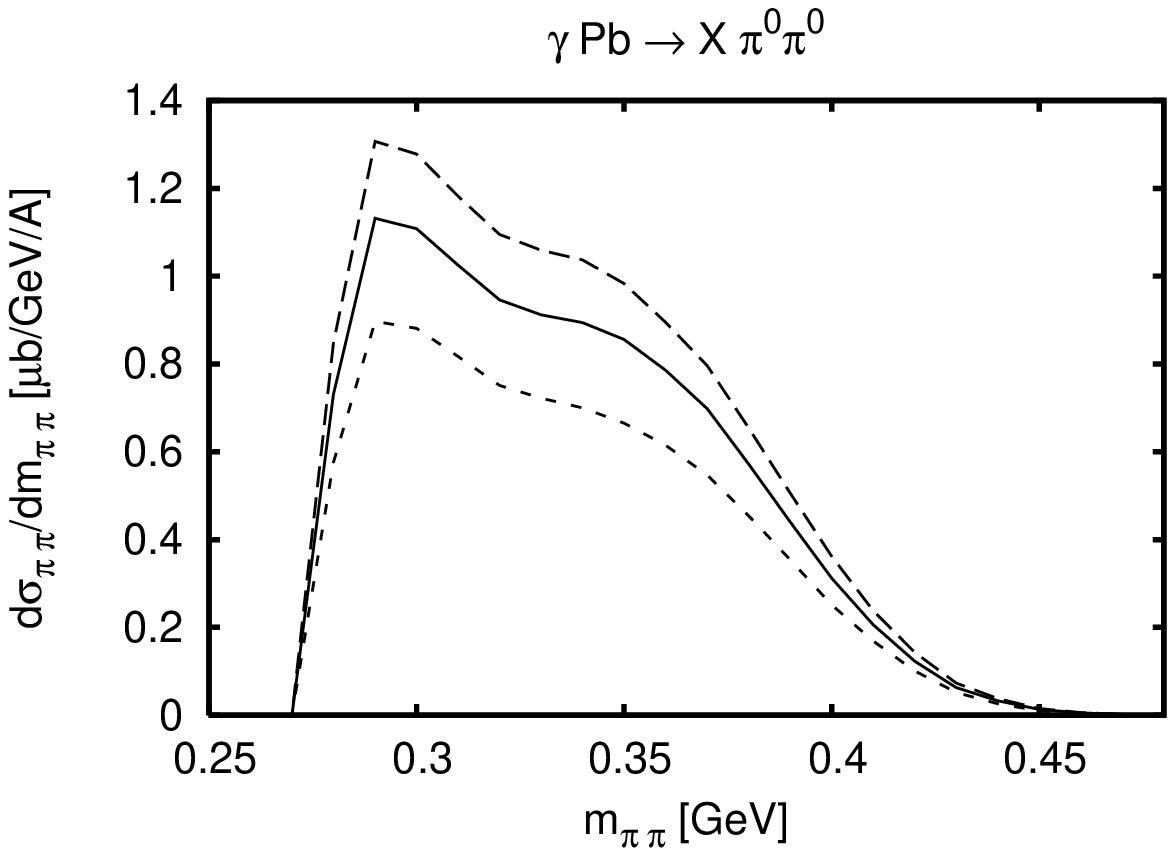}
\includegraphics[width=7cm]{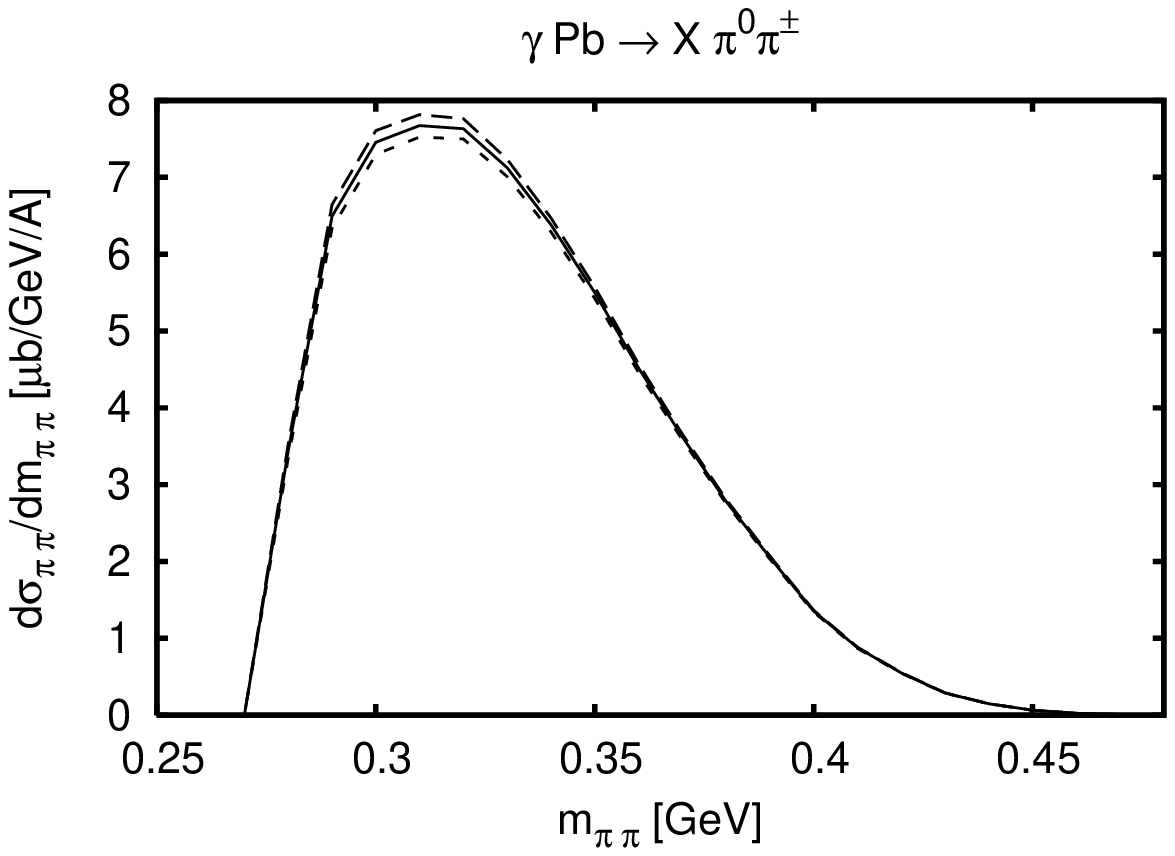}
\caption{Illustration of the uncertainty due to the elementary $\pi \pi$ production process. We show standard calculations without hadronic potential for the pion for $E_{\gamma}=400-500\; \MeV$. Only the cross sections used as input for $\gamma N\rightarrow N \pi \pi$ are varied between experimental data (solid lines), experimental upper (long dashed lines) and experimental lower bounds (short dashed lines) of all cross sections.}
\label{two_pi_error}
\end{center}
\end{figure}

\cor{The main} theoretical uncertainties in our calculation stem from two different sources - pion reaction mechanism and initial production cross section. \cor{The pion FSI reaction mechanism incorporated in the model and its uncertainties have been discussed in the preceding sections.}\del{was evaluated by means of comparison of the mean free path and the absorption cross sections, still the theoretical uncertainty is highly non-negligible due to the semi-classical treatment. } The influence of the uncertainty in the elementary cross sections is illustrated in \fig{\ref{two_pi_error}}. The three curves \cor{shown} in each plot use as input either the given data points or, simultaneously for all data sets, the lower and upper bounds.\del{ To evaluate the bounds we utilized the errorbars (systematic and statistical errors linearly summed up) reported by the experiments.} \cor{One has to note, that $\gamma n \rightarrow p \pi^- \pi^0$, $\gamma p \rightarrow p \pi^+ \pi^-$ and especially $\gamma n\rightarrow n \pi^0 \pi^0 $ \cite{Kleber:2000qs} are sources of uncertainty in the elementary cross sections. We emphasize, that this scenario of simultaneous lower and upper bounds instead of a computational intensive $\chi$-squared-analysis is only meant to give an estimate for the error.} The statistical error due to the Monte-Carlo method was estimated to be about $2\%$ by comparing parallel runs, being therefore negligible.

To summarize this section we emphasize that final state interactions of the pions are strong and tend to shift the maximum of the $\pi\pi$ mass distribution in all channels towards lower masses. This effect considerably complicates drawing a link between the experimental data and a possible softening of the in-medium $I=0$ channel. 
Any theory aiming to describe the observed effect on the basis of a  partial \red{chiral symmetry restoration} or an in-medium modification of the $\pi\pi$ production process must take the final state effects into account.

\section{Summary}
 In this work we have presented the treatment of low-energy pions within the GiBUU transport model. In particular, we have studied the mean free path and compared to optical model results. Studying the long wave-length region, we have pointed out the problem with pion energies below $20 \MeV$ due to the semi-classical treatment in our model. As a benchmark, we have investigated pion absorption cross sections with very good agreement to experimental data from light to heavy nuclei. At low energies, we have found a strong dependence of this observable on the Coulomb potential and the pion hadronic potential. Finally, we have shown predictions for $\gamma A\to \pi \pi A$ in the regime of $400\MeV\leq E_{\gamma} \leq 500 \MeV$ and also new calculations for $400\MeV\leq E_{\gamma} \leq 460 \MeV$. We emphasize, that the link from $\pi\pi$ photoproduction experiments to underlying effects like a possible softening of the in-medium $\pi\pi$ interaction is considerably complicated by the necessary presence of pion final state interactions which also shifts the $\pi\pi$ invariant mass distributions towards lower masses.

\section{Acknowledgments}
\begin{acknowledgments}
One of the authors (OB) is very grateful to R. Shyam, Saha Institute of Nuclear Physics, for many helpful discussions and support. We thank the GiBUU group, especially K. Gallmeister, A. Larionov and T. Falter. Thanks to M. Post and J. Lehr for fruitful discussions.

This work was supported by Deutsche Forschungsgemeinschaft (DFG). One of us, LAR, has been supported by the Alexander von Humboldt Foundation. 
\end{acknowledgments}

\bibliography{literatur}

\begin{thebibliography}{10}

\bibitem{Drukarev:1988kd}
E.~G. Drukarev and E.~M. Levin,
\newblock Nucl. Phys. {\bf A511}, 679 (1990).

\bibitem{Cohen:1991nk}
T.~D. Cohen, R.~J. Furnstahl and D.~K. Griegel,
\newblock Phys. Rev. {\bf C45}, 1881 (1992).

\bibitem{Brockmann:1996iv}
R.~Brockmann and W.~Weise,
\newblock Phys. Lett. {\bf B367}, 40 (1996).

\bibitem{Bernard:1987im}
V.~Bernard, U.~G. Meissner and I.~Zahed,
\newblock Phys. Rev. Lett. {\bf 59}, 966 (1987).

\bibitem{Hatsuda:1999kd}
T.~Hatsuda, T.~Kunihiro and H.~Shimizu,
\newblock Phys. Rev. Lett. {\bf 82}, 2840 (1999).

\bibitem{PDGdata}
K.~Hagiwara {\em et~al.},
\newblock {Phys. Rev. D} {\bf 66}, 010001+ (2002).

\bibitem{Bonutti1}
F.~Bonutti {\em et~al.},
\newblock Phys. Rev. Lett. {\bf 77}, 603 (1996).

\bibitem{Bonutti2}
F.~Bonutti {\em et~al.},
\newblock Nucl. Phys. {\bf A677}, 213 (2000).

\bibitem{Messch}
J.~G. Messchendorp {\em et~al.},
\newblock Phys. Rev. Lett. {\bf 89}, 222302 (2002), [nucl-ex/0205009].

\bibitem{SchadmandHabil}
S.~Schadmand,
\newblock Photoproduction from nuclei in the resonance region,
\newblock Habilitationsschrift, 2005,
\newblock http://www.fz-juelich.de/ikp/copolt/susan/opusii.pdf.

\bibitem{Schadmand:2005ji}
S.~Schadmand,
\newblock nucl-ex/0505023.

\bibitem{Schadmand:2005xy}
S.~Schadmand,
\newblock nucl-ex/0504012.

\bibitem{Muhlich:2004zj}
P.~M\"uhlich, L.~Alvarez-Ruso, O.~Buss and U.~Mosel,
\newblock Phys. Lett. {\bf B595}, 216 (2004), [nucl-th/0401042].

\bibitem{osetSimulation}
L.~L. Salcedo, E.~Oset, M.~J. Vicente-Vacas and C.~Garcia-Recio,
\newblock Nucl. Phys. {\bf A484}, 557 (1988).

\bibitem{Engel:1993jh}
A.~Engel, W.~Cassing, U.~Mosel, M.~Schafer and G.~Wolf,
\newblock Nucl. Phys. {\bf A572}, 657 (1994), [nucl-th/9307008].

\bibitem{kadanoffBaym}
L.~Kadanoff and G.~Baym,
\newblock {\em Quantum Statistical Mechanics} (Addison Wesley Publishing
  Company, 1994).

\bibitem{Effenberger:1996rc}
M.~Effenberger, A.~Hombach, S.~Teis and U.~Mosel,
\newblock Nucl. Phys. {\bf A614}, 501 (1997), [nucl-th/9610022].

\bibitem{Teis:1996kx}
S.~Teis, W.~Cassing, M.~Effenberger, A.~Hombach, U.~Mosel and G.~Wolf,
\newblock Z. Phys. {\bf A356}, 421 (1997), [nucl-th/9609009].

\bibitem{GiBUUWebpage}
The {GiBUU} website,
\newblock http://theorie.physik.uni-giessen.de/GiBUU.

\bibitem{diplom}
O.~Buss,
\newblock {\em Low-energy pions in a Boltzmann-Uehling-Uhlenbeck transport
  simulation} (Diploma thesis, Justus-Liebig-Universit\"at Giessen, 2004),
\newblock http://theorie.physik.uni-giessen.de/documents/diplom/buss.pdf.

\bibitem{F2003}
{JTC 1/SC 22/WG 5},
\newblock {\em {ISO/IEC 1539-1:2004}} (International Organization for
  Standardization, 2004).

\bibitem{subversion}
B.~Collins-Sussman, B.~W. Fitzpatrick and C.~M. Pilato,
\newblock {\em Version Control with Subversion} (O'Reilly, 2004).

\bibitem{ManleySaleski}
D.~M. Manley and E.~M. Saleski,
\newblock Phys. Rev. {\bf D45}, 4002 (1992).

\bibitem{effeDoktor}
M.~Effenberger,
\newblock {\em {Eigenschaften von Hadronen in einem vereinheitlichten
  Transportmodell}},
\newblock PhD thesis, JLU Giessen, Institut f\"ur theoretisches Physik I, 1999.

\bibitem{Carter:1971tj}
A.~A. Carter, J.~R. Williams, D.~V. Bugg, P.~J. Bussey and D.~R. Dance,
\newblock Nucl. Phys. {\bf B26}, 445 (1971).

\bibitem{Landoldt}

\newblock {Landoldt and Boernstein}, editor{\em New Series} Vol.~12 (Springer
  Verlag, Berlin, 1988).

\bibitem{Davidson:1972ky}
D.~Davidson, T.~Bowen, P.~K. Caldwell, E.~W. Jenkins, R.~M. Kalbach, D.~V.
  Petersen, A.~E. Pifer and R.~E. Rothschild,
\newblock Phys. Rev. {\bf D6}, 1199 (1972).

\bibitem{Kriss:1999cv}
B.~J. Kriss {\em et~al.},
\newblock Phys. Rev. {\bf C59}, 1480 (1999).

\bibitem{Andreev:1988tv}
V.~P. Andreev {\em et~al.},
\newblock Z. Phys. {\bf A329}, 371 (1988).

\bibitem{Daum:2001yh}
M.~Daum {\em et~al.},
\newblock Eur. Phys. J. {\bf C23}, 43 (2002), [nucl-ex/0108008].

\bibitem{Hardie:1997mg}
J.~G. Hardie {\em et~al.},
\newblock Phys. Rev. {\bf C56}, 20 (1997).

\bibitem{Tsuboyama:1988mq}
T.~Tsuboyama, N.~Katayama, F.~Sai and S.~S. Yamamoto,
\newblock Nucl. Phys. {\bf A486}, 669 (1988).

\bibitem{Shimizu:1982dx}
F.~Shimizu, Y.~Kubota, H.~Koiso, F.~Sai, S.~Sakamoto and S.~S. Yamamoto,
\newblock Nucl. Phys. {\bf A386}, 571 (1982).

\bibitem{Bondar:1995zv}
A.~Bondar {\em et~al.},
\newblock Phys. Lett. {\bf B356}, 8 (1995).

\bibitem{ericsonWeise}
T.~Ericson and W.~Weise,
\newblock {\em Pions and nuclei} (Clarendon Press,Oxford, 1988).

\bibitem{Peters:1998mb}
W.~Peters, H.~Lenske and U.~Mosel,
\newblock Nucl. Phys. {\bf A640}, 89 (1998), [nucl-th/9803009].

\bibitem{osetSpreading}
E.~Oset and L.~L. Salcedo,
\newblock Nucl. Phys. {\bf A468}, 631 (1987).

\bibitem{osetlow}
J.~Nieves, E.~Oset and C.~Garcia-Recio,
\newblock Nucl. Phys. {\bf A554}, 554 (1993).

\bibitem{Geissel:2002ur}
H.~Geissel {\em et~al.},
\newblock Phys. Rev. Lett. {\bf 88}, 122301 (2002).

\bibitem{Kohl:2001fx}
A1, M.~Kohl {\em et~al.},
\newblock Phys. Lett. {\bf B530}, 67 (2002), [nucl-ex/0104004].

\bibitem{MehremRadi}
R.~A. Mehrem, H.~M.~A. Radi and J.~O. Rasmussen,
\newblock Phys. Rev. {\bf C30}, 301 (1984).

\bibitem{Hecking}
P.~Hecking,
\newblock Phys. Lett. {\bf B103}, 401 (1981).

\bibitem{Cassing}
W.~Cassing, V.~Metag, U.~Mosel and K.~Niita,
\newblock Phys. Rep. {\bf 188} (1990).

\bibitem{stricker}
K.~Stricker, H.~McManus and J.~A. Carr,
\newblock Phys. Rev. {\bf C19}, 929 (1979).

\bibitem{ChaiRiska}
J.~Chai and D.~O. Riska,
\newblock Nucl. Phys. {\bf A329}, 429 (1979).

\bibitem{Sadler:2004yq}
Crystal Ball, M.~E. Sadler {\em et~al.},
\newblock Phys. Rev. {\bf C69}, 055206 (2004), [nucl-ex/0403040].

\bibitem{ashery}
D.~Ashery, I.~Navon, G.~Azuelos, H.~K. Walter, H.~J. Pfeiffer and F.~W.
  Schlep\"utz,
\newblock Phys. Rev. {\bf C23}, 2173 (1981).

\bibitem{Friedman:1991it}
E.~Friedman, A.~Goldring, R.~R. Johnson, O.~Meirav, D.~Vetterli, P.~Weber and
  A.~Altman,
\newblock Phys. Lett. {\bf B257}, 17 (1991).

\bibitem{nakai}
K.~Nakai, T.~Kobayashi, T.~Numao, T.~A. Shibata, J.~Chiba and K.~Masutani,
\newblock Phys. Rev. Lett. {\bf 44}, 1446 (1980).

\bibitem{byfield}
H.~Byfield {\em et~al.},
\newblock Phys. Rev. {\bf 86}, 17 (1952).

\bibitem{Carroll:1976hj}
A.~S. Carroll {\em et~al.},
\newblock Phys. Rev. {\bf C14}, 635 (1976).

\bibitem{Clough:1974qt}
A.~S. Clough {\em et~al.},
\newblock Nucl. Phys. {\bf B76}, 15 (1974).

\bibitem{Wilkin:1973xd}
C.~Wilkin {\em et~al.},
\newblock Nucl. Phys. {\bf B62}, 61 (1973).

\bibitem{osetPionicAtoms}
J.~Nieves, E.~Oset and C.~Garcia-Recio,
\newblock Nucl. Phys. {\bf A554}, 509 (1993).

\bibitem{twoPi23}
J.~C. Nacher, E.~Oset, M.~J. Vicente and L.~Roca,
\newblock Nucl. Phys. {\bf A695}, 295 (2001), [nucl-th/0012065].

\bibitem{twoPi24}
J.~A.~G. Tejedor and E.~Oset,
\newblock Nucl. Phys. {\bf A600}, 413 (1996).

\bibitem{Roca:2002vd}
L.~Roca, E.~Oset and M.~J. Vicente~Vacas,
\newblock Phys. Lett. {\bf B541}, 77 (2002), [nucl-th/0201054].

\bibitem{Zabrodin:1997xd}
A.~Zabrodin {\em et~al.},
\newblock Phys. Rev. {\bf C55}, 1617 (1997).

\bibitem{Braghieri:1994rf}
A.~Braghieri {\em et~al.},
\newblock Phys. Lett. {\bf B363}, 46 (1995).

\bibitem{Kleber:2000qs}
V.~Kleber {\em et~al.},
\newblock Eur. Phys. J. {\bf A9}, 1 (2000).

\bibitem{Wolf:2000qt}
M.~Wolf {\em et~al.},
\newblock Eur. Phys. J. {\bf A9}, 5 (2000).

\bibitem{Kotulla:2003cx}
M.~Kotulla {\em et~al.},
\newblock Phys. Lett. {\bf B578}, 63 (2004), [nucl-ex/0310031].

\end{thebibliography}

\end{document}